\documentclass[notitlepage,aps,floats,prd,preprintnumbers,nofootinbib,superscriptaddress,longbibliography, eqsecnum]{revtex4-1}

\usepackage{amssymb}
\usepackage{amsmath}

\usepackage{graphicx}
\usepackage{graphics}
\usepackage{dcolumn}
\usepackage[dvipsnames]{xcolor}
\usepackage{fancyhdr} 
\usepackage{graphicx}
\usepackage{float}
\usepackage{multirow}
\usepackage{cases}

\usepackage{subfig}
\usepackage[colorlinks=true,linktocpage=true,urlcolor=blue,citecolor=blue,linkcolor=blue]{hyperref}

\usepackage[utf8]{inputenc}

\usepackage{dsfont}

\usepackage[justification=centerlast, format=plain, labelfont=bf]{caption}

\makeatletter
\renewcommand\tableofcontents{%
    \@starttoc{toc}%
}
\makeatother

\DeclareCaptionJustification{justified}{\leftskip=0pt \rightskip=0pt \parfillskip=0pt plus 1fil}

    \newcommand{\be}{\begin{equation}}
  \newcommand{\ee}{\end{equation}}
    \newcommand{\ba}{\begin{align}}
  \newcommand{\ea}{\end{align}}

\newcommand{\Msun}{M_{\odot}}

\makeatletter
\def\doauthor#1#2#3{%
  \ignorespaces#1\unskip
  \begingroup
   #3%
  \@if@empty{#2}{\@listcomma\endgroup{}{}}{\endgroup{\comma@space}{}\frontmatter@footnote{#2}}%
  \space \@listand
}%
\makeatother

\makeatletter
\def\@ssect@ltx#1#2#3#4#5#6[#7]#8{%
  \def\H@svsec{\phantomsection}%
  \@tempskipa #5\relax
  \@ifdim{\@tempskipa>\z@}{%
    \begingroup
      \interlinepenalty \@M
      #6{%
       \@ifundefined{@hangfroms@#1}{\@hang@froms}{\csname @hangfroms@#1\endcsname}%
       {\hskip#3\relax\H@svsec}{#8}%
      }%
      \@@par
    \endgroup
    \@ifundefined{#1smark}{\@gobble}{\csname #1smark\endcsname}{#7}%
  }{%
    \def\@svsechd{%
      #6{%
       \@ifundefined{@runin@tos@#1}{\@runin@tos}{\csname @runin@tos@#1\endcsname}%
       {\hskip#3\relax\H@svsec}{#8}%
      }%
      \@ifundefined{#1smark}{\@gobble}{\csname #1smark\endcsname}{#7}%
      \addcontentsline{toc}{#1}{\protect\numberline{}#8}%
    }%
  }%
  \@xsect{#5}%
}%
\makeatother

\begin{document}

\preprint{KCL-2021-74}

\title{GALLUMI: A Galaxy Luminosity Function Pipeline for Cosmology and Astrophysics}

\author{Nashwan Sabti$^{\mathds{S},}$}
\affiliation{Department of Physics, King's College London, Strand, London WC2R 2LS, UK}
\author{Julian B. Mu\~{n}oz$^{\mathds{M},}$}
\affiliation{Harvard-Smithsonian Center for Astrophysics, Cambridge, MA 02138, USA}
\author{Diego Blas$^{\mathds{B},}$}
\affiliation{Grup de F\'isica Te\`orica, 
Departament  de  F\'isica, Universitat  Aut\`onoma  de  Barcelona,   Bellaterra, 08193 Barcelona, Spain}
\affiliation{Institut de Fisica d’Altes Energies (IFAE), The Barcelona Institute of Science and Technology,\\ Campus UAB, 08193 Bellaterra  (Barcelona), Spain}

\def\thefootnote{$\mathds{S}$\hspace{0.7pt}}\footnotetext{\href{mailto:nashwan.sabti@kcl.ac.uk}{nashwan.sabti@kcl.ac.uk}}
\def\thefootnote{$\mathds{M}$\hspace{-0.9pt}}\footnotetext{\href{mailto:julianmunoz@cfa.harvard.edu}{julianmunoz@cfa.harvard.edu}}
\def\thefootnote{$\mathds{B}$}\footnotetext{\href{mailto:dblas@ifae.es}{dblas@ifae.es}}
\setcounter{footnote}{0}
\def\thefootnote{\arabic{footnote}}

\begin{abstract}
    \noindent Observations of high-redshift galaxies have provided us with a rich tool to study the physics at play during the epoch of reionisation. The luminosity function (LF) of these objects is an indirect tracer of the complex processes that govern galaxy formation, including those of the first dark-matter structures. In this work, we present an extensive analysis of the UV galaxy LF at high redshifts to extract cosmological and astrophysical parameters. We provide a number of phenomenological approaches in modelling the UV LF and take into account various sources of uncertainties and systematics in our analysis, including cosmic variance, dust extinction, scattering in the halo-galaxy connection, and the Alcock-Paczy\'{n}ski effect. Using UV LF measurements from the Hubble Space Telescope together with external data on the matter density, we derive the large-scale matter clustering amplitude to be $\sigma_8=0.76^{+0.12}_{-0.14}$, after marginalising over the unknown astrophysical parameters. We find that with current data this result is only weakly sensitive to our choice of astrophysical modelling, as well as the calibration of the underlying halo mass function. As a cross check, we run our analysis pipeline with mock data from the IllustrisTNG hydrodynamical simulations and find consistent results with their input cosmology. In addition, we perform a simple forecast for future space telescopes, where an improvement of roughly 30\% upon our current result is expected. Finally, we obtain constraints on astrophysical parameters and the halo-galaxy connection for the models considered here. All methods discussed in this work are implemented in the form of a versatile likelihood code, \texttt{GALLUMI}, which we make public.
\\\\
{\centering\noindent \href{https://github.com/NNSSA/GALLUMI_public}{\raisebox{-1pt}{\includegraphics[width=9pt]{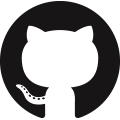}}}\hspace{2pt} The \texttt{GALLUMI} code and scripts for making the plots in this paper can be found \href{https://github.com/NNSSA/GALLUMI_public}{here}.}
\end{abstract}
\enlargethispage{1cm}
\maketitle

\vspace{-0.4cm}

{\linespread{0.99}\small\tableofcontents}

\newpage

%%%%%%%%%%%%%%%%%%%%%%%%%%%%%%%%%%
\section{Introduction}
\label{sec:introduction}
%%%%%%%%%%%%%%%%%%%%%%%%%%%%%%%%%%
A promising avenue in observational cosmology today involves the exploration of the epochs of cosmic dawn and reionisation, which broadly cover the redshift range $30 \gtrsim z \gtrsim 5$. The Universe found itself in an interesting state during this time: the Cosmic Microwave Background (CMB) photons had long decoupled from the baryons, allowing the latter to fall into the dark-matter (DM) potential wells that had been building up for about a hundred million years. It was during this era that baryons could cool down enough to collapse inside DM mini-halos and the first stars formed~\cite{Couchman:1986en, Tegmark:1996yt}. As the growth of DM structure continued over several stellar lifetime cycles, the mass of star-forming regions followed suit, which eventually lead to the formation of larger galaxies whose emission reionised the hydrogen in the intergalactic medium. With the advent of modern galaxy surveys, we are about to enter a novel era of cosmology where these first few generations of galaxies could be directly observed~\cite{Gardner:2006ky, Bromm:2011cw, Finkelstein2016}. These objects hold a wealth of information about astrophysics and cosmology during a relatively uncharted epoch. \\

The extraction of such information requires understanding galaxy formation and evolution at high redshifts.
While many questions remain, great progress has been made thanks to a decades-long observational campaign with the Hubble Space Telescope (HST) that has culminated in a number of catalogues spanning the redshift range $z = 4-10$ and which has led to the discovery of tens of thousands of UV-bright galaxies, see e.g.~\cite{Bouwens:2014fua,Finkelstein_2015,Atek:2015axa,Livermore:2016mbs,Bouwens_2017asdasd,Mehta_2017,Ishigaki_2018,Oesch_2018,Atek:2018nsc,Rojas_Ruiz_2020, Bouwens_2021}. 
The key observable we will focus on is the UV galaxy luminosity function (UV LF), which follows the abundance of galaxies as a function of their UV luminosity (or magnitude). This quantity depends on the complex physics that governs galaxy formation and therefore directly captures the interplay between a number of cosmological and astrophysical effects~\cite{Dayal:2018hft}. In particular,  galaxies are formed inside DM halos, which means that their abundance depends on the halo mass function and thus the physics of DM structure formation~\cite{Berlind:2002rn}. One of the main challenges to unearth this information involves the poorly understood connection between (invisible) DM halos and (visible) galaxies, which is highly dependent on the astrophysics at play~\cite{Wechsler:2018pic}. For example, one expects a smaller number of very bright galaxies, since the abundance of the heaviest DM halos is exponentially suppressed. Nevertheless, quasar feedback in high-mass galaxies is known to reduce the efficiency with which star formation occurs, thereby diminishing the abundance of the brightest galaxies as well~\cite{Weinberger:2017bbe}. 
Likewise, the faint end of the UV LF probes the smallest DM halos, where deviations from the standard cold-DM paradigm would leave the most striking impact on their abundance~\cite{Weinberg:2013aya, DelPopolo:2016emo,Bullock:2017xww, deMartino:2020gfi}. It is, however, also in this range where feedback mechanisms such as supernovae shocks and stellar winds are expected to reduce the brightness of low-mass galaxies~\cite{Kay:2001hq, Efstathiou:2000xp, Naab_2017}.\\

It is, therefore, crucial to separate the cosmological and astrophysical contributions to the UV LF. Several different approaches have been utilised in the literature to model the UV LF, ranging from analytic approaches~\cite{mashian2015empirical, Yung_2018, Park:2018ljd, Gillet:2019fjd}, to more complex models based on N-body simulations~\cite{Cai:2014fja, Sun_2016, Tacchella:2018qny, Behroozi:2019kql}, and full-scale hydrodynamical simulations~\cite{Salvaterra:2010nb, Jaacks, Dayal:2012ah,oshea2015probing, Vogelsberger_dust2020}. The uncertainty in the cosmology part is well-controlled, as the process of how DM halos collapse and form is reasonably well established, especially given the guidance from N-body simulations. In contrast, the baryonic physics is much harder to model, often necessitating  a trade-off between complexity and analytical understanding. Here we will follow a simple, semi-analytic method (SAM) in modelling the relationship between galaxies and their host halos. This has two advantages: firstly, a minimal set of assumptions provides a greater amount of flexibility to let the data guide our model, and secondly, SAMs can swiftly evaluate the UV LF without computationally expensive simulations. The main drawback of SAMs is that their functional forms are not derived \emph{ab initio}, and thus require simulations or data to be calibrated. Nevertheless, the two advantages outlined above allow for a quick way to use the UV LF to learn about cosmology as well as astrophysics. The UV LF has been used in previous works to measure the star-formation efficiency at early times~\cite{Sun_2016, Bouwens_2021}, improve our knowledge of reionisation~\cite{Naidu:2019gvi, Mason:2015cna}, and determine the clustering amplitude $\sigma_8$ in combination with other data sets~\cite{Sahlen:2021bqt}. Beyond the Standard Model, there exists a rich literature on exploiting the UV LF to study reionisation in alternative cosmologies~\cite{Romanello:2021gnp}, and to derive constraints on warm- and fuzzy-DM models~\cite{Bozek:2014uqa, Schultz:2014eia, Dayal:2014nva, Corasaniti:2016epp, Menci:2017nsr, Menci:2018lis, Rudakovskyi:2021jyf}, dynamical dark energy~\cite{Menci:2020ybl}, the primordial matter power spectrum~\cite{Yoshiura:2020soa}, as well as small-scale non-Gaussianities beyond the reach of the CMB~\cite{Chevallard:2014sxa, Sabti:2020ser}.\\

The goal of this work is to establish a versatile pipeline to model the UV LF, with the aim of extracting cosmological and astrophysical parameters in conjunction with other data sets. We do this by implementing our methods as a likelihood code, \texttt{GALLUMI}, in the publicly available MCMC sampler \texttt{MontePython}\footnote{\href{https://github.com/brinckmann/montepython_public}{https://github.com/brinckmann/montepython\_public}}~\cite{Audren:2012wb, Brinckmann:2018cvx} (which interfaces with the Boltzmann solver \texttt{CLASS}\footnote{\href{https://github.com/lesgourg/class_public}{https://github.com/lesgourg/class\_public}}~\cite{Lesgourgues:2011re,Blas:2011rf}). We describe the code in detail in the main text, but summarise some of its salient features below:
\begin{itemize}
    \item The code contains different astrophysical models for the halo-galaxy connection. Given the large uncertainties on the astrophysical conditions of the early Universe, we implement three different approaches to connect the magnitudes of galaxies to the mass of their host halo, and compare between them. This allows for a direct way to test the impact of a number of assumptions that enter our modelling.

    \item From the data side, we include the latest compilation of the UV LF obtained with the Hubble Space Telescope~\cite{Oesch_2018, Bouwens_2021}. Besides this real data, we also analyse data from the hydrodynamical simulation IllustrisTNG~\cite{Vogelsberger_dust2020}, as a way to check the accuracy of our parameter extraction technique and the consistency with real data.
    
    \item  We consider a number of corrections to the data that account for cosmic variance, the attenuation caused by dust and the Alcock-Paczy\'{n}ski effect. In addition, we allow for scatter in the halo-galaxy connection.

    \item  The modular design of our code makes it straightforward to add new features to it, some of which we have already included (e.g., different halo mass functions and dust calibrations).
\end{itemize}

As a demonstration, we perform several analyses to measure cosmological and astrophysical parameters, focusing in particular on the clustering amplitude $\sigma_8$. In our companion work~\cite{Sabti:2021unj}, we will use \texttt{GALLUMI} to measure the small-scale clustering of matter, highlighting the complementarity of the UV luminosity function with other cosmological probes.\\

The remainder of this paper is structured as follows:
In Sec.~\ref{sec:models}, we describe our fiducial model for the UV galaxy luminosity function. In Sec.~\ref{sec:data}, we give an overview of the data we use, together with an explanation of a number of corrections and errors that we take into account. We present our code and analysis pipeline in Sec.~\ref{sec:gallumi}, summarise our main results in Sec.~\ref{sec:results}, and study the dependence of our results on the astrophysical modelling in Sec.~\ref{sec:astro_modelling}. Finally, we conclude in Sec.~\ref{sec:conclusions}. Complementary results and checks are included in the Appendices~\ref{app:robustness_checks}$-$\ref{app:Hubble_TNG_data}. 

Throughout this paper, we will assume a flat $\Lambda$CDM cosmology and, unless otherwise stated, fix the total sum of neutrino masses to $\sum m_\nu = 0.06\,\mathrm{eV}$.

%%%%%%%%%%%%%%%%%%%%%%
% Formalism and Models
%%%%%%%%%%%%%%%%%%%%%%
\section{Formalism and Models}
\label{sec:models}

In this section, we will lay out a phenomenological description to model the UV luminosity function at high redshifts. An important part of this approach will involve the connection between the properties of galaxies and those of their host halo. In what follows, we will first detail the characteristics of our fiducial model for the halo-galaxy connection. Several alternative approaches are then provided in Sec.~\ref{sec:astro_modelling}, together with a discussion on their implications for cosmological-parameter extraction.\\

The UV LF is defined as the (comoving) number density of galaxies per unit magnitude: $\Phi_\mathrm{UV} = \mathrm{d}n_\mathrm{gal}/\mathrm{d}M_\mathrm{UV}$, where $M_\mathrm{UV}$ is the absolute UV magnitude of the galaxies (usually defined at 1500\,\r{A} in the rest frame~\cite{Williams_2018}).
From this definition, we can then decompose $\Phi_\mathrm{UV}$ as a product of three separate parts: \emph{1)} the halo mass function (HMF), \emph{2)} the halo occupation distribution (HOD) and \emph{3)} the halo-galaxy connection (the relation between halo mass and the magnitude of the galaxies residing inside the halo). The HOD specifies the average occupation number of galaxies inside each halo, which to good approximation can be assumed to be unity for the halos and redshifts considered in this work ($M_\mathrm{h}\sim 10^{10}-10^{12}\,M_\odot$ and $z=4-10$)~\cite{Bhowmick:2019nnj}. That is, there is only one central galaxy in each halo. With such a choice, $n_\mathrm{gal} = n_\mathrm{halo} \equiv n_\mathrm{h}$ and the UV LF can subsequently be written as\footnote{Our LF does not include a turn-over mass that accounts for the inefficiency of galaxies formed in small halos~\cite{Gillet:2019fjd}. This is justified for our Hubble Legacy Fields data (see Sec.~\ref{subsec:hst_observations}), as they cover halo masses roughly above $10^{10}\,M_\odot$ (see Fig.~\ref{fig:posteriors_sigmas_v1v2v3}), for which the turn-over is not relevant. Nevertheless, we plan to extend \texttt{GALLUMI} to include this term.}:
\begin{align}
    \label{eq:phi_UV}
    \Phi_\mathrm{UV} = \frac{\mathrm{d}n_\mathrm{h}}{\mathrm{d}M_\mathrm{h}}\times\frac{\mathrm{d}M_\mathrm{h}}{\mathrm{d}M_\mathrm{UV}}\ ,
\end{align}
where $\mathrm{d}n_\mathrm{h}/\mathrm{d}M_\mathrm{h}$ is the halo mass function. We will now discuss both components of $\Phi_\mathrm{UV}$ in detail.

\subsection{Halo Mass Function}
\label{subsec:HMF}

The halo mass function describes the mass distribution of dark-matter halos and contains information about the initial conditions of matter density perturbations in the early Universe and their subsequent evolution. We adopt the usual form for the HMF~\cite{Jenkins:2000bv}:
\vspace*{-0.1cm}
\begin{align}
    \label{eq:HMF}
    \frac{\mathrm{d}n_\mathrm{h}}{\mathrm{d}M_\mathrm{h}} = \frac{\overline{\rho}_\mathrm{m}}{M_\mathrm{h}}\frac{\mathrm{d}\ln\sigma_{M_\mathrm{h}}^{-1}}{\mathrm{d}M_\mathrm{h}}f(\sigma_{M_\mathrm{h}})\ ,
\end{align}
where $\overline{\rho}_\mathrm{m}$ is the average comoving matter energy density (including both dark matter and baryons) and $\sigma_{M_\mathrm{h}}$ is the root-mean-square of the density field smoothed over a mass scale $M_\mathrm{h}$:
\begin{align}
    \label{eq:sigmasq_M}
    \sigma^2_{M_\mathrm{h}}(z) &= \int\frac{\mathrm{d}^3k}{(2\pi)^3}W_{M_\mathrm{h}}^2(k)T^2_\zeta(k,z)P_\zeta(k)\ ,
\end{align}
with $W_{M_\mathrm{h}}$ a window function (real-space spherical top hat in our case), $T_\zeta$ the transfer function of the comoving curvature perturbation $\zeta$ and $P_\zeta$ its primordial power spectrum. This last quantity is given by:
\begin{align}
 P_\zeta(k) = \frac{2\pi^2}{k^3}A_\mathrm{s}\left(\frac{k}{k_\mathrm{pivot}}\right)^{n_\mathrm{s}-1}\ .
\end{align}
In this equation, $A_\mathrm{s}$ is the amplitude of the primordial power spectrum at the pivot scale $k_\mathrm{pivot}$ and $n_\mathrm{s}$ is the spectral tilt. The transfer function $T_\zeta$ can be easily obtained from a cosmological Boltzmann solver, such as \texttt{CLASS}~\cite{Lesgourgues:2011re,Blas:2011rf} or \texttt{CAMB}~\cite{Lewis:1999bs}. The function $f(\sigma_{M_\mathrm{h}})$ in Eq.~\eqref{eq:HMF} is the fraction of mass that has collapsed into halos per $\ln\sigma_{M_\mathrm{h}}^{-1}$ interval and integrates up to unity. Within the spherical-collapse formalism of~\cite{Press:1973iz}, the threshold above which matter density perturbations collapse is mass-independent, allowing for the function $f(\sigma_{M_\mathrm{h}})$ to be analytically derived~\cite{Press:1973iz, Bond:1990iw}. This results in a description that qualitatively agrees with what is found in N-body simulations, but shows significant quantitative differences otherwise (see e.g.~\cite{Jenkins:2000bv} and references therein). On the other hand, excursion-set approaches based on ellipsoidal collapse of dark matter halos, in which the critical density varies with the peak height (i.e., with a mass-dependent barrier height), are in better agreement with numerical simulations~\cite{Sheth:1999mn, Sheth:1999su, Jenkins:2000bv}. The function $f(\sigma_{M_\mathrm{h}})$ in this case takes a more general form and has a number of free parameters that are then calibrated to the results of numerical simulations. In our fiducial model, we will use the Sheth-Tormen mass function, given by~\cite{Sheth:2001dp}:
\begin{align}
    \label{eq:ST_HMF}
    f_\mathrm{ST}(\sigma_M) = & A_\mathrm{ST}\sqrt{\frac{2a_\mathrm{ST}}{\pi}}\left[1+\left(\frac{\sigma_{M_\mathrm{h}}^2}{a_\mathrm{ST}\delta_\mathrm{ST}^2}\right)^{p_\mathrm{ST}}\right]\frac{\delta_\mathrm{ST}}{\sigma_{M_\mathrm{h}}}\exp\left(-\frac{a_\mathrm{ST}\delta_\mathrm{ST}^2}{2\sigma_{M_\mathrm{h}}^2}\right)\ ,
\end{align}

\enlargethispage{0.2cm}
where $A_\mathrm{ST} = 0.3222$, $a_\mathrm{ST} = 0.707$, $p_\mathrm{ST} = 0.3$ and $\delta_\mathrm{ST} = 1.686$. Note that, while this HMF is calibrated at redshift $z = 0$, comparisons with simulations at higher redshifts have shown that the prescription above still allows for a good fit, see e.g.~\cite{Lukic:2007fc, Schneider:2014rda, Tacchella:2018qny}. Besides the Sheth-Tormen HMF, we will also consider the Reed HMF in a check to assess the impact of the HMF choice on our results (see App.~\ref{app:other_HMF}). This question has also interesting consequences for models of cosmic dawn, see e.g.~\cite{Mirocha:2020qto} for a discussion.

\subsection{Halo-Galaxy Connection}
\label{subsec:halo-galaxy}

A number of properties of galaxies are known to be correlated with those of their host halo~\cite{Wechsler:2018pic,behroozi2021observational, Zhang:2021aau}. For instance, heavier halos tend to accommodate more massive and luminous galaxies compared to lower-mass halos. Here we follow a simple phenomenological approach to relate the absolute magnitude of galaxies to the mass of their host halo.  We start with the assumption that the UV emission in high-redshift galaxies is dominated by young, massive stars. As a consequence, the UV luminosity of a galaxy is related to its star-formation rate (SFR)~\cite{Kennicutt1983,Salim:2007is}, which depends on the halo mass amongst other properties.
As we are only interested in the abundance of galaxies as a function of their luminosity (or host halo mass), we will assume that the dependence of the SFR on other galaxy properties mostly averages out (see also the end of Sec.~\ref{subsubsec:Mhalo_MUV_connection} for a discussion on this). Numerical simulations, together with observational data, show that the average SFR exhibits a peak for halos like that of the Milky Way, see e.g.~\cite{Moster_2018, Wechsler:2018pic, Behroozi:2019kql, Sun_2016} and references therein. Star formation tends to be less efficient in heavier halos, since those often host active galactic nuclei whose emission can significantly impact the surrounding environment~\cite{Fabian:2012xr}. A similar picture also applies to lower-mass halos, where various feedback processes, including supernovae shocks, also reduce the star-formation efficiency~\cite{Kay:2001hq}. A simple function that captures this property is a double-power law relation, either between the average star-formation rate $\dot{M_*}$ and the halo accretion rate $\dot{M_\mathrm{h}}$ \emph{or} between the average stellar mass $M_*$ and the halo mass $M_\mathrm{h}$. In this work, we will take an agnostic look at this choice and consider both cases through three models for the halo-galaxy connection. For the sake of clarity, in this section we will only describe the halo-galaxy connection used in our main analysis (which we refer to as \textbf{model I}) and elaborate on the other two models in Sec.~\ref{sec:astro_modelling}. 

\subsubsection{Relating Halo Masses to Magnitudes}
\label{subsubsec:Mhalo_MUV_connection}

\begin{figure}[h!]
    \centering
    \includegraphics[width=0.7\linewidth]{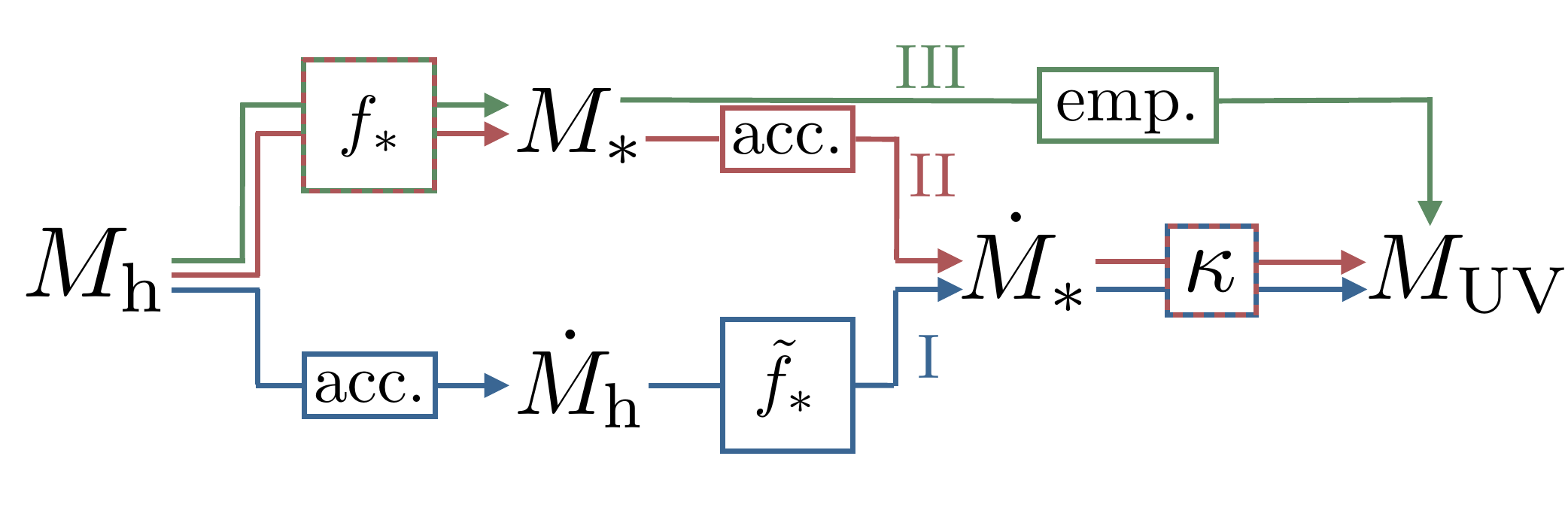}
    \caption{Schematic overview of our approaches to connect the UV magnitude $M_\mathrm{UV}$ of galaxies to the mass of their host halo $M_\mathrm{h}$. The blue, red and green lines correspond to models I$-$III, respectively, described in this section and  Sec.~\ref{sec:astro_modelling}. The quantities $\tilde{f}_*$ and $f_*$ are the double-power law relations in Eqs.~\eqref{eq:ftilde} and~\eqref{eq:fstar}. The boxes with `acc' stand for the accretion models in Eqs.~\eqref{eq:Mhdot_EPS} and~\eqref{eq:Mstar_Mstardot}, the box with `$\kappa$' indicates the relation between star-formation rate and UV luminosity in Eq.~\eqref{eq:Mstardot_LUV}, and the box with `emp' denotes the empirically determined connection between stellar mass and UV magnitude in model III. These different ways of linking $M_\mathrm{UV}$ to $M_\mathrm{h}$ allow us to test the robustness of our results against a number of assumptions in our modelling.
    }
    \label{fig:astro_diagram}
\end{figure}

In our fiducial model (\textbf{model I}), we relate the star- and halo-mass accretion rates via a double-power law:
\begin{align}
    \label{eq:ftilde}
    \widetilde{f_*} = \dfrac{\dot{M}_*}{\dot{M}_\mathrm{h}} = \dfrac{\epsilon_*}{\left(\dfrac{M_\mathrm{h}}{M_c}\right)^{\alpha_*}+\left(\dfrac{M_\mathrm{h}}{M_c}\right)^{\beta_*}}\ ,
\end{align}
where $\alpha_* \leq 0$, $\beta_* \geq 0$, $\epsilon_*\geq 0$ and $M_c \geq 0$ are free parameters which we will fit for with the data. The parameters $\alpha_*$ and $\beta_*$ regulate the slope of the faint and bright end of the UV LF, respectively, while $\epsilon_*$ controls its amplitude (i.e., the star-formation efficiency) and $M_c$ sets the mass at which the star-formation rate peaks. In principle, these astrophysical parameters can evolve with redshift -- we will discuss this matter below in Sec.~\ref{subsec:z_evol_astro}. The star-formation rate $\dot{M_*}$ of a galaxy determines its UV luminosity $L_\mathrm{UV}$ (or equivalently its UV magnitude $M_\mathrm{UV}$) through~\cite{Madau:1997pg,Kennicutt:1998zb}:
\begin{align}
    \label{eq:Mstardot_LUV}
    \dot{M}_* = \kappa_\mathrm{UV}L_\mathrm{UV}\ ,
\end{align}
where $\kappa_\mathrm{UV} = 1.15\times 10^{-28}\, M_\odot\mathrm{\,s\, erg}^{-1}\mathrm{yr}^{-1}$ is a conversion factor that is derived from a stellar synthesis population model with a Salpeter initial stellar mass function in the range of $0.1-100\,M_\odot$, a constant star-formation rate and an evolving stellar metallicity~\cite{Madau:2014bja} (see also~\cite{Furlanetto_2017} for a comment on this factor). The luminosity and absolute magnitude are related as usual within the AB magnitude system via~\cite{Oke:1983nt}:
\begin{align}
    \label{eq:LUV}
    \log_{10}\left(\frac{L_\mathrm{UV}}{\mathrm{erg \, s^{-1}}}\right) = 0.4\, (51.63 - M_\mathrm{UV})\ .
\end{align}
We account for dust extinction directly at the level of the data, which we will detail in Sec.~\ref{subsec:dust_corr}. As for $\dot{M}_\mathrm{h}$ in Eq.~\eqref{eq:ftilde}, we use the extended Press-Schechter formalism and obtain the accretion rate as~\cite{Neistein:2006ak,Correa:2014xma}:
\begin{align}
    \label{eq:Mhdot_EPS}
    \dot{M}_\mathrm{h} = -\sqrt{\frac{2}{\pi}}\frac{(1+z)H(z)M_\mathrm{h}}{\sqrt{\sigma_{M_\mathrm{h}}^2(Q) - \sigma_{M_\mathrm{h}}^2}}\frac{1.686}{D^2(z)}\frac{\mathrm{d}D(z)}{\mathrm{d}z}\ ,
\end{align}
with $D(z)$ the linear growth factor, $\sigma_{M_\mathrm{h}}^2$ the variance of the smoothed density field evaluated at redshift $z$ (see Eq.~\eqref{eq:sigmasq_M}) and $\sigma_{M_\mathrm{h}}^2(Q) = \sigma^2(M_\mathrm{h}/Q)$, where $Q$ is a free parameter that we fit for with the data using priors motivated by simulations (see Table~\ref{tab:priors}). This equation for $\dot{M}_\mathrm{h}$ provides an excellent agreement with results from N-body simulations for the mass range of interest in this work ($M_\mathrm{h} \sim 10^{10} - 10^{12}\, M_\odot$)~\cite{Correa:2014xma}. It describes an exponential mass accretion of dark-matter halos at high redshifts, which is then followed by a less efficient power-law accretion at low redshifts, when dark energy starts to dominate and accelerate the expansion of the Universe. As a final step, we plug Eq.~\eqref{eq:Mhdot_EPS} in Eq.~\eqref{eq:ftilde} and solve for $M_\mathrm{UV}$ as a function of $M_\mathrm{h}$. We provide a schematic overview of our three astrophysical models in Fig.~\ref{fig:astro_diagram}, and refer to Sec.~\ref{sec:astro_modelling} for a description of and comparison with models II and III.
\enlargethispage{0.5cm}

An important note that should be made here is that the above prescription is not applicable at the level of individual halos and galaxies, since it does not depend on their unique formation history. For example, baryonic processes, such as starbursts, can cause the $M_\mathrm{UV} - M_\mathrm{h}$ relation to exhibit scatter~\cite{White:1991mr}. Instead, the galaxy properties derived here only depend on the halo mass and, therefore, should be thought of as \emph{average} quantities. This then also means that there is some stochasticity in our predictions of the $M_\mathrm{UV} - M_\mathrm{h}$ relation that should be taken into account. An estimate of this error can be obtained from simulations that track the star formation and evolution within galaxies, see e.g.~\cite{Tacchella:2018qny} where an uncertainty of roughly $20-30\%$ is found. In our work, we will account for this stochasticity by assuming that the magnitudes of the galaxies that can be hosted by DM halos are distributed according to a Gaussian: 
\begin{align}
\label{eq:gaussian_MUV}
    P(M_\mathrm{UV}) = \frac{1}{\sqrt{2\pi}\sigma_{M_\mathrm{UV}}}\exp\left[-\frac{\left(M_\mathrm{UV} - \langle M_\mathrm{UV}\rangle\right)^2}{2\sigma_{M_\mathrm{UV}}^2}\right]\ ,
\end{align}
where $\langle M_\mathrm{UV}\rangle$ is the average prediction from our halo-galaxy prescription and $\sigma_{M_\mathrm{UV}}$ is a free parameter that we will marginalise over, see Section~\ref{sec:results} for more details. This latter quantity is degenerate with $\beta_*$ and ignoring it would therefore introduce a bias on $\beta_*$~\cite{ren2019}. We also make a comparison with the scatter found in the IllustrisTNG hydrodynamical simulations~\cite{Vogelsberger_dust2020} in App.~\ref{app:Hubble_TNG_data}.\\

\begin{figure}[h!]
    \centering
    \includegraphics[width=\linewidth]{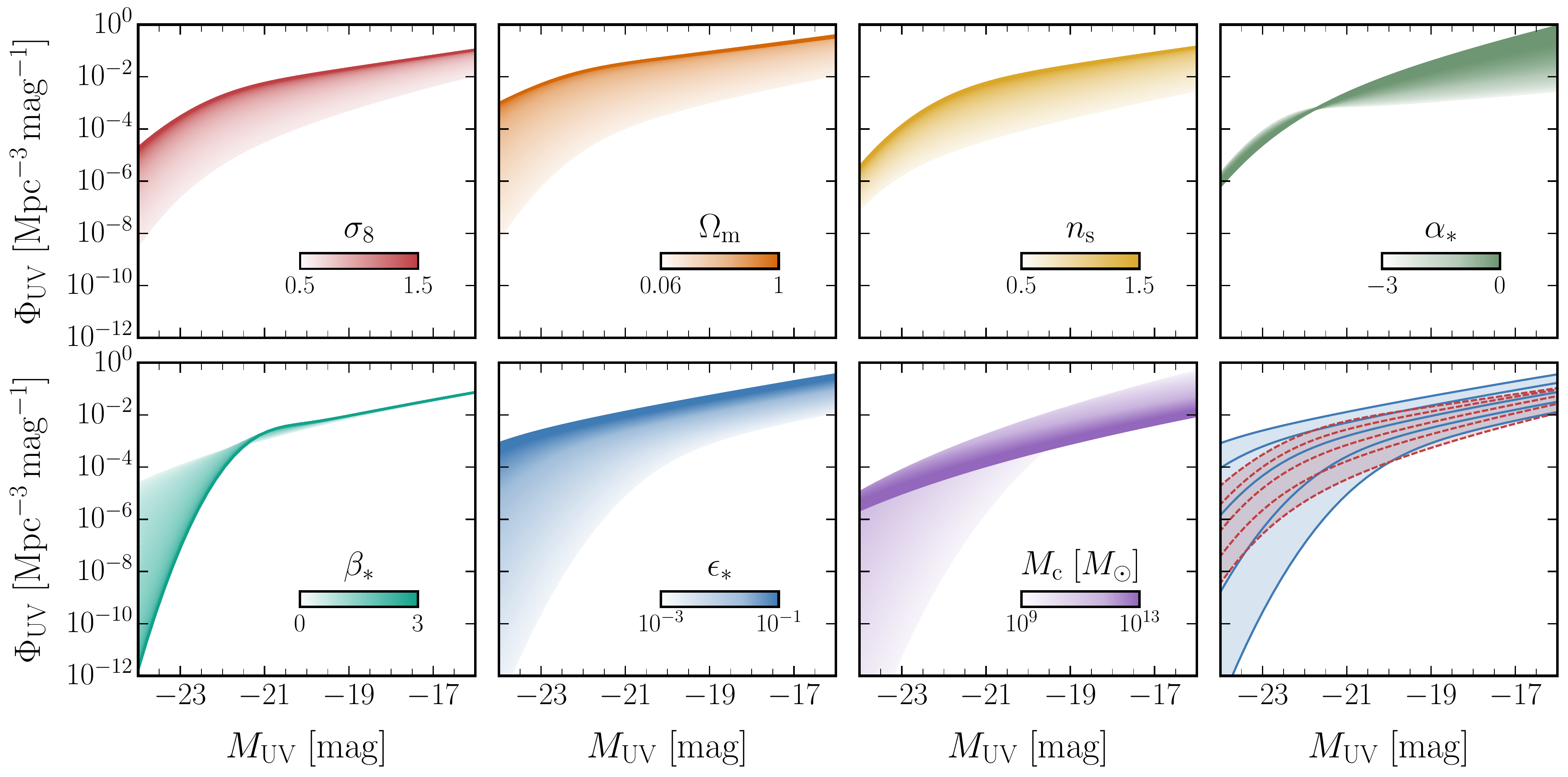}
    \caption{An illustration of the dependence of the UV galaxy luminosity function $\Phi_{\rm UV}$ on the different relevant parameters at $z=6$.
    The first three panels show the influence of the main cosmological parameters, which are the clustering amplitude $\sigma_8$, the matter-density parameter $\Omega_\mathrm{m}$ and the tilt of the primordial power spectrum $n_\mathrm{s}$. The next four display the effect of the astrophysical parameters $\{\alpha_*, \beta_*, \epsilon_*, M_c\}$ (see Eq.~\eqref{eq:ftilde} for their definitions). The panel on the lower right shows the $\sigma_8$ and $\epsilon_*$ curves overlaid, which highlights the strong -- but not exact -- degeneracy between these two parameters. 
    }
    \label{fig:UVLF_param_depend}
\end{figure}

After combining all ingredients together, we obtain the UV LF as a function of cosmological parameters (through the HMF) and astrophysical parameters (through the halo-galaxy connection). In Fig.~\ref{fig:UVLF_param_depend}, we show the parameter dependencies of the UV LF within our model I at $z=6$. In each panel, only one parameter is varied, while all the others are fixed. In the case of varying $\sigma_8$, this is equivalent to changing the amplitude $A_\mathrm{s}$ of the primordial power spectrum, which simply increases or decreases the amount of formed halos and thus also galaxies. 
The situation is similar when varying
$\Omega_\mathrm{m}$ and $n_\mathrm{s}$.
The UV LF increases with larger $n_\mathrm{s}$ due to the additional power at small scales. The rest of the panels highlight how feedback processes (through $\alpha_*$ and $\beta_*$), an altering star-formation efficiency (through $\epsilon_*$), and a different turn-over mass ($M_c$) change the UV LF. This figure also shows the strong degeneracies that exist between astrophysical and comological parameters at a single redshift (see lower-right panel). Nonetheless, the degeneracy direction at each redshift slice is slightly different, such that this problem can be alleviated by combining data at different redshifts. This degeneracy can also be further broken by assuming a parametric redshift evolution for the astrophysical parameters.

\enlargethispage{0.8cm}

\subsubsection{Redshift Evolution of Astrophysical Parameters}
\label{subsec:z_evol_astro}
The approach described above, in which we relate the magnitude of galaxies to the mass of their host halo, can be readily applied at any given redshift. However, since we have access to data over a wide range of redshifts, we are also interested in the time evolution of the halo-galaxy connection. Therefore, there still remains the question of how the four astrophysical parameters in Eq.~\eqref{eq:ftilde} evolve with redshift. A number of works in the literature have studied this question, where some find that the observed UV LF can be well described assuming that the astrophysical parameters are redshift-independent~\cite{Gillet:2019fjd, Mirocha:2020sid}. Other works allow (some of) the astrophysical parameters to evolve with redshift, for example via a power-law relation~\cite{Sahlen:2021bqt}, or obtain a $z$-evolution through abundance matching or similar techniques~\cite{Sun_2016, Furlanetto_2017}.\\

Rather than forcing a certain parametrisation for the redshift evolution, we will start with the most conservative approach and assign a set of $\{\alpha_*,\beta_*,\epsilon_*,M_c\}$ parameters at each redshift slice. Such a choice will then cover any possible redshift evolution, but comes with the downside of having to deal with an additional 28 free parameters in the analysis (4 at each redshift in the range $z=4-10$). While this does take into account our ignorance regarding the redshift behaviour, it may prove to be excessively conservative. As such, given the results from this conservative model, we can come up with a \emph{minimal} model that still fits the data well, but where the amount of free parameters is appreciably reduced. Besides making the analysis more computationally friendly, this will also signify the most optimistic case when deriving cosmological parameters. All other redshift parametrisations will then give a result somewhere between those of the conservative and minimal models. \\

\begin{figure}[h!]
    \centering
    \includegraphics[width=0.8\textwidth]{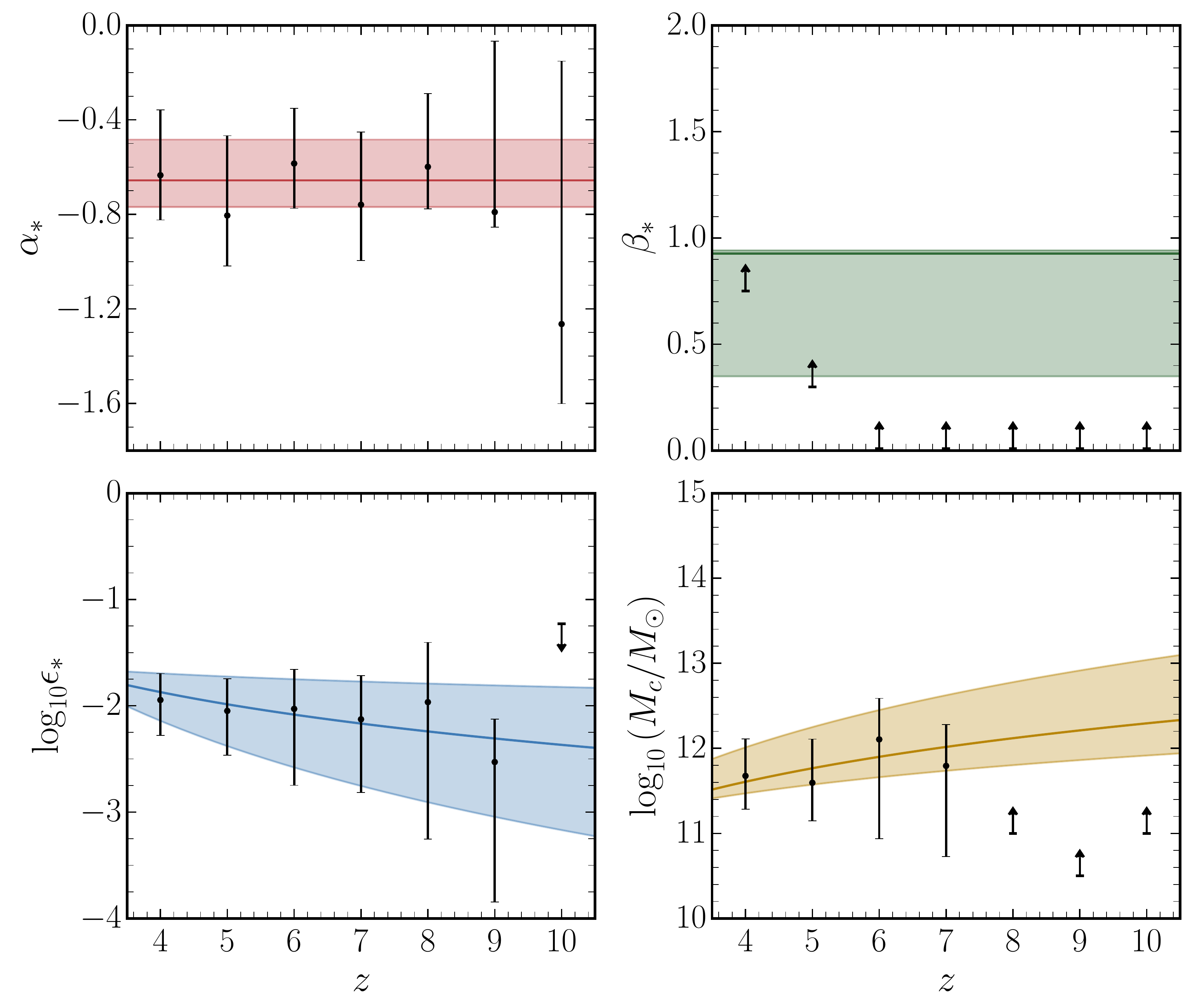}
    \caption{Redshift evolution of the astrophysical parameters $\alpha_*, \beta_*, \epsilon_*\ \mathrm{and}\ M_c$ (defined in Eq.~\eqref{eq:ftilde}) of galaxies using LF data from the Hubble Space Telescope (see Sec.~\ref{subsec:hst_observations} and Fig.~\ref{fig:v2_posteriors} for details). The black dots with error-bars are obtained within the most conservative approach, where the parameters are inferred at each redshift independently, as in Eq.~\eqref{eq:conservative_astro}. The dots represent their average values, the error-bars the 68\% confidence levels, and the arrows are upper or lower limits also at 68\% CL. Similarly, the solid curves and shaded regions show the average evolution and 68\% confidence intervals of the astrophysical parameters using our fiducial model, which assumes a power-law behaviour with $z$ for $\epsilon_*$ and $M_c$, see Eq.~\eqref{eq:fiducial_astro}. Note that in obtaining these plots, cosmological parameters are marginalised over.
    }
    \label{fig:astro_z_evo}
    \vspace{-0.3cm}
\end{figure}

The redshift evolution of the astrophysical parameters in the conservative model is shown in Fig.~\ref{fig:astro_z_evo}, where the dots indicate the average values, the error bars denote the 68\% confidence limits and the arrows show upper or lower limits also at 68\% CL. Given these fits, we can examine which parameters evolve with redshift, and which stay constant within their uncertainties. The index $\alpha_*$ remains roughly flat, whereas $\beta_*$ in particular cannot be well measured in the HST data. This is simply because $\beta_*$ controls the slope of the bright-end tail of the UV LF and the Poisson errors dominate in that region. The other two, however, may appear to evolve. In order to establish our minimal model for the redshift evolution of the astrophysical parameters, we have tested two other assumptions for each of the parameters, where we either keep them constant with $z$ or evolve them as power laws\footnote{We note that a $z$-independent parametrisation is less preferred than our minimal model, with $\Delta\chi^2 \approx 4.6$ for two degrees of freedom difference, and results in a $\sigma_8$ that is not consistent with the Planck CMB value at more than $3\sigma$ (within $\Lambda$CDM).}. We have found that keeping the indices $\alpha_*$ and $\beta_*$ fixed, but letting $\epsilon_*$ and $M_c$ evolve as power laws with $z$ (or, equivalently, linear in log-space for the logarithm of these quantities) produces the best fit to the data with the smallest amount of free parameters. This choice gives a reduced $\chi^2$ per degree of freedom of ${\sim}1.2$. We show in Fig.~\ref{fig:astro_z_evo} the redshift evolution of the astrophysical parameters in this minimal model. From a physical point of view (and looking at the correlation between $\epsilon_*$ and $M_c$ in Fig.~\ref{fig:v2_posteriors}), an increase of $\epsilon_*$ with time could indicate a gradual growth of the star-formation efficiency at lower $z$, and the decrease of $M_c$ could be pointing towards a build-up of quasar feedback, which reduces star formation in high-mass halos.
\\

In summary, we propose a conservative and a fiducial parametrisation of the astrophysical parameters of the following form:
\begin{align}
    \mathrm{Conservative} = &\ 
    \label{eq:conservative_astro}
        \pmb{\bigl\{}
        \text{Independent set of}\ \{\alpha_*, \beta_*, \epsilon_*, M_c\}\ \text{at each}\ z\\\nonumber\\
    \mathrm{Fiducial} = & 
    \label{eq:fiducial_astro}
    \begin{cases}
        \alpha_*(z) = \alpha_* \\
        \beta_*(z) = \beta_* \\
        \log_{10}\epsilon_*(z) = \epsilon_*^\mathrm{s}\times \log_{10}\left(\frac{1+z}{1+6}\right) + \epsilon_*^\mathrm{i} \\
        \log_{10}\frac{M_c(z)}{M_\odot} = M_c^\mathrm{s}\times \log_{10}\left(\frac{1+z}{1+6}\right) + M_c^\mathrm{i}
    \end{cases}\ ,
\end{align}

where the fiducial one has only six free parameters ($\alpha_*$, $\beta_*$, $\epsilon_*^\mathrm{s}$, $\epsilon_*^\mathrm{i}$, $M_c^\mathrm{s}$ and $M_c^\mathrm{i}$). In the rest of this work, we will perform our main analysis with the fiducial model\footnote{This redshift evolution of the astrophysical parameters is specific to the model chosen for the halo-galaxy connection, so care must be taken when using models II and III outlined in Sec.~\ref{sec:astro_modelling} rather than our fiducial model I).} and show the results as obtained with the conservative method in App.~\ref{app:results_z_general}.

%%%%%%%%%%%%%%%%%%%%%%
% Data
%%%%%%%%%%%%%%%%%%%%%%
\section{Data}
\label{sec:data}

\subsection{HST Observations}
\label{subsec:hst_observations}
The rest-frame UV light emitted by young stars at high redshifts can be observed today with optical and (near-)infrared telescopes. Over the years, a number of surveys have been conducted with a variety of instruments, including, but not limited to, deep field surveys by the Hubble Space Telescope~\cite{Bouwens:2014fua,Finkelstein_2015,Atek:2015axa,Livermore:2016mbs,Bouwens_2017asdasd,Mehta_2017,Ishigaki_2018,Oesch_2018,Atek:2018nsc,Rojas_Ruiz_2020, Bouwens_2021}, VIMOS-VLT~\cite{Khusanova:2019cxr} and the Subaru/CFHT/Isaac Newton observatories~\cite{van_der_Burg_2010, Harikane2021, Santos_2021}. In this work, we will use data obtained with the HST, as it currently allows for determinations of the UV LF over the widest redshift range ($z = 4-10$). With the HST, there are, broadly speaking, two different observing strategies that can be utilised to identify and study the younger generations of galaxies. One is based on blank fields observations, where the survey area is devoid of any foreground objects (e.g. stars) down to a certain threshold magnitude. These are the {\it Hubble Legacy Fields} (HLF). The other approach takes advantage of the power of gravitational lensing, and uses fields where strong-lensing clusters magnify the flux of faint, background galaxies enough to become observable. These are known as the {\it Hubble Frontier Fields} (HFF). Each of these methods has its own benefits -- for example, blank fields tend to cover more area and thus allow to probe the bright end of the LF, while lensing fields are particularly useful for determining the faint end of the LF~\cite{Maizy:2009df}. On the other hand, both types of observations are potentially prone to different systematics, such as lensing uncertainties in the modelling of the cluster lenses~\cite{Bouwens_2017asdasd}, magnification bias in the blank field surveys~\cite{Wyithe:2011gh,Mason:2015wla}, and contamination of the signal due to the contribution of quasars to the bright end of the LF~\cite{Harikane2021}. In our analysis, we will use the HLF and HFF data from~\cite{Oesch_2018, Bouwens_2021}, which are obtained from blank and parallel field surveys and cover the magnitude range $-23<M_\mathrm{UV}<-16$. That is, no lensing fields are included. In addition, we will use mock data from the IllustrisTNG hydrodynamical simulation~\cite{Vogelsberger_dust2020} to cross-check the possible impact of data systematics on our results (see Sec.~\ref{subsec:TNG_mock}). The HST UV LF data together with our best-fit models are shown in App.~\ref{app:Hubble_TNG_data}.

\subsection{Cosmic Variance}
\label{subsec:cosmic_variance}
One of the most relevant contributions to the total error budget in the observed luminosity functions is cosmic variance, which arises from the underlying fluctuations in the matter density field. While at the bright end of the luminosity function -- where the number of observed galaxies is small -- Poisson error dominates, cosmic variance will remain an unavoidable error that should be taken into account everywhere else. This is particularly important at high redshifts and when the observed areas covered by surveys are small~\cite{Trapp_2020} (in the case of the HST, these range from a few tens to a few hundreds of arcmin$^2$, see e.g.~\cite{Bouwens_2021}). Previous works have obtained an estimate for the cosmic variance in the UV LFs, either through semi-analytic methods~\cite{Newman:2001ca, Somerville:2003bq, Stark:2007pv, Moster:2010hf, Trapp_2020} or through N-body simulations~\cite{Kitzbichler:2006ec, Trenti:2007dh, Bhowmick:2019nnj, Ucci_2021}. As an example, for the magnitude range of interest ($-23<M_\mathrm{UV}<-16$) and a survey area of 500 arcmin$^2$ over a redshift width $\Delta z=1$, the cosmic variance amounts to an error of at most $20\%-40\%$ at $z=6$. Given that the Hubble LFs we use are constructed from a collection of patches, each observed to different depths, we take a simple approach here and impose a 20\% minimal error on each data point individually and uncorrelated with the rest. For most magnitude and redshift ranges of interest, we find that this choice corresponds to a conservative estimate of the cosmic variance. We leave a more sophisticated approach to take into account cosmic variance patch by patch in our modelling for future work~\cite{Trapp_2020}.

\subsection{Dust Correction}
\label{subsec:dust_corr}
Massive galaxies can accumulate dust in their interstellar medium. This dust preferentially absorbs the UV flux and reprocesses it to be subsequently emitted in the infrared. As a consequence, the UV LF of the brightest galaxies at low redshifts can be significantly attenuated~\cite{Yung_2018}. A common approach in modelling dust extinction involves the so-called IRX-$\beta$ relation from~\cite{Meurer:1999jj}: by looking at the correlation between the absorbed UV flux and the excess infrared flux in the same star-forming galaxies, an estimate for the dust attenuation can be obtained. So far, this approach has only been applied to low-redshift objects, and thus it should be noted that the underlying assumptions behind this method and its applicability at high redshifts are still actively debated topics (see e.g.~\cite{Ma:2019hwa, Mirocha_2020, Liang:2020aay} and references therein). Assuming that the UV spectrum of galaxies follows a power law $f_\lambda\sim\lambda^\beta$, a fit can be found between the average dust extinction parameter $\langle A_\mathrm{UV}\rangle$ and the average of $\beta$~\cite{Smit:2012nf,Vogelsberger_dust2020}:
\begin{align}
    \label{eq:dust_general}
    \langle A_\mathrm{UV}\rangle = C_0 + 0.2\ln(10)\sigma_\beta^2C_1^2+C_1\langle\beta\rangle\ ,
\end{align}
where $C_0$ and $C_1$ are free parameters, and at each magnitude the slope $\beta$ is assumed to follow a Gaussian distribution with standard deviation $\sigma_\beta = 0.34$~\cite{Bouwens:2011yy}. The average value of $\beta$ is fitted to observations by following the prescription in~\cite{Trenti:2014hka}:
\begin{align}
    \label{eq:beta_dust}
    \langle\beta(z,M_\mathrm{UV})\rangle =
    \begin{cases}
    a(z)e^{-\frac{b(z)}{a(z)}}+c & M_\mathrm{UV}\geq M_0\\
    a(z) + b(z) + c & M_\mathrm{UV} < M_0
    \end{cases}\ ,
\end{align}
where $a(z) = \beta_{M_0}(z)-c$, $b(z) = (M_\mathrm{UV}-M_0) \mathrm{d}\beta/\mathrm{d}M_0(z)$, $c = -2.33$, $M_0 = -19.5$ and the values of $\beta_{M_0}$ and $\mathrm{d}\beta/\mathrm{d}M_0$ are taken from~\cite{Bouwens:2013hxa}. The exponent at magnitudes $M_\mathrm{UV}\geq M_0$ prevents the dust extinction from becoming negative. Note that the observations of $ \langle\beta\rangle$ from~\cite{Bouwens:2013hxa} only cover the redshift range $z\leq8$. Nevertheless, at $z > 8$ the dust extinction is significantly smaller than in the lower redshift range, and thus, following~\cite{Yung_2018}, we simply neglect it. Eq.~\eqref{eq:dust_general} has been calibrated using different low-redshift probes, see e.g.~\cite{Siana:2009um, Overzier:2010aa, Casey:2014cqa, Castellano:2014lua, reddy2015mosdef}. In our main analysis, we adopt the calibration from~\cite{Overzier:2010aa} with $C_0 = 4.54$ and $C_1 = 2.07$. This reference obtained the IRX-$\beta$ relation by remeasuring the same sample of galaxies as in the pioneering work~\cite{Meurer:1999jj}, but with an instrument that has a larger aperture size. This then allowed for a more complete photometric sample and eliminated a number of systematic errors previously introduced in the calibration of the IRX-$\beta$ relation in~\cite{Meurer:1999jj}. To check how our results depend on the different calibrations of $C_0$ and $C_1$, we rerun our analysis with the fits from~\cite{Casey:2014cqa} and~\cite{J_Bouwens_2016}, and show the results in App.~\ref{app:alternative_dust}. We now can obtain the desired average extinction as:
\begin{align}
    \label{eq:dust}
    \langle A_\mathrm{UV}\rangle(z, M_\mathrm{UV}) = 4.54 + 0.86\ln(10)\sigma_\beta^2 + 2.07\langle\beta\rangle(z, M_\mathrm{UV})\ ,
\end{align}
and the observed UV magnitudes can be converted to intrinsic magnitudes. In addition, given that the dust extinction term $\langle A_\mathrm{UV}\rangle$ is a function of magnitude, this correction will also change the magnitude bin sizes -- by altering both ends of the bins by different amounts -- and thus the UV LF itself~\cite{Corasaniti:2016epp}. In summary, the dust attenuation changes the magnitude $M_\mathrm{UV}$, the bin size $\Delta M_\mathrm{UV}$, the luminosity function $\Phi_\mathrm{UV}$ and its error $\sigma_{\Phi_\mathrm{UV}}$ as follows~\cite{Corasaniti:2016epp}:
\begin{align}
    M_\mathrm{UV}^\mathrm{new} & =  M_\mathrm{UV}^\mathrm{old} - \langle A_\mathrm{UV}\rangle(z, M_\mathrm{UV})\,,\\
    \Delta M_\mathrm{UV}^\mathrm{new} & = \Delta M_\mathrm{UV}^\mathrm{old} + \langle A_\mathrm{UV}\rangle\left(z,\, M_\mathrm{UV} - \frac{\Delta M_\mathrm{UV}}{2}\right) - \langle A_\mathrm{UV}\rangle\left(z,\, M_\mathrm{UV} + \frac{\Delta M_\mathrm{UV}}{2}\right)\,,\\
    \Phi_\mathrm{UV}^\mathrm{new} & = \Phi_\mathrm{UV}^\mathrm{old}\frac{\Delta M_\mathrm{UV}^\mathrm{old}}{\Delta M_\mathrm{UV}^\mathrm{new}}\,,\\
    \sigma_{\Phi_\mathrm{UV}}^\mathrm{new} & =  \sigma_{\Phi_\mathrm{UV}}^\mathrm{old}\frac{\Delta M_\mathrm{UV}^\mathrm{old}}{\Delta M_\mathrm{UV}^\mathrm{new}}\ .
\end{align}

The bin sizes of the HST data reported in~\cite{Bouwens_2021} are $\Delta M_\mathrm{UV}^\mathrm{old} = 0.5$ for $z = 4-8$ and $\Delta M_\mathrm{UV}^\mathrm{old} = 0.8$ for $z = 9-10$.

\subsection{Alcock-Paczy\'{n}ski Effect}
\label{subsec:AP_effect}
The observed luminosity function is obtained by dividing the number of detected galaxies in a magnitude bin by the volume of the survey area. This volume is computed by assuming a certain cosmology and, in order to avoid introducing artificial biases, should be recomputed accordingly when a different cosmology is assumed~\cite{Alcock:1979mp}. Therefore, since in our analysis we also vary cosmological parameters, we need to rescale the data (LF \emph{and} its errors) as follows\footnote{Note that the fraction of volumes here is the inverse of the one shown in~\cite{Sahlen:2021bqt}. The UV LF is defined per unit volume and thus has to be multiplied by a volume element in the fiducial cosmology (i.e. $V_\mathrm{old}$) in order to get the number of galaxies in it.}:
    \begin{align}
        \Phi_\mathrm{UV}^\mathrm{new} = \Phi_\mathrm{UV}^\mathrm{old} \times \frac{V_\mathrm{old}}{V_\mathrm{new}}\ , \qquad \sigma_{\Phi_\mathrm{UV}}^\mathrm{new} =  \sigma_{\Phi_\mathrm{UV}}^\mathrm{old}\times \frac{V_\mathrm{old}}{V_\mathrm{new}}\ ,
    \end{align}
    where $V$ is a comoving volume given by:
    \begin{align}
        V(z) &= \frac{4\pi}{3}\left[r_\mathrm{com}^3\left(z+\frac{\Delta z}{2}\right) - r_\mathrm{com}^3\left(z-\frac{\Delta z}{2}\right)\right]\,,\\
        r_\mathrm{com}(z) &= \int_0^{z} \frac{\mathrm{d}z'}{H_0\sqrt{\Omega_\mathrm{m}(1+z')^3+1-\Omega_\mathrm{m}}}\ ,
    \end{align}
    where $r_\mathrm{com}$ is the comoving distance and $H_0$ is the Hubble parameter today. The cosmology assumed in the HST data from~\cite{Bouwens_2021} is a flat $\Lambda$CDM cosmology with $H_0 = 70\,\mathrm{km}\,\mathrm{s}^{-1}\mathrm{Mpc}^{-1}$ and $\Omega_\mathrm{m} = 0.3$, and the data was binned using $\Delta z = 1$. In addition to the rescaling of the LF data, a different cosmology changes the luminosity distance to a galaxy and thus also the derived absolute magnitude from a given apparent magnitude. Keeping in mind that the apparent magnitude is the observed quantity, and thus ought to remain the same, the deduced absolute magnitude shifts as:
    \begin{align}
        M_\mathrm{UV}^\mathrm{new} = M_\mathrm{UV}^\mathrm{old} - 5\log_{10}\left(\frac{d_\mathrm{L}^\mathrm{new}}{d_\mathrm{L}^\mathrm{old}}\right) = M_\mathrm{UV}^\mathrm{old} - 5\log_{10}\left(\frac{r_\mathrm{com}^\mathrm{new}}{r_\mathrm{com}^\mathrm{old}}\right)\ .
    \end{align}
    Note that this shift in magnitudes is linear, and thus does not affect the magnitude bin sizes. We apply the three corrections to the data (cosmic variance, dust and the Alcock-Paczy\'{n}ski effect) in the same order as discussed in this section.

\subsection{Mock Data from IllustrisTNG}
\label{subsec:TNG_mock}
Besides real data from the Hubble Space Telescope, we also perform an analysis with mock data from the IllustrisTNG hydrodynamical simulations~\cite{Vogelsberger_dust2020}. This makes for an important check of our approach, as it allows us to verify that our method is actually able to extract the underlying cosmology. In particular, it cross-checks that any potential systematics/contaminations in the real data, such as contributions of quasars to the bright end of the UV LF~\cite{Harikane2021}, do not significantly affect our results. We summarise the main points in extracting the LF and performing the analysis below, and refer the interested reader to the corresponding script on the \href{https://github.com/NNSSA/GALLUMI_public}{\texttt{GitHub}} page for more details. We also show the mock data together with our best-fit model in App.~\ref{app:Hubble_TNG_data}.

\begin{itemize}
    \item We make use of all three simulation boxes (TNG50, TNG100, TNG300)~\cite{Marinacci:2017wew, Naiman2018, Nelson:2017cxy, Pillepich:2017fcc, Springel:2017tpz, Nelson:2019jkf, Pillepich:2019bmb} and stitch the luminosity functions together. In particular, we first remove all the points where the simulations lose accuracy due to resolution effects\footnote{This manifests itself as a decrease in the number of galaxies at fainter magnitudes. If we denote $M_\mathrm{UV}^{[i]}$ as the magnitude bin at which the number of galaxies peaks, here we remove all points with magnitudes  $M_\mathrm{UV} \geq M_\mathrm{UV}^{[i-1]}$ to account for this effect.}, and then compute the total LF $\Phi_\mathrm{UV}^\mathrm{tot}$ using inverse error weighting:
    \begin{align}
        \left(\frac{\Phi_\mathrm{UV}^\mathrm{tot}}{\sigma_\mathrm{UV}^\mathrm{tot}}\right)^2 &= \left(\frac{\Phi_\mathrm{TNG50}}{\sigma_\mathrm{TNG50}}\right)^2 + \left(\frac{\Phi_\mathrm{TNG100}}{\sigma_\mathrm{TNG100}}\right)^2 + \left(\frac{\Phi_\mathrm{TNG300}}{\sigma_\mathrm{TNG300}}\right)^2\,,\\
        \left(\frac{1}{\sigma_\mathrm{UV}^\mathrm{tot}}\right)^2 &= \left(\frac{1}{\sigma_\mathrm{TNG50}}\right)^2 + \left(\frac{1}{\sigma_\mathrm{TNG100}}\right)^2 + \left(\frac{1}{\sigma_\mathrm{TNG300}}\right)^2\ ,
    \end{align}
    where the $\sigma$s indicate the errors in the LF (see below).
    
    \item Each of the errors $\sigma_\mathrm{TNG50/100/300}$ above has two contributions: Poisson error and ``cosmic" (sample) variance. We include Poisson errors simply by taking the square root of the number of galaxies in each bin.
    As for cosmic variance, we use the \texttt{galcv} code\footnote{\href{https://github.com/adamtrapp/galcv}{https://github.com/adamtrapp/galcv}}~\cite{Trapp_2020}, which utilises a bias approach to find the LF in a region overdense by $\delta_\mathrm{b}$ to be:
    \begin{align}
        \Phi_\mathrm{UV}\left[1 + \epsilon_\mathrm{cv}\frac{\delta_\mathrm{b}}{\sigma_\mathrm{PB}}\right]\ ,
    \end{align}
    where  $\epsilon_\mathrm{cv}$ is the relative cosmic variance of the LF (the output of the code), and ${\sigma^2_\mathrm{PB}}$ is the variance of $\delta_\mathrm{b}$, which depends on the geometry of the region (a square box for the IllustrisTNG simulations). By assuming that $\delta_\mathrm{b}$ follows a Gaussian distribution of width $\sigma_\mathrm{PB}$, the 1-sigma error due to cosmic variance is then simply given by $\Phi_\mathrm{UV}\epsilon_\mathrm{cv}$. The error in the UV LF is then obtained as $\sigma^2 = \sigma^2_\mathrm{Poisson} + \sigma^2_\mathrm{cosmic-variance}$. Finally, as in our HST data analysis, we impose a minimal error of 20\% on each data point in the LF of each box, this time to account for intrinsic statistical deviations.

    \item The UV LF is corrected for dust by using the original empirical model of~\cite{Meurer:1999jj}:
    \begin{align}
        \label{eq:original_dust}
        \langle A_\mathrm{UV}\rangle(z, M_\mathrm{UV}) = 4.43 + 0.79\ln(10)\sigma_\beta^2 + 1.99\langle\beta\rangle(z, M_\mathrm{UV})\ .
    \end{align}
    In this case, the exponential tail in Eq.~\eqref{eq:beta_dust} is not included, i.e., $\langle\beta\rangle$ evolves linearly with $M_\mathrm{UV}$, and $A_\mathrm{UV}$ is set equal to 0 once it becomes negative.
    
    \item No massive neutrinos are included in the analysis. Compared with the sum of neutrino masses $\Sigma m_\nu = 0.06\,\mathrm{eV}$ used everywhere else in this work, this does not have a significant effect (well below $1\sigma$) on the inferred cosmological parameters.
    
\end{itemize}

%%%%%%%%%%%%%%%%%%%%%%
% GALLUMI
%%%%%%%%%%%%%%%%%%%%%%
\section{\texttt{GALLUMI}}
\label{sec:gallumi}

In this section, we briefly overview our analysis pipeline. 
Every model detailed in this paper is implemented as a likelihood code in the publicly available MCMC sampler \texttt{MontePython}~\cite{Brinckmann:2018cvx, Audren:2012wb}, which itself makes use of the Boltzmann code \texttt{CLASS}~\cite{Lesgourgues:2011re, Blas:2011rf} to obtain predictions of the matter power spectrum and its integrated quantities. 
We dub this package of galaxy luminosity function likelihoods {\tt GALLUMI}\footnote{\href{https://github.com/NNSSA/GALLUMI_public}{https://github.com/NNSSA/GALLUMI\_public}}.
A major advantage of {\tt GALLUMI} is that the UV LF analysis can be readily run together with other cosmological and astrophysical analyses, including the Planck CMB~\cite{Planck:2019nip} and the Pantheon~\cite{Scolnic:2017caz, Jones:2017udy} or JLA~\cite{SDSS:2014iwm} supernovae likelihoods. In addition, the MCMC chains can also be quickly analysed with the built-in functions of \texttt{MontePython}. Finally, this implementation also allows for a user-friendly way to adapt, modify, and extend the UV LF code.\\

All quantities are computed as detailed in Secs.~\ref{sec:models} and~\ref{sec:data}, with the exception that we use the discrete form of Eq.~\eqref{eq:phi_UV}, given by:
\begin{align}
    \Phi_\mathrm{UV}(z, M_\mathrm{UV}, \boldsymbol{\theta}) = \frac{1}{\Delta M_\mathrm{UV}}\int\limits_0^\infty \mathrm{d}M_\mathrm{h} \left[\frac{\mathrm{d}n_\mathrm{h}}{\mathrm{d}M_\mathrm{h}}(z, M_\mathrm{h}, \boldsymbol{\theta}) \int\limits_{M_\mathrm{UV,1}}^{M_\mathrm{UV,2}}\mathrm{d}M_\mathrm{UV}'P(M_\mathrm{UV}', z, M_\mathrm{h}, \boldsymbol{\theta})\right]\ ,
\end{align}
where $M_\mathrm{UV}$ is the centre of the magnitude bin, $\Delta M_\mathrm{UV}$ is the bin width, $M_\mathrm{UV,1/2} = M_\mathrm{UV}\pm \Delta M_\mathrm{UV}/2$ are the edges of the bin, $\boldsymbol{\theta}$ represents a vector of the free parameters (astrophysical and cosmological) in our model, and $P(M_\mathrm{UV}', z, M_\mathrm{h}, \boldsymbol{\theta})$ is the probability distribution for a halo of mass $M_\mathrm{h}$ hosting a galaxy of magnitude $M_\mathrm{UV}'$ at redshift $z$, and is given in Eq.~\eqref{eq:gaussian_MUV}. In this last quantity, the halo mass enters the computation of $\langle M_\mathrm{UV}\rangle(z, M_\mathrm{h}, \boldsymbol{\theta})$. In our fiducial analysis, $\boldsymbol{\theta} = \{\sigma_8,n_\mathrm{s},\omega_\mathrm{b}, \omega_\mathrm{m}, \alpha_*, \beta_*, \epsilon_*^\mathrm{s}, \epsilon_*^\mathrm{i}, M_c^\mathrm{s}, M_c^\mathrm{i}, Q, \sigma_{M_\mathrm{UV}}$\}, see Sec.~\ref{sec:models}.\\

We assume that the data is Gaussian distributed, such that the log-likelihood reads:
    \begin{align}
        \label{eq:chisq}
        -2\ln\mathcal{L} = \chi^2(z,\boldsymbol{\theta}) = \underset{M_\mathrm{UV}}{\sum}\left(\frac{\Phi_\mathrm{model}(z,M_\mathrm{UV},\boldsymbol{\theta})-\Phi_\mathrm{data}(z,M_\mathrm{UV}, \boldsymbol{\theta})}{\sigma_\Phi^\mathrm{data}(z,M_\mathrm{UV}, \boldsymbol{\theta})}\right)^2\ .
    \end{align}
In this equation the  $\Phi_\mathrm{data}$ and $\sigma_\Phi^\mathrm{data}$ terms both depend on the cosmological parameters (via $\boldsymbol{\theta}$) due to the Alcock-Paczy\'{n}ski effect (see Sec.~\ref{subsec:AP_effect}). 
While Eq.~\eqref{eq:chisq} is a good approximation at fainter magnitudes, where the number of observed galaxies is sufficiently large, it may break down at the brighter end, as the number of galaxies there will follow a Poisson distribution with a small mean. Nevertheless, since these magnitude bins have the fewest amount of galaxies, they also have the largest (Poisson) errors. As such, they do not dominate the constraining power of the UV LF. We have explicitly checked that removing the brightest data points with magnitudes $M_\mathrm{UV} \leq -21$ alters our results for $\sigma_8$ by roughly three percent, well below our uncertainties. This shows that our results are not driven by the brightest objects (and that quasar contamination is likely unimportant in our case).

%%%%%%%%%%%%%%%%%%%%%%
% Results
%%%%%%%%%%%%%%%%%%%%%%
\section{Results}
\label{sec:results}

Having discussed our fiducial model, the data we use and our analysis pipeline, we can now present our results. We remind the reader that all results shown in this section are obtained with our \textbf{model I} for the halo-galaxy connection, as described in Sec.~\ref{subsec:halo-galaxy}.

\subsection{Priors and External Data}
\label{subsec:priors}

First of all, an important choice in the analysis comes in the form of which parameters we assume are fixed, and which priors are imposed on the free parameters. We summarise and justify our choices below, which hold for all analyses performed in this paper:

\begin{itemize}
    \item  We fix the angular scale of the sound horizon $\theta_\mathrm{s}$ to the Planck CMB value of $\theta_\mathrm{s} = 1.04110$~\cite{Aghanim:2018eyx}. This is a purely geometrical quantity that is determined by the angular scales of the acoustic peaks in the temperature and polarisation spectra. It is extremely well measured by CMB observations (down to ${\sim}0.03\%$~\cite{Aghanim:2018eyx}) and its determination is largely independent of the underlying physics of the CMB era. Fixing this parameter versus letting it vary within its minuscule error makes nearly no difference in our results.
    
    \item The spectral tilt $n_\mathrm{s}$ is restricted to $0.7 \leq n_\mathrm{s} \leq 1.3$. This is because UV LFs are not able to constrain $n_\mathrm{s}$ well, and thus we limit the range over which it can vary, following a similar -- but less restrictive -- approach as in cosmic-shear analyses, see e.g.~\cite{HSC:2018mrq, KiDS:2020suj, DES:2021wwk}.
    
    \item We use a Gaussian prior on the baryon density parameter $\omega_\mathrm{b} = \Omega_\mathrm{b}h^2$ from measurements of the primordial deuterium abundance. In practice, we add the following term to the $\chi^2$ in Eq.~\eqref{eq:chisq}:
    \begin{align}
        \chi^2_\mathrm{BBN}(\omega_\mathrm{b}) = \left(\frac{\omega_\mathrm{b} - \omega_\mathrm{b}^\mathrm{data}}{\sigma_{\omega_\mathrm{b}}}\right)^2\ ,
    \end{align}
    where $\omega_\mathrm{b}^\mathrm{data} = 0.02233$ and $\sigma_{\omega_\mathrm{b}} = 0.00036$~\cite{Pisanti:2020efz}.
    
    \item We make use of distance moduli measurements from the Pantheon supernovae (SNe) type-Ia  catalogue~\cite{Scolnic:2017caz, Jones:2017udy}. In principle, this likelihood will constrain the total-matter abundance $\Omega_\mathrm{m}$, but not the Hubble constant $H_0$ as these SNe have not been calibrated (e.g.~with Cepheids~\cite{Riess:2019cxk}).
    
    \item We fix the sum of neutrino masses to $\sum m_\nu = 0.06\,\mathrm{eV}$ and the effective number of relativistic species to $N_\mathrm{eff} = 3.044$. This is the same choice made in the main Planck CMB and cosmic-shear analyses, which allows us to make a proper comparison (see also our companion paper~\cite{Sabti:2021unj}). For the runs with the IllustrisTNG mock data, we do not include neutrino masses, in order to match the input of the simulations. We find the shift to $\sigma_8$ induced by ignoring the small neutrino masses $\sum m_\nu = 0.06\,\mathrm{eV}$ to be well below our accuracy.
\end{itemize}

It is worth mentioning that by fixing $\theta_\mathrm{s}$, we essentially also fix the quantity $\Omega_\mathrm{m}h^{3.4}$~\cite{2dFGRSTeam:2002tzq, Kable:2018bgg}. This means that together with the BBN and Pantheon priors, we automatically obtain a limit on the Hubble parameter $h$ within the standard $\Lambda$CDM model assumed here. This by far dominates the posteriors on $h$ that we will show, as the UV LF is not highly sensitive to the value of this variable. For the rest of the parameters, we adapt the flat priors in Table~\ref{tab:priors}.

\begin{table}[t!]
\begin{center}
{\renewcommand{\arraystretch}{1.35}
\begin{tabular}{p{1.25cm}|p{1.25cm}|p{1.25cm}|p{1.25cm}|p{1.25cm}|p{1.25cm}|p{1.25cm}|p{1.25cm}|p{1.25cm}|p{1.25cm}|p{1.25cm}|p{1.25cm}}
\hline\hline
\hfil $\boldsymbol{\sigma_8}$ & 
\hfil $\boldsymbol{\omega_\mathrm{m}}$ & 
\hfil $\boldsymbol{100\omega_\mathrm{b}}$ & 
\hfil $\boldsymbol{n_\mathrm{s}}$ & 
\hfil $\boldsymbol{\alpha_*}$ & 
\hfil $\boldsymbol{\beta_*}$ & 
\hfil $\boldsymbol{\epsilon_*^\mathrm{s}}$ &
\hfil $\boldsymbol{\epsilon_*^\mathrm{i}}$ &
\hfil $\boldsymbol{M_c^\mathrm{s}}$ &
\hfil $\boldsymbol{M_c^\mathrm{i}}$ &
\hfil $\boldsymbol{Q}$ &
\hfil $\boldsymbol{\sigma_{M_\mathrm{UV}}}$
\\
\hline\hline
\hfil [0.1, 3] &
\hfil [0, 1] &
\hfil [0, 1] &
\hfil [0.7, 1.3] &
\hfil [-3, 0] &
\hfil [0, 3] &
\hfil [-3, 3] &
\hfil [-3, 3] &
\hfil [-3, 3] &
\hfil [7, 15] &
\hfil [1.5, 2.5] &
\hfil [$10^{-3}$, 3]
\\
\hline\hline
\end{tabular}
}
\end{center}
\caption{The parameters that we vary in our fiducial analysis, together with their (flat) priors. The astrophysical parameters $\alpha_*$, $\beta_*$, $\epsilon_*^\mathrm{s}$, $\epsilon_*^\mathrm{i}$, $M_c^\mathrm{s}$ and $M_c^\mathrm{i}$ enter the halo-galaxy connection in Eqs.~\eqref{eq:ftilde} and~\eqref{eq:fiducial_astro}. The parameter $Q$ enters the accretion rate in Eq.~\eqref{eq:Mhdot_EPS}, and its prior range encompasses the several different calibrations found for this parameter in the literature, see e.g.~\cite{Neistein:2006ak,Schneider:2020xmf}. The variable $\sigma_{M_\mathrm{UV}}$ is the width of the distribution in Eq.~\eqref{eq:gaussian_MUV}, and controls the scatter in the halo-galaxy connection. Note that its prior range stops at $10^{-3}$, in order to avoid division by zero. This does not have any impact on our results, as the data bins are much larger in size than this value of $\sigma_{M_\mathrm{UV}}$, see Sec.~\ref{subsec:dust_corr}. We also impose an upper limit on the matter density parameter of $\Omega_\mathrm{m} \leq 1$, to ensure a flat Universe.}
\label{tab:priors}
\end{table}

\subsection{Posteriors on Cosmological and Astrophysical Parameters}
\label{subsec:posteriors}

\begin{figure}[t!]
    \vspace{-0.3cm}
    \centering
    \includegraphics[width=\textwidth]{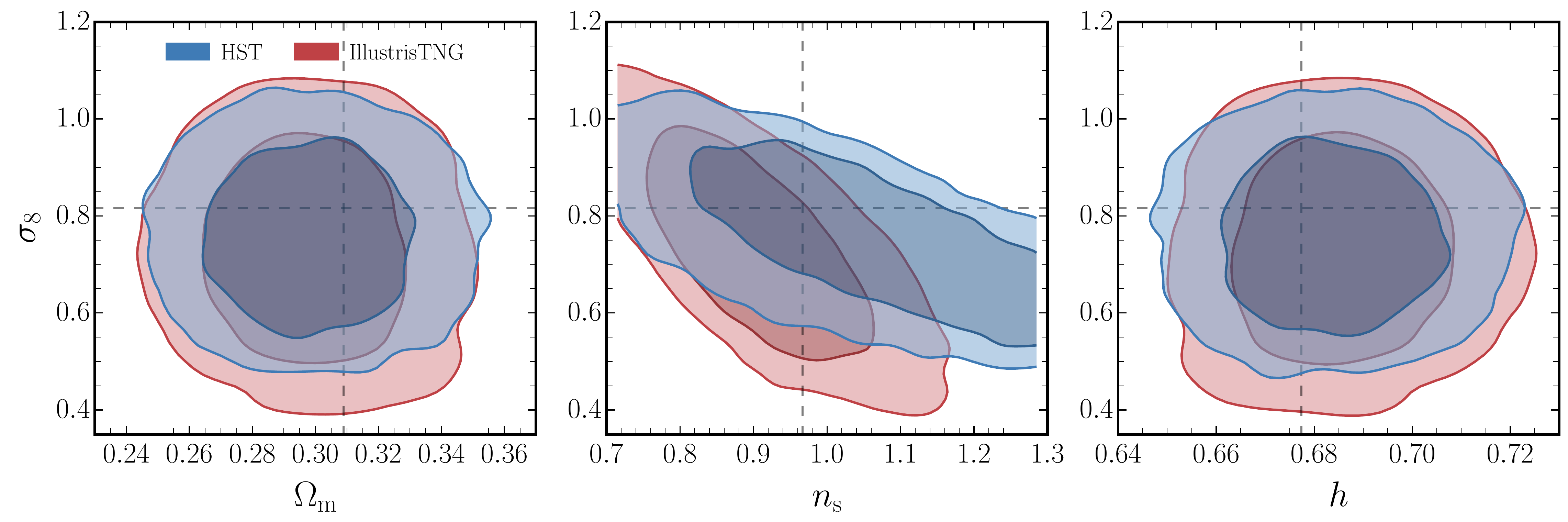}
    \caption{Posteriors for the clustering amplitude $\sigma_8$, the matter density parameter $\Omega_\mathrm{m}$, the spectral tilt $n_\mathrm{s}$, and the Hubble parameter $h$, after marginalising over the unknown astrophysical parameters and the rest of the cosmological parameters. The blue and red colours indicate where we used the HST data from~\cite{Oesch_2018, Bouwens_2021} and mock data from IllustrisTNG~\cite{Vogelsberger_dust2020}, respectively. The inner (outer) contours depict the 68\% (95\%) confidence levels. The grey, dashed lines correspond to the fiducial values used in the IllustrisTNG simulations, which are recovered with our approach.
    }
    \label{fig:v2_posteriors_cosmo}
\end{figure}

The main results of our analysis are shown in Fig.~\ref{fig:v2_posteriors_cosmo}, where the blue and red contours indicate the posteriors obtained with the HST and IllustrisTNG mock data respectively. These are obtained by varying all cosmological and astrophysical parameters specified in Sec.~\ref{sec:gallumi}. We begin our discussion of the results when using the HST data from~\cite{Oesch_2018, Bouwens_2021}. We find that UV galaxy luminosity functions are able to measure the clustering amplitude to be:
\begin{align}
    \label{eq:sigma8_fiducial}
    \sigma_8 = 0.76^{+0.12}_{-0.14}\ ,
\end{align}
at 68\% CL. We also find that this measurement of $\sigma_8$ is only very weakly dependent (well below the 1$\sigma$ level) on the choice of the halo mass function and the calibration of the dust correction, see App.~\ref{app:robustness_checks}. 
While the inferred values of $\sigma_8$ from the UV LFs (which reach $15-20\%$ precision) are not as strong as those obtained with cosmic-shear measurements (which are at the ${\sim}6\%$ level, see e.g.~\cite{DES:2021wwk}) or CMB observations (at the sub-percent level~\cite{Aghanim:2018eyx}), they are highly complementary. This is because the UV LFs cover different ranges of physical scales and redshifts. As we explore in our companion paper~\cite{Sabti:2021unj}, the UV LFs are a powerful probe of the clustering of matter at small scales. That is, one can measure the matter power spectrum at comoving scales of order $R\sim 1\, \mathrm{Mpc}$ with this probe. Therefore, the UV LF can provide us with useful insights on the clustering of matter at scales beyond the reach of current CMB and LSS observations~\cite{Sabti:2021unj} (see also~\cite{Chevallard:2014sxa,Sabti:2020ser} for discussions on this topic within the context of small-scale non-Gaussianity).\\

Compared to the results of a recent work that also reported measurements on $\sigma_8$~\cite{Sahlen:2021bqt}, our relative error on $\sigma_8$ is a factor ${\sim}2.3$ larger, which can be traced back to a number of differences in our approaches. Firstly, as opposed to~\cite{Sahlen:2021bqt}, we do not fix the spectral tilt $n_\mathrm{s}$, which we know correlates with $\sigma_8$ (see middle panel of Fig.~\ref{fig:v2_posteriors_cosmo}). Secondly, in~\cite{Sahlen:2021bqt} only one parameter ($\epsilon_*$) is allowed to evolve with redshift through a power law. In our fiducial model, we allow both $\epsilon_*$ and $M_c$ to evolve with redshift, as is preferred by current HST data. 
This increases the total error budget in $\sigma_8$. Lastly, a handful of measurements of the galaxy magnitude as a function of halo mass are added in~\cite{Sahlen:2021bqt}. These were reported in~\cite{Harikane:2017lcw} and obtained by analysing the small-scale correlation function of galaxies. 
The results are, however, strongly dependent on the cosmology assumed. For instance, varying $\sigma_8$  modifies the abundance of halos, and therefore the clustering model employed to fit the data, which would change the inferred relationship between $\left< M_{\rm UV}\right>$ and $M_{\rm h}$. We therefore do not consider such data points in our analysis, whereas they are included in~\cite{Sahlen:2021bqt} (despite that cosmological parameters are being varied over in their analysis). We have explicitly checked that these three contributions each lead to a stricter model and thus smaller errors. Regarding the third one, it would further bias the results towards the fiducial cosmology of~\cite{Harikane:2017lcw}, which set $\sigma_8=0.8159$.\\

We also perform our full analysis on the mock data from IllustrisTNG, where we find a comparable measurement of $\sigma_8 = 0.73_{-0.16}^{+0.15}$. The errors here are slightly larger because of cosmic (sample) variance, which especially dominates the smallest simulation box. So far, all the results reported here are obtained using our fiducial astrophysical model, see Sec.~\ref{subsec:z_evol_astro}. If instead we assume the conservative model, where the astrophysical parameters vary independently at each redshift (as in Eq.~\eqref{eq:conservative_astro}), we find a measurement of $\sigma_8 = 0.74^{+0.14}_{-0.27}$ using the HST data. Here the lower error has relaxed by a factor of ${\sim}2$ compared to our fiducial model result in Eq.~\eqref{eq:sigma8_fiducial}, see also App.~\ref{app:results_z_general} for more details.\\

\begin{figure}[t!]
    \vspace{-0.4cm}
    \centering
    \includegraphics[width=\textwidth]{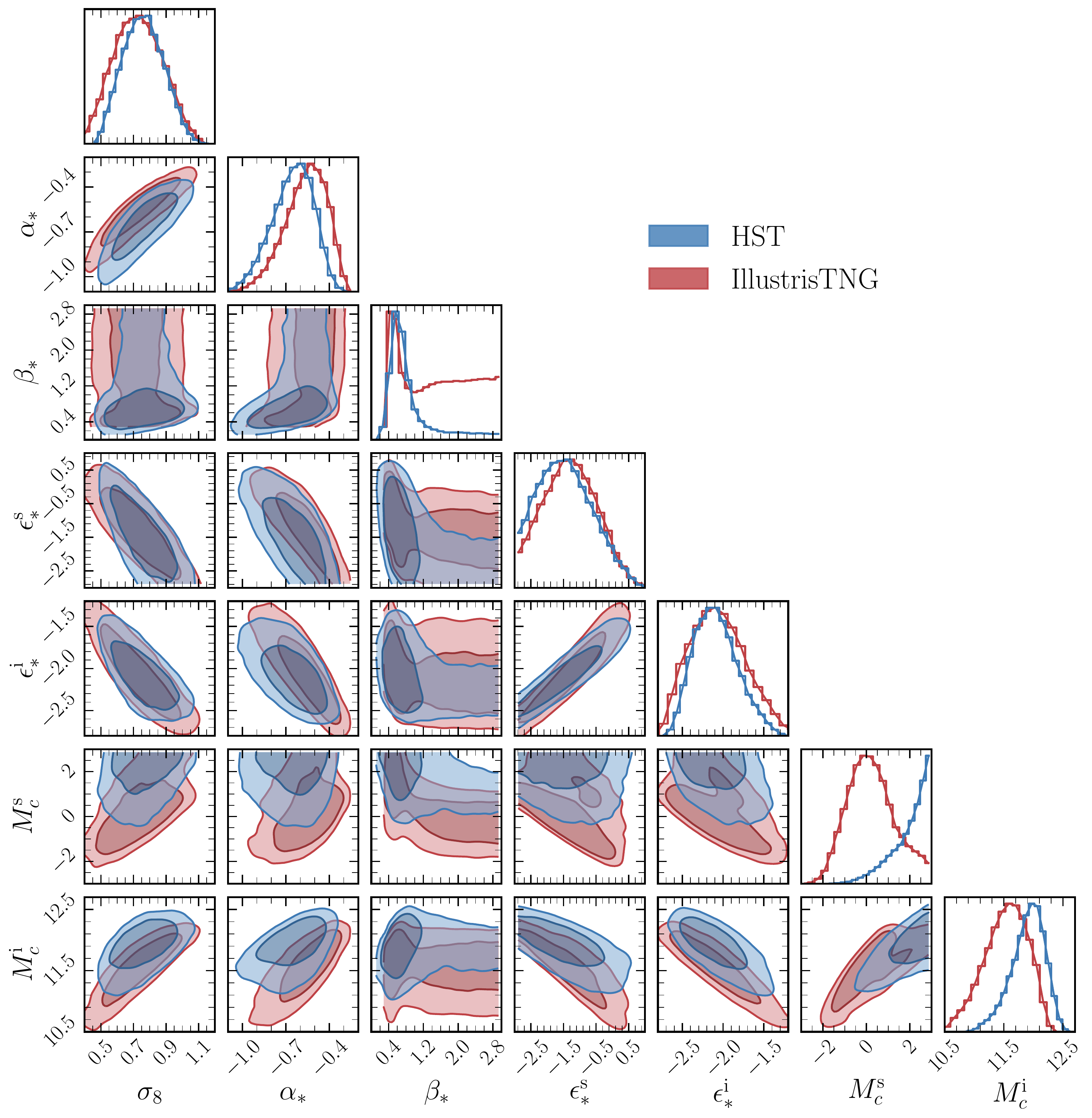}
    \caption{Posteriors for the clustering amplitude $\sigma_8$ and the astrophysical parameters within our fiducial model, see Eqs.~\eqref{eq:ftilde} and~\eqref{eq:fiducial_astro}. Note that the parameters $\epsilon_*^\mathrm{s}$ and $M_c^\mathrm{s}$ control the redshift evolution of $\epsilon_*$ and $M_c$. The blue and red colours indicate where we used the HST data from~\cite{Oesch_2018, Bouwens_2021} and mock data from IllustrisTNG~\cite{Vogelsberger_dust2020}, respectively. The inner (outer) contour in the 2D posteriors depicts the 68\% (95\%) confidence level and the diagonal panels show the marginalised 1D posterior for each parameter. Using the IllustrisTNG data, we obtain smaller values for the $M_c^\mathrm{s}$ and $M_c^\mathrm{i}$ parameters than with the HST data. This is probably because the IllustrisTNG simulation tends to overpredict the amount of quasars at high redshifts~\cite{Weinberger:2017bbe}, which results in stronger feedback at the bright end of the luminosity function and can cause a decrease of $M_c$.}
    \label{fig:v2_posteriors}
    \vspace{-0.7cm}
\end{figure}

\enlargethispage{0.2cm}

In addition to our determination of $\sigma_8$, the posteriors in Fig.~\ref{fig:v2_posteriors_cosmo} show measurements of the matter density parameter $\Omega_\mathrm{m}$ and the Hubble parameter $h$. As explained in Sec.~\ref{subsec:priors}, this is a direct consequence of the fact that we use Pantheon SNe Ia data (which gives us $\Omega_\mathrm{m}$) and that we fix the angular size of the sound horizon (which restricts a certain combination of $\Omega_\mathrm{m}$ and $h$). Also, as expected, UV LFs are only weakly sensitive to the spectral tilt $n_\mathrm{s}$, since they probe a relatively narrow range of scales. We find good agreement (within 68\% CL) between our posterior for $n_\mathrm{s}$ and the IllustrisTNG input ($n_\mathrm{s}=0.9667$).\\

Besides our cosmological-parameter constraints, we are also able to measure astrophysical parameters with the UV LFs. Within the IllustrisTNG framework, our astrophysical parameters map to effective parameters that capture the sub-grid physics (e.g. supernovae shocks, AGN feedback) implemented in the simulations and thus cannot be pre-computed. In Fig.~\ref{fig:v2_posteriors}, we show the posteriors for the astrophysical parameters in our model I (see Eqs.~\eqref{eq:ftilde} and~\eqref{eq:fiducial_astro}) jointly with that for $\sigma_8$. We note that these posteriors are obtained while also varying cosmological parameters, as opposed to the usual approach of fixing the cosmology (see e.g.~\cite{Gillet:2019fjd}). In brief, we find that the star-formation efficiency $\epsilon_*$ can be well measured, as well as its redshift evolution. The turn-over mass $M_c$, at which the star-formation rate peaks, is roughly measured as $M_c\approx 10^{12}\,\Msun$, and its redshift behaviour is only bound to be growing with redshift, since $M_c^\mathrm{s} > 0$. We find an excellent measurement of the faint-end slope $\alpha_*=-0.65_{-0.12}^{+0.16}$ at 68\% CL, whereas the bright-end slope is roughly unconstrained at 95\% CL (though the 68\% CL region is restricted to $\beta_*\approx 0-1$). These results are contextualised in Fig.~\ref{fig:astro_z_evo}, where we show the redshift evolution of each astrophysical parameter.

\subsection{Future Data}
\label{subsec:forecast}
Upcoming space missions, including the James Webb Space Telescope (JWST)~\cite{Gardner:2006ky} and the Nancy Grace Roman Space Telescope (NGRST)~\cite{Spergel:2015sza}, will expand upon the course set by the Hubble Space Telescope. For instance, the surveys conducted by the JWST will reach at least similar depths (i.e., apparent magnitude limits) as those by the HST, but critically, cover much redder bands (e.g. 1-5 microns in the case of the JWST JADES survey~\cite{bunker_2019} versus a maximum of 1.6 micron with the HST) and attain higher resolution~\cite{Gardner:2006ky}. This will improve the quality of the data in many different ways, one of the reasons being that it will provide a better handle on how to deal with contamination issues, for example due to emission lines from nebulae, see e.g.~\cite{Williams_2018}. In a complementary way, the NGRST has the potential to significantly reduce the statistical errors in the UV LF by covering large regions of the sky. This means that the next-generation space telescopes will not only observe galaxies at higher redshifts and with dimmer magnitudes than currently possible, but also with superior accuracy. In this section, we are interested in understanding how improved measurements of the faint and bright end of the UV LF will lead to tighter limits on the clustering amplitude $\sigma_8$ as compared to the one derived using HST data (see Eq.~\eqref{eq:sigma8_fiducial}).\\

For our forecasts, we will follow the same approach as detailed in Sec.~5 of~\cite{Sabti:2020ser} to generate the mock data. We consider two different types of surveys, a deep- and a wide-field one. For the former, we assume a PANORAMIC-like survey by the JWST~\cite{PANORAMIC_JWST}, with a depth similar to that of current HST data~\cite{Bouwens_2021} and a survey area of $1500\,\mathrm{arcmin}^2$. For the latter one, we assume a High-Latitude-like survey by the NGRST~\cite{Spergel:2015sza}, with a depth of $m_\mathrm{UV} = 26.5$ and a survey area of $2000\,\mathrm{deg}^2$, see also~\cite{Mason:2015cna}. In both cases, we use a minimum absolute magnitude of $M_\mathrm{UV}^\mathrm{min} \approx -24$, a magnitude bin size of $\Delta M_\mathrm{UV} = 0.5$ and a redshift width of $\Delta z = 1$. The mock data covers the redshift range\footnote{We have checked that including higher redshifts does not improve the prospects significantly, mainly because of the large Poisson errors.} $z = 4-10$ and is created using the best-fit parameter values in our fiducial model (obtained using HST data): $\{h,\, \sigma_8,\, n_\mathrm{s},\, \Omega_\mathrm{m},\, \omega_\mathrm{b},\, \alpha_*$,\, $\beta_*$,\, $\epsilon_*^\mathrm{s}$,\, $\epsilon_*^\mathrm{i}$,\, $M_c^\mathrm{s}$,\, $M_c^\mathrm{i},\, \sigma_{M_\mathrm{UV}},\, Q\} = \{0.676,\, 0.878,\, 0.995,\, 0.307,\, 0.0221,\,-0.594,\, 0.611,\, -1.96, \, -2.17,\, 2.95,\, 12.1,\, 0.00250,\, 2.47\}$. The number of galaxies $N_\mathrm{gal}$ in each bin is obtained by sampling from a Poisson distribution with an average given by the number of galaxies in the same bin within the fiducial model. The Poisson errors are then calculated as $\sqrt{N_\mathrm{gal}}$. We do not impose a minimal error on the data as done in~\cite{Sabti:2020ser}, but instead use the \texttt{galcv}~\cite{Trapp_2020} to estimate the cosmic variance. A script to generate the mock data is provided on the \href{https://github.com/NNSSA/GALLUMI_public}{GitHub} page of the code.\\

\begin{figure}[t!]
    \centering
    \includegraphics[width=0.55\textwidth]{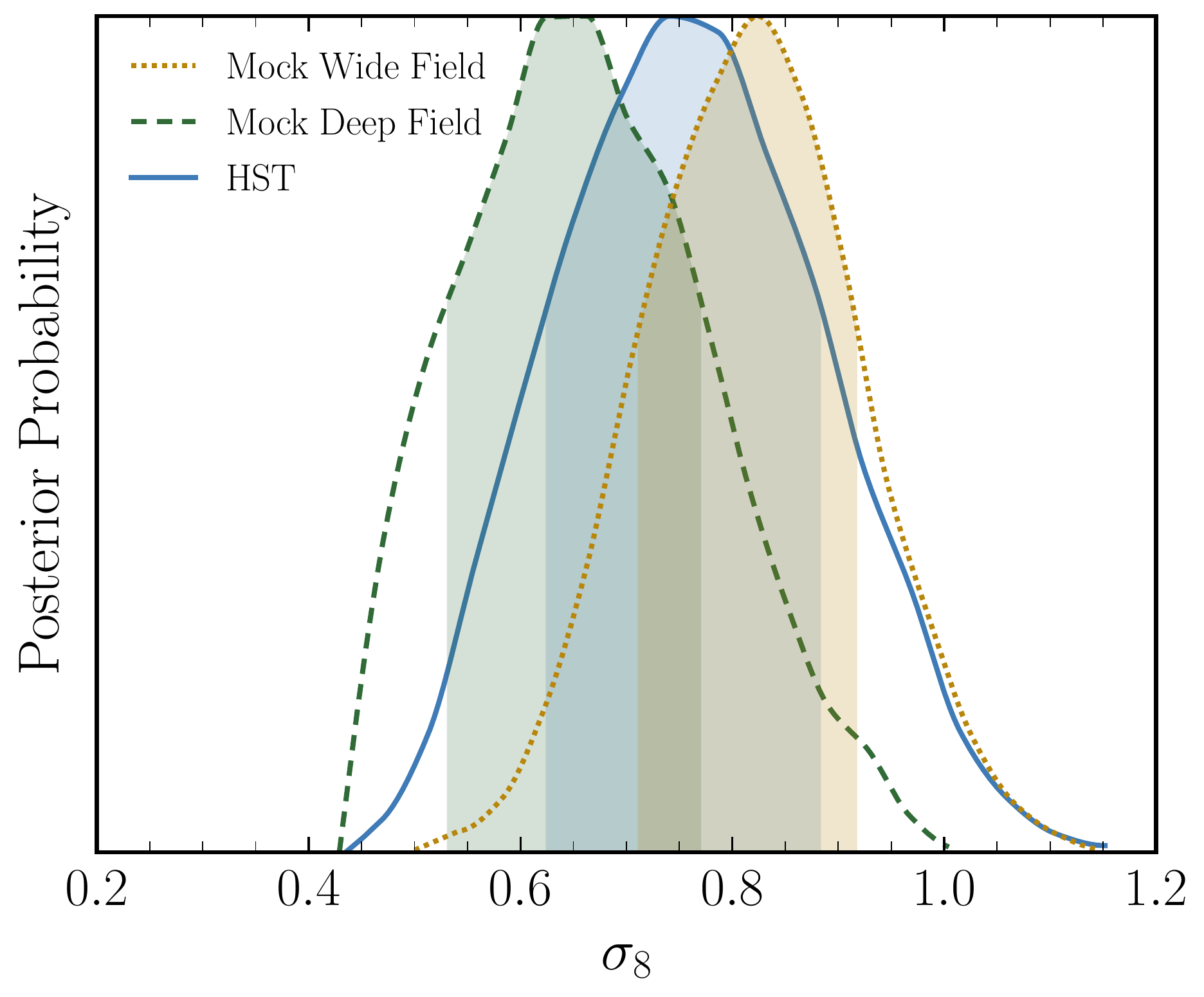}
    \caption{Prospects for improved determinations of the clustering amplitude $\sigma_8$ using future wide- (yellow, dotted) and deep-field (green, dashed) surveys. The shaded areas indicate the 68\% CL. Upcoming wide-field surveys will refine determinations of the bright-end of the UV LF and, therefore, have the potential to reduce the error in $\sigma_8$ from ${\sim}17\%$ to ${\sim}12\%$. This is in contrast to future deep-field surveys, that are not expected to improve much upon the current result (shown in blue).
    }
    \label{fig:future_data}
\end{figure}

With the mock data ready, we can now run the same analysis as before using our model I. The 1D posteriors for $\sigma_8$ are shown in Fig.~\ref{fig:future_data}. Note that the posteriors for the forecasts are not centered around the fiducial value of $\sigma_8 = 0.878$, because the mock data is sampled from a distribution. We find that a future deep-field survey, which mostly probes the mid and faint end of the UV LF, would not appreciably improve upon the current results. In contrast, the larger area covered by a future wide-field survey would allow for improved determinations of the bright end of the UV LF and reduce the error in $\sigma_8$ from ${\sim}17\%$ to ${\sim}12\%$. This difference between the two types of surveys is as expected, since $\sigma_8$ is a large-scale quantity, which when varied affects mostly the abundance of the heaviest halos and thus the brightest galaxies, see Fig.~\ref{fig:UVLF_param_depend}.

%%%%%%%%%%%%%%%%%%%%%%
% Dependence of results on astrophysics
%%%%%%%%%%%%%%%%%%%%%%
\section{Dependence of Results on Astrophysical Modelling}
\label{sec:astro_modelling}
Up to now, all our results are obtained using a semi-analytic model for the halo-galaxy connection, where the UV luminosity of a galaxy is linked to the mass of its host halo by assuming a double-power law relation between the stellar and halo-mass accretion rates (\textbf{model I}, described in Sec.~\ref{subsubsec:Mhalo_MUV_connection}). A number of assumptions are made during the construction of this model and, therefore, it is important to determine the robustness of our results against the various ways in which the astrophysics can be modelled. As such, we will consider two additional models for the halo-galaxy connection:

\begin{itemize}
    \item \textbf{Model II} -- Here we relate the average stellar mass $M_*$ to the halo mass $M_\mathrm{h}$ via a double-power law:
    \begin{align}
        \label{eq:fstar}
        f_* = \frac{M_*}{M_\mathrm{h}} = \frac{\epsilon_*}{\left(\frac{M_\mathrm{h}}{M_c}\right)^{\alpha_*}+\left(\frac{M_\mathrm{h}}{M_c}\right)^{\beta_*}}\ ,
    \end{align}
    where the astrophysical parameters $\{\alpha_*, \beta_*, \epsilon_*,M_c\}$ have a similar interpretation as in Eq.~\eqref{eq:ftilde}. We further assume that the average stellar mass can be written in terms of the star-formation rate as~\cite{Park:2018ljd, Gillet:2019fjd}:
    \begin{align}
        \label{eq:Mstar_Mstardot}
        M_* = t_*\frac{\dot{M}_*}{H(z)}\ ,
    \end{align}
    where  $H(z)$ is the Hubble expansion rate and $t_*$ is a free (dimensionless) parameter that modulates the timescales. This equation roughly implies that the stellar-accretion rate is proportional to the dynamical time of dark-matter halos, since in a matter dominated Universe $t_\mathrm{dyn}\sim \overline{\rho}_\mathrm{m}^{-1/2}\sim H^{-1}(z)$. In contrast to model I, where the equivalent of $t_*$ can be obtained by plugging Eq.~\eqref{eq:Mhdot_EPS} in Eq.~\eqref{eq:ftilde}, the $t_*$ here is independent of halo mass. Note that $t_*$ is fully degenerate with $\epsilon_*$ in Eq.~\eqref{eq:fstar}. As such, for this model we will use the ratio of the two, which we denote by $r_* \equiv \epsilon_*/t_*$. The stellar accretion rate $\dot M_*$ can then be related to the UV magnitude using Eq.~\eqref{eq:Mstardot_LUV} as before. From this, we finally obtain $M_\mathrm{UV}$ as a function of $M_\mathrm{h}$.
    
    \item \textbf{Model III} -- In this case we relate $M_*$ and $M_\mathrm{h}$ through a double-power law as in Eq.~\eqref{eq:fstar}, but use an empirically determined relation between $M_*$ and $M_\mathrm{UV}$ to obtain $M_\mathrm{h}$. 
    This relation is based on observations of galaxies in the (near-)infrared, which provide a relatively robust determination (roughly within a factor of ${\sim}2$) of their stellar mass-to-light ratio across a wide range of star-formation histories~\cite{Rix_Rieke, Kauffmann:1998gm, Bell:2000jt, Bell:2003cj}. An estimate of the stellar mass is then obtained by multiplying the stellar mass-to-light ratio by the luminosity in the corresponding bandpass, which is similarly obtained using a library of stellar synthesis population models fitted on spectral energy distributions, see e.g.~\cite{Drory:2004eh}. Finally, $M_*$ is then related to a UV magnitude by using observations of the same galaxies in the (rest-frame) UV with e.g. the Hubble fields, see also~\cite{stark2009, gonzalez2011, duncan2014, grazian2015, song2016, Bhatawdekar2019,Kikuchihara2020, stefanon2021galaxy}. In our analysis, we will use the $M_*-M_\mathrm{UV}$ relations from~\cite{song2016} at $z=4-5$ and~\cite{stefanon2021galaxy} at $z = 6-10$ to construct our third model for the halo-galaxy connection. As opposed to the two models considered before, this one assumes an empirical relation to link stellar mass to UV luminosity, rather than a theoretical one.
\end{itemize}

We provide a schematic overview of our approaches in Fig.~\ref{fig:astro_diagram}, where each model is represented by a different color. By comparing the results obtained with these models, we can assess the impact of a number of assumptions in our approach. In particular, with models I and II we can test whether assuming a double-power law in the form of $\widetilde{f}_*$ (Eq.~\eqref{eq:ftilde}) or $f_*$ (Eq.~\eqref{eq:fstar}) makes a difference, while with model III we can check the relation between $\dot{M}_*$ and $M_\mathrm{UV}$ in Eq.~\eqref{eq:Mstardot_LUV}.\\

\begin{figure}[h!]
    \centering
    \includegraphics[width=\textwidth]{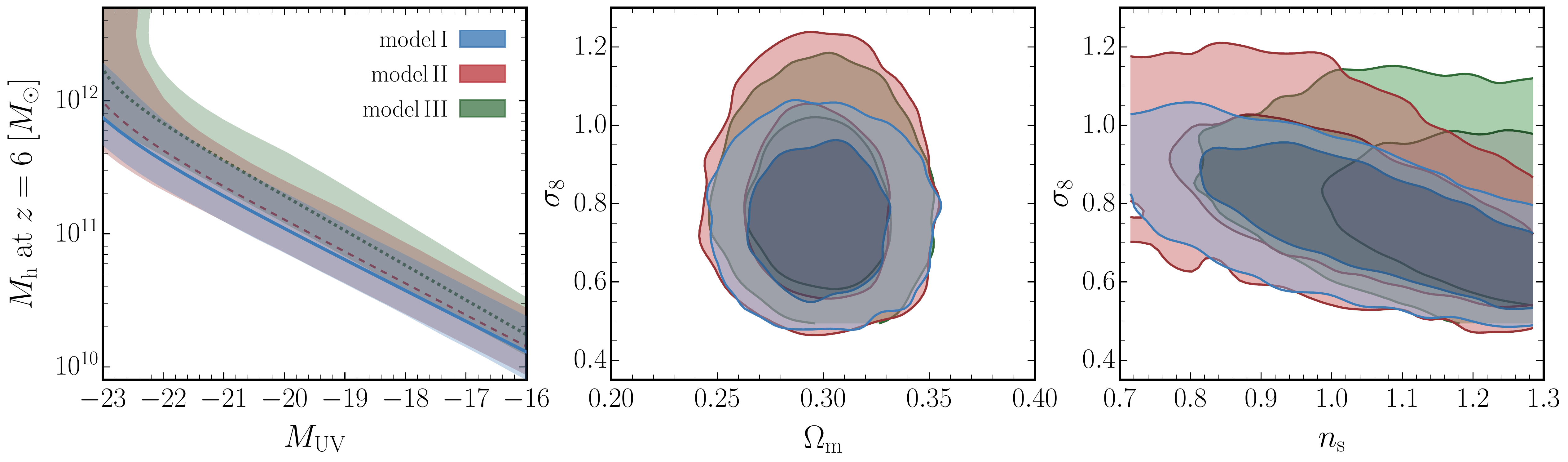}
    \caption{Impact of the modelling of the halo-galaxy connection on our inferences of cosmological parameters. \emph{Left panel}: The halo-galaxy connection at redshift $z=6$ for our models I$-$III, see Sec.~\eqref{subsec:halo-galaxy} and surrounding text. The shaded areas represent the 68\% CL and the lines inside are the averages. \emph{Middle and Right panels}: 2D posteriors on the clustering amplitude $\sigma_8$, the matter density parameter $\Omega_\mathrm{m}$ and the spectral tilt $n_\mathrm{s}$. The inner (outer) contour in the 2D posteriors indicates the 68\% (95\%) confidence level.}
    \label{fig:posteriors_sigmas_v1v2v3}
\end{figure}

We rerun our analysis with models II and III, using the HST data from~\cite{Oesch_2018, Bouwens_2021} and the same redshift evolution parametrisation for the astrophysical parameters as in model I (Eq.~\eqref{eq:fiducial_astro}). We show the connection between host-halo mass and galaxy UV magnitude in the left panel of Fig.~\ref{fig:posteriors_sigmas_v1v2v3} for each of the three models, where the errors are at 68\% CL. Our fiducial parametrisation of the redshift-evolution of the astrophysical parameters (used for model I throughout the text) gives a similar result for the $M_\mathrm{h}-M_\mathrm{UV}$ relation with the two other models. We note, in passing, that within models II and III the data allows $\beta_*$ to be large enough that the formation of bright galaxies is quenched (which results in the uptick for $M_{\rm UV}\lesssim-22$ in the left panel of Fig.~\ref{fig:posteriors_sigmas_v1v2v3}). In the middle panel of this figure, we find that the determination of $\sigma_8$ is consistent across all three models. This is an important cross check that our results are largely immune to choices in the astrophysical parametrisation, and instead are driven by the cosmology. The posteriors on $n_\mathrm{s}$ vary more within the models, as the UV LF cannot measure this quantity well. The small deviation of the $n_\mathrm{s}$ posterior in model III is likely due to the large scatter present in this model, especially at higher redshifts~\cite{stefanon2021galaxy}.\\

%%%%%%%%%%%%%%%%%%%%%%
% Conclusions
%%%%%%%%%%%%%%%%%%%%%%
\section{Conclusions}
\label{sec:conclusions}
The exploration of the eras of cosmic dawn and reionisation ($z\sim 5-30$) is the next frontier in precision cosmology. A particularly interesting objective of upcoming surveys will involve improved measurements of the abundance and power spectrum of high-redshift galaxies. It is therefore timely to establish a framework with a pipeline that enables us to easily and efficiently perform analyses to extract cosmological and astrophysical parameters with these probes. In this work, we present a first step towards this goal in the form of \texttt{GALLUMI}: an accessible likelihood code that models the UV galaxy luminosity function and is readily usable in conjunction with other cosmological/astrophysical analyses. The main feature of \texttt{GALLUMI} is its flexibility to astrophysical parametrisations. \texttt{GALLUMI} includes three default astrophysical models to connect halo masses to UV luminosities, and other approaches can be straightforwardly added to the code following the template therein. We discuss the current UV LF models in Secs.~\ref{sec:models} and~\ref{sec:astro_modelling}, and our analysis pipeline in Sec.~\ref{sec:gallumi}. Upcoming extensions of \texttt{GALLUMI} will include a turn-over term at faint magnitudes and different window functions.\\

We elaborate on the broader implications of \texttt{GALLUMI} for cosmology in our companion paper~\cite{Sabti:2021unj}. Here, instead, we have examined the robustness of cosmological constraints to the assumptions about astrophysics. As a a demonstration, we have combined UV LF data together with SNe type-Ia distance moduli measurements and primordial abundance determinations to measure cosmological and astrophysical parameters within $\Lambda$CDM. In particular, we focused on the clustering amplitude $\sigma_8$, which we found that with current data can be measured at the level of $\sigma_8 = 0.76^{+0.12}_{-0.14}$, where the errors are at 68\% CL. We have checked this measurement against the three default models in our code, different choices for the halo mass function, alternative calibrations of the dust extinction, and found consistent results across the board. Besides using state-of-the-art measurements of the UV LF by the Hubble Space Telescope, we also ran our analysis with mock data from the IllustrisTNG hydrodynamical simulations and obtained consistent results (see Sec.~\ref{sec:data}, where we provide an overview of both data sets). In addition to cosmology, we have also obtained measurements on a number of astrophysical parameters, including those that regulate feedback processes, the star-formation efficiency, and the mass at which the star-formation peaks. Finally, we have performed a simple forecast for future telescopes, focusing on surveys from JWST and NGRST, where we found that an improvement of ${\sim}30\%$ upon the current limit on $\sigma_8$ can be expected. A full account of the results of this analysis is presented in Sec.~\ref{sec:results} and a study of the dependence of our results on the astrophysical modelling is included in Sec.~\ref{sec:astro_modelling}.\\

In summary, UV galaxy luminosity functions provide us with a new handle to measure astrophysical and cosmological effects in synergy with other probes such as the CMB. Here we have established a solid, fast, and versatile framework to exploit such data. With upcoming space missions, such as the James Webb Space Telescope and Nancy Grace Roman Space Telescope, being able to further map out galaxies at even higher redshifts and fainter magnitudes than currently possible, the UV luminosity function will play an increasingly important role in our understanding of the Universe during the eras of cosmic dawn and reionisation.

%%%%%%%%%%%%%%%%%%%%%%%%%%%%%%%%%%%%%%%%%%%%%%%%%%%%%%
\section*{acknowledgements}
%%%%%%%%%%%%%%%%%%%%%%%%%%%%%%%%%%%%%%%%%%%%%%%%%%%%%%
We thank Daniel Eisenstein, Ben Johnson, Sandro Tacchella, and Charlotte Mason for insightful comments on several aspects of this project. We are grateful to Steven Furlanetto and Adam Trapp for their assistance with the $\texttt{galcv}$ code, Xuejian Shen and Mark Vogelsberger for providing us with LF data from IllustrisTNG, and Linda Xu for helping out with \texttt{MontePython} queries. NS is a recipient of a King's College London NMS Faculty Studentship. JBM is supported by a Clay fellowship at the Smithsonian Astrophysical Observatory. DB is supported by a `Ayuda Beatriz Galindo Senior' from the Spanish `Ministerio de Universidades', grant BG20/00228. IFAE is partially funded by the CERCA program of the Generalitat de Catalunya. The research leading to these results has received funding from the Spanish Ministry of Science and Innovation (PID2020-115845GB-I00/AEI/10.13039/501100011033). We acknowledge the use of the public cosmological codes \texttt{CLASS}~\cite{Blas:2011rf,Lesgourgues:2011re}, \texttt{MontePython}~\cite{Audren:2012wb, Brinckmann:2018cvx} and \texttt{galcv}~\cite{Trapp_2020}. The simulations in this work were performed on the Rosalind research computing facility at King’s College London, and the FASRC Cannon cluster supported by the FAS Division of Science Research Computing Group at Harvard University.

\bibliographystyle{apsrev4-1}
\bibliography{biblio}

\begin{thebibliography}{156}%
\makeatletter
\providecommand \@ifxundefined [1]{%
 \@ifx{#1\undefined}
}%
\providecommand \@ifnum [1]{%
 \ifnum #1\expandafter \@firstoftwo
 \else \expandafter \@secondoftwo
 \fi
}%
\providecommand \@ifx [1]{%
 \ifx #1\expandafter \@firstoftwo
 \else \expandafter \@secondoftwo
 \fi
}%
\providecommand \natexlab [1]{#1}%
\providecommand \enquote  [1]{``#1''}%
\providecommand \bibnamefont  [1]{#1}%
\providecommand \bibfnamefont [1]{#1}%
\providecommand \citenamefont [1]{#1}%
\providecommand \href@noop [0]{\@secondoftwo}%
\providecommand \href [0]{\begingroup \@sanitize@url \@href}%
\providecommand \@href[1]{\@@startlink{#1}\@@href}%
\providecommand \@@href[1]{\endgroup#1\@@endlink}%
\providecommand \@sanitize@url [0]{\catcode `\\12\catcode `\$12\catcode
  `\&12\catcode `\#12\catcode `\^12\catcode `\_12\catcode `\%12\relax}%
\providecommand \@@startlink[1]{}%
\providecommand \@@endlink[0]{}%
\providecommand \url  [0]{\begingroup\@sanitize@url \@url }%
\providecommand \@url [1]{\endgroup\@href {#1}{\urlprefix }}%
\providecommand \urlprefix  [0]{URL }%
\providecommand \Eprint [0]{\href }%
\providecommand \doibase [0]{http://dx.doi.org/}%
\providecommand \selectlanguage [0]{\@gobble}%
\providecommand \bibinfo  [0]{\@secondoftwo}%
\providecommand \bibfield  [0]{\@secondoftwo}%
\providecommand \translation [1]{[#1]}%
\providecommand \BibitemOpen [0]{}%
\providecommand \bibitemStop [0]{}%
\providecommand \bibitemNoStop [0]{.\EOS\space}%
\providecommand \EOS [0]{\spacefactor3000\relax}%
\providecommand \BibitemShut  [1]{\csname bibitem#1\endcsname}%
\let\auto@bib@innerbib\@empty
%</preamble>
\bibitem [{\citenamefont {Couchman}\ and\ \citenamefont
  {Rees}(1986)}]{Couchman:1986en}%
  \BibitemOpen
  \bibfield  {author} {\bibinfo {author} {\bibfnamefont {H.~M.~P.}\
  \bibnamefont {Couchman}}\ and\ \bibinfo {author} {\bibfnamefont {M.~J.}\
  \bibnamefont {Rees}},\ }\href@noop {} {\bibfield  {journal} {\bibinfo
  {journal} {Mon. Not. Roy. Astron. Soc.}\ }\textbf {\bibinfo {volume} {221}},\
  \bibinfo {pages} {53} (\bibinfo {year} {1986})}\BibitemShut {NoStop}%
\bibitem [{\citenamefont {Tegmark}\ \emph {et~al.}(1997)\citenamefont
  {Tegmark}, \citenamefont {Silk}, \citenamefont {Rees}, \citenamefont
  {Blanchard}, \citenamefont {Abel},\ and\ \citenamefont
  {Palla}}]{Tegmark:1996yt}%
  \BibitemOpen
  \bibfield  {author} {\bibinfo {author} {\bibfnamefont {M.}~\bibnamefont
  {Tegmark}}, \bibinfo {author} {\bibfnamefont {J.}~\bibnamefont {Silk}},
  \bibinfo {author} {\bibfnamefont {M.~J.}\ \bibnamefont {Rees}}, \bibinfo
  {author} {\bibfnamefont {A.}~\bibnamefont {Blanchard}}, \bibinfo {author}
  {\bibfnamefont {T.}~\bibnamefont {Abel}}, \ and\ \bibinfo {author}
  {\bibfnamefont {F.}~\bibnamefont {Palla}},\ }\href {\doibase 10.1086/303434}
  {\bibfield  {journal} {\bibinfo  {journal} {Astrophys. J.}\ }\textbf
  {\bibinfo {volume} {474}},\ \bibinfo {pages} {1} (\bibinfo {year} {1997})},\
  \Eprint {http://arxiv.org/abs/astro-ph/9603007} {arXiv:astro-ph/9603007}
  \BibitemShut {NoStop}%
\bibitem [{\citenamefont {Gardner}\ \emph {et~al.}(2006)\citenamefont {Gardner}
  \emph {et~al.}}]{Gardner:2006ky}%
  \BibitemOpen
  \bibfield  {author} {\bibinfo {author} {\bibfnamefont {J.~P.}\ \bibnamefont
  {Gardner}} \emph {et~al.},\ }\href {\doibase 10.1007/s11214-006-8315-7}
  {\bibfield  {journal} {\bibinfo  {journal} {Space Sci. Rev.}\ }\textbf
  {\bibinfo {volume} {123}},\ \bibinfo {pages} {485} (\bibinfo {year}
  {2006})},\ \Eprint {http://arxiv.org/abs/astro-ph/0606175}
  {arXiv:astro-ph/0606175} \BibitemShut {NoStop}%
\bibitem [{\citenamefont {Bromm}\ and\ \citenamefont
  {Yoshida}(2011)}]{Bromm:2011cw}%
  \BibitemOpen
  \bibfield  {author} {\bibinfo {author} {\bibfnamefont {V.}~\bibnamefont
  {Bromm}}\ and\ \bibinfo {author} {\bibfnamefont {N.}~\bibnamefont
  {Yoshida}},\ }\href {\doibase 10.1146/annurev-astro-081710-102608} {\bibfield
   {journal} {\bibinfo  {journal} {Ann. Rev. Astron. Astrophys.}\ }\textbf
  {\bibinfo {volume} {49}},\ \bibinfo {pages} {373} (\bibinfo {year} {2011})},\
  \Eprint {http://arxiv.org/abs/1102.4638} {arXiv:1102.4638 [astro-ph.CO]}
  \BibitemShut {NoStop}%
\bibitem [{\citenamefont {Finkelstein}(2016)}]{Finkelstein2016}%
  \BibitemOpen
  \bibfield  {author} {\bibinfo {author} {\bibfnamefont {S.~L.}\ \bibnamefont
  {Finkelstein}},\ }\href {\doibase 10.1017/pasa.2016.26} {\bibfield  {journal}
  {\bibinfo  {journal} {Publications of the Astronomical Society of Australia}\
  }\textbf {\bibinfo {volume} {33}} (\bibinfo {year} {2016}),\
  10.1017/pasa.2016.26}\BibitemShut {NoStop}%
\bibitem [{\citenamefont {Bouwens}\ \emph {et~al.}(2015)\citenamefont {Bouwens}
  \emph {et~al.}}]{Bouwens:2014fua}%
  \BibitemOpen
  \bibfield  {author} {\bibinfo {author} {\bibfnamefont {R.}~\bibnamefont
  {Bouwens}} \emph {et~al.},\ }\href {\doibase 10.1088/0004-637X/803/1/34}
  {\bibfield  {journal} {\bibinfo  {journal} {Astrophys. J.}\ }\textbf
  {\bibinfo {volume} {803}},\ \bibinfo {pages} {34} (\bibinfo {year} {2015})},\
  \Eprint {http://arxiv.org/abs/1403.4295} {arXiv:1403.4295 [astro-ph.CO]}
  \BibitemShut {NoStop}%
\bibitem [{\citenamefont {Finkelstein}\ \emph {et~al.}(2015)\citenamefont
  {Finkelstein} \emph {et~al.}}]{Finkelstein_2015}%
  \BibitemOpen
  \bibfield  {author} {\bibinfo {author} {\bibfnamefont {S.~L.}\ \bibnamefont
  {Finkelstein}} \emph {et~al.},\ }\href {\doibase 10.1088/0004-637x/810/1/71}
  {\bibfield  {journal} {\bibinfo  {journal} {The Astrophysical Journal}\
  }\textbf {\bibinfo {volume} {810}},\ \bibinfo {pages} {71} (\bibinfo {year}
  {2015})}\BibitemShut {NoStop}%
\bibitem [{\citenamefont {Atek}\ \emph {et~al.}(2015)\citenamefont {Atek} \emph
  {et~al.}}]{Atek:2015axa}%
  \BibitemOpen
  \bibfield  {author} {\bibinfo {author} {\bibfnamefont {H.}~\bibnamefont
  {Atek}} \emph {et~al.},\ }\href {\doibase 10.1088/0004-637X/814/1/69}
  {\bibfield  {journal} {\bibinfo  {journal} {Astrophys. J.}\ }\textbf
  {\bibinfo {volume} {814}},\ \bibinfo {pages} {69} (\bibinfo {year} {2015})},\
  \Eprint {http://arxiv.org/abs/1509.06764} {arXiv:1509.06764 [astro-ph.GA]}
  \BibitemShut {NoStop}%
\bibitem [{\citenamefont {Livermore}\ \emph {et~al.}(2017)\citenamefont
  {Livermore}, \citenamefont {Finkelstein},\ and\ \citenamefont
  {Lotz}}]{Livermore:2016mbs}%
  \BibitemOpen
  \bibfield  {author} {\bibinfo {author} {\bibfnamefont {R.}~\bibnamefont
  {Livermore}}, \bibinfo {author} {\bibfnamefont {S.}~\bibnamefont
  {Finkelstein}}, \ and\ \bibinfo {author} {\bibfnamefont {J.}~\bibnamefont
  {Lotz}},\ }\href {\doibase 10.3847/1538-4357/835/2/113} {\bibfield  {journal}
  {\bibinfo  {journal} {Astrophys. J.}\ }\textbf {\bibinfo {volume} {835}},\
  \bibinfo {pages} {113} (\bibinfo {year} {2017})},\ \Eprint
  {http://arxiv.org/abs/1604.06799} {arXiv:1604.06799 [astro-ph.GA]}
  \BibitemShut {NoStop}%
\bibitem [{\citenamefont {Bouwens}\ \emph {et~al.}(2017)\citenamefont
  {Bouwens}, \citenamefont {Oesch}, \citenamefont {Illingworth}, \citenamefont
  {Ellis},\ and\ \citenamefont {Stefanon}}]{Bouwens_2017asdasd}%
  \BibitemOpen
  \bibfield  {author} {\bibinfo {author} {\bibfnamefont {R.~J.}\ \bibnamefont
  {Bouwens}}, \bibinfo {author} {\bibfnamefont {P.~A.}\ \bibnamefont {Oesch}},
  \bibinfo {author} {\bibfnamefont {G.~D.}\ \bibnamefont {Illingworth}},
  \bibinfo {author} {\bibfnamefont {R.~S.}\ \bibnamefont {Ellis}}, \ and\
  \bibinfo {author} {\bibfnamefont {M.}~\bibnamefont {Stefanon}},\ }\href
  {\doibase 10.3847/1538-4357/aa70a4} {\bibfield  {journal} {\bibinfo
  {journal} {The Astrophysical Journal}\ }\textbf {\bibinfo {volume} {843}},\
  \bibinfo {pages} {129} (\bibinfo {year} {2017})}\BibitemShut {NoStop}%
\bibitem [{\citenamefont {Mehta}\ \emph {et~al.}(2017)\citenamefont {Mehta}
  \emph {et~al.}}]{Mehta_2017}%
  \BibitemOpen
  \bibfield  {author} {\bibinfo {author} {\bibfnamefont {V.}~\bibnamefont
  {Mehta}} \emph {et~al.},\ }\href {\doibase 10.3847/1538-4357/aa6259}
  {\bibfield  {journal} {\bibinfo  {journal} {The Astrophysical Journal}\
  }\textbf {\bibinfo {volume} {838}},\ \bibinfo {pages} {29} (\bibinfo {year}
  {2017})},\ \Eprint {http://arxiv.org/abs/1702.06953} {arXiv:1702.06953
  [astro-ph.GA]} \BibitemShut {NoStop}%
\bibitem [{\citenamefont {Ishigaki}\ \emph {et~al.}(2018)\citenamefont
  {Ishigaki}, \citenamefont {Kawamata}, \citenamefont {Ouchi}, \citenamefont
  {Oguri}, \citenamefont {Shimasaku},\ and\ \citenamefont
  {Ono}}]{Ishigaki_2018}%
  \BibitemOpen
  \bibfield  {author} {\bibinfo {author} {\bibfnamefont {M.}~\bibnamefont
  {Ishigaki}}, \bibinfo {author} {\bibfnamefont {R.}~\bibnamefont {Kawamata}},
  \bibinfo {author} {\bibfnamefont {M.}~\bibnamefont {Ouchi}}, \bibinfo
  {author} {\bibfnamefont {M.}~\bibnamefont {Oguri}}, \bibinfo {author}
  {\bibfnamefont {K.}~\bibnamefont {Shimasaku}}, \ and\ \bibinfo {author}
  {\bibfnamefont {Y.}~\bibnamefont {Ono}},\ }\href {\doibase
  10.3847/1538-4357/aaa544} {\bibfield  {journal} {\bibinfo  {journal} {The
  Astrophysical Journal}\ }\textbf {\bibinfo {volume} {854}},\ \bibinfo {pages}
  {73} (\bibinfo {year} {2018})}\BibitemShut {NoStop}%
\bibitem [{\citenamefont {Oesch}\ \emph {et~al.}(2018)\citenamefont {Oesch},
  \citenamefont {Bouwens}, \citenamefont {Illingworth}, \citenamefont
  {Labb\'{e}},\ and\ \citenamefont {Stefanon}}]{Oesch_2018}%
  \BibitemOpen
  \bibfield  {author} {\bibinfo {author} {\bibfnamefont {P.~A.}\ \bibnamefont
  {Oesch}}, \bibinfo {author} {\bibfnamefont {R.~J.}\ \bibnamefont {Bouwens}},
  \bibinfo {author} {\bibfnamefont {G.~D.}\ \bibnamefont {Illingworth}},
  \bibinfo {author} {\bibfnamefont {I.}~\bibnamefont {Labb\'{e}}}, \ and\
  \bibinfo {author} {\bibfnamefont {M.}~\bibnamefont {Stefanon}},\ }\href
  {\doibase 10.3847/1538-4357/aab03f} {\bibfield  {journal} {\bibinfo
  {journal} {The Astrophysical Journal}\ }\textbf {\bibinfo {volume} {855}},\
  \bibinfo {pages} {105} (\bibinfo {year} {2018})}\BibitemShut {NoStop}%
\bibitem [{\citenamefont {Atek}\ \emph {et~al.}(2018)\citenamefont {Atek},
  \citenamefont {Richard}, \citenamefont {Kneib},\ and\ \citenamefont
  {Schaerer}}]{Atek:2018nsc}%
  \BibitemOpen
  \bibfield  {author} {\bibinfo {author} {\bibfnamefont {H.}~\bibnamefont
  {Atek}}, \bibinfo {author} {\bibfnamefont {J.}~\bibnamefont {Richard}},
  \bibinfo {author} {\bibfnamefont {J.-P.}\ \bibnamefont {Kneib}}, \ and\
  \bibinfo {author} {\bibfnamefont {D.}~\bibnamefont {Schaerer}},\ }\href
  {\doibase 10.1093/mnras/sty1820} {\bibfield  {journal} {\bibinfo  {journal}
  {Mon. Not. Roy. Astron. Soc.}\ }\textbf {\bibinfo {volume} {479}},\ \bibinfo
  {pages} {5184} (\bibinfo {year} {2018})},\ \Eprint
  {http://arxiv.org/abs/1803.09747} {arXiv:1803.09747 [astro-ph.GA]}
  \BibitemShut {NoStop}%
\bibitem [{\citenamefont {Rojas-Ruiz}\ \emph {et~al.}(2020)\citenamefont
  {Rojas-Ruiz}, \citenamefont {Finkelstein}, \citenamefont {Bagley},
  \citenamefont {Stevans}, \citenamefont {Finkelstein}, \citenamefont {Larson},
  \citenamefont {Mechtley},\ and\ \citenamefont {Diekmann}}]{Rojas_Ruiz_2020}%
  \BibitemOpen
  \bibfield  {author} {\bibinfo {author} {\bibfnamefont {S.}~\bibnamefont
  {Rojas-Ruiz}}, \bibinfo {author} {\bibfnamefont {S.~L.}\ \bibnamefont
  {Finkelstein}}, \bibinfo {author} {\bibfnamefont {M.~B.}\ \bibnamefont
  {Bagley}}, \bibinfo {author} {\bibfnamefont {M.}~\bibnamefont {Stevans}},
  \bibinfo {author} {\bibfnamefont {K.~D.}\ \bibnamefont {Finkelstein}},
  \bibinfo {author} {\bibfnamefont {R.}~\bibnamefont {Larson}}, \bibinfo
  {author} {\bibfnamefont {M.}~\bibnamefont {Mechtley}}, \ and\ \bibinfo
  {author} {\bibfnamefont {J.}~\bibnamefont {Diekmann}},\ }\href {\doibase
  10.3847/1538-4357/ab7659} {\bibfield  {journal} {\bibinfo  {journal} {The
  Astrophysical Journal}\ }\textbf {\bibinfo {volume} {891}},\ \bibinfo {pages}
  {146} (\bibinfo {year} {2020})}\BibitemShut {NoStop}%
\bibitem [{\citenamefont {Bouwens}\ \emph {et~al.}(2021)\citenamefont
  {Bouwens}, \citenamefont {Oesch}, \citenamefont {Stefanon}, \citenamefont
  {Illingworth}, \citenamefont {Labbé}, \citenamefont {Reddy}, \citenamefont
  {Atek}, \citenamefont {Montes}, \citenamefont {Naidu}, \citenamefont
  {Nanayakkara},\ and\ \citenamefont {et~al.}}]{Bouwens_2021}%
  \BibitemOpen
  \bibfield  {author} {\bibinfo {author} {\bibfnamefont {R.~J.}\ \bibnamefont
  {Bouwens}}, \bibinfo {author} {\bibfnamefont {P.~A.}\ \bibnamefont {Oesch}},
  \bibinfo {author} {\bibfnamefont {M.}~\bibnamefont {Stefanon}}, \bibinfo
  {author} {\bibfnamefont {G.}~\bibnamefont {Illingworth}}, \bibinfo {author}
  {\bibfnamefont {I.}~\bibnamefont {Labbé}}, \bibinfo {author} {\bibfnamefont
  {N.}~\bibnamefont {Reddy}}, \bibinfo {author} {\bibfnamefont
  {H.}~\bibnamefont {Atek}}, \bibinfo {author} {\bibfnamefont {M.}~\bibnamefont
  {Montes}}, \bibinfo {author} {\bibfnamefont {R.}~\bibnamefont {Naidu}},
  \bibinfo {author} {\bibfnamefont {T.}~\bibnamefont {Nanayakkara}}, \ and\
  \bibinfo {author} {\bibnamefont {et~al.}},\ }\href {\doibase
  10.3847/1538-3881/abf83e} {\bibfield  {journal} {\bibinfo  {journal} {The
  Astronomical Journal}\ }\textbf {\bibinfo {volume} {162}},\ \bibinfo {pages}
  {47} (\bibinfo {year} {2021})}\BibitemShut {NoStop}%
\bibitem [{\citenamefont {Dayal}\ and\ \citenamefont
  {Ferrara}(2018)}]{Dayal:2018hft}%
  \BibitemOpen
  \bibfield  {author} {\bibinfo {author} {\bibfnamefont {P.}~\bibnamefont
  {Dayal}}\ and\ \bibinfo {author} {\bibfnamefont {A.}~\bibnamefont
  {Ferrara}},\ }\href {\doibase 10.1016/j.physrep.2018.10.002} {\bibfield
  {journal} {\bibinfo  {journal} {Phys. Rept.}\ }\textbf {\bibinfo {volume}
  {780-782}},\ \bibinfo {pages} {1} (\bibinfo {year} {2018})},\ \Eprint
  {http://arxiv.org/abs/1809.09136} {arXiv:1809.09136 [astro-ph.GA]}
  \BibitemShut {NoStop}%
\bibitem [{\citenamefont {Berlind}\ \emph {et~al.}(2003)\citenamefont
  {Berlind}, \citenamefont {Weinberg}, \citenamefont {Benson}, \citenamefont
  {Baugh}, \citenamefont {Cole}, \citenamefont {Dave}, \citenamefont {Frenk},
  \citenamefont {Jenkins}, \citenamefont {Katz},\ and\ \citenamefont
  {Lacey}}]{Berlind:2002rn}%
  \BibitemOpen
  \bibfield  {author} {\bibinfo {author} {\bibfnamefont {A.~A.}\ \bibnamefont
  {Berlind}}, \bibinfo {author} {\bibfnamefont {D.~H.}\ \bibnamefont
  {Weinberg}}, \bibinfo {author} {\bibfnamefont {A.~J.}\ \bibnamefont
  {Benson}}, \bibinfo {author} {\bibfnamefont {C.~M.}\ \bibnamefont {Baugh}},
  \bibinfo {author} {\bibfnamefont {S.}~\bibnamefont {Cole}}, \bibinfo {author}
  {\bibfnamefont {R.}~\bibnamefont {Dave}}, \bibinfo {author} {\bibfnamefont
  {C.~S.}\ \bibnamefont {Frenk}}, \bibinfo {author} {\bibfnamefont
  {A.}~\bibnamefont {Jenkins}}, \bibinfo {author} {\bibfnamefont
  {N.}~\bibnamefont {Katz}}, \ and\ \bibinfo {author} {\bibfnamefont {C.~G.}\
  \bibnamefont {Lacey}},\ }\href {\doibase 10.1086/376517} {\bibfield
  {journal} {\bibinfo  {journal} {Astrophys. J.}\ }\textbf {\bibinfo {volume}
  {593}},\ \bibinfo {pages} {1} (\bibinfo {year} {2003})},\ \Eprint
  {http://arxiv.org/abs/astro-ph/0212357} {arXiv:astro-ph/0212357} \BibitemShut
  {NoStop}%
\bibitem [{\citenamefont {Wechsler}\ and\ \citenamefont
  {Tinker}(2018)}]{Wechsler:2018pic}%
  \BibitemOpen
  \bibfield  {author} {\bibinfo {author} {\bibfnamefont {R.~H.}\ \bibnamefont
  {Wechsler}}\ and\ \bibinfo {author} {\bibfnamefont {J.~L.}\ \bibnamefont
  {Tinker}},\ }\href {\doibase 10.1146/annurev-astro-081817-051756} {\bibfield
  {journal} {\bibinfo  {journal} {Ann. Rev. Astron. Astrophys.}\ }\textbf
  {\bibinfo {volume} {56}},\ \bibinfo {pages} {435} (\bibinfo {year} {2018})},\
  \Eprint {http://arxiv.org/abs/1804.03097} {arXiv:1804.03097 [astro-ph.GA]}
  \BibitemShut {NoStop}%
\bibitem [{\citenamefont {Weinberger}\ \emph {et~al.}(2018)\citenamefont
  {Weinberger} \emph {et~al.}}]{Weinberger:2017bbe}%
  \BibitemOpen
  \bibfield  {author} {\bibinfo {author} {\bibfnamefont {R.}~\bibnamefont
  {Weinberger}} \emph {et~al.},\ }\href {\doibase 10.1093/mnras/sty1733}
  {\bibfield  {journal} {\bibinfo  {journal} {Mon. Not. Roy. Astron. Soc.}\
  }\textbf {\bibinfo {volume} {479}},\ \bibinfo {pages} {4056} (\bibinfo {year}
  {2018})},\ \Eprint {http://arxiv.org/abs/1710.04659} {arXiv:1710.04659
  [astro-ph.GA]} \BibitemShut {NoStop}%
\bibitem [{\citenamefont {Weinberg}\ \emph {et~al.}(2015)\citenamefont
  {Weinberg}, \citenamefont {Bullock}, \citenamefont {Governato}, \citenamefont
  {Kuzio~de Naray},\ and\ \citenamefont {Peter}}]{Weinberg:2013aya}%
  \BibitemOpen
  \bibfield  {author} {\bibinfo {author} {\bibfnamefont {D.~H.}\ \bibnamefont
  {Weinberg}}, \bibinfo {author} {\bibfnamefont {J.~S.}\ \bibnamefont
  {Bullock}}, \bibinfo {author} {\bibfnamefont {F.}~\bibnamefont {Governato}},
  \bibinfo {author} {\bibfnamefont {R.}~\bibnamefont {Kuzio~de Naray}}, \ and\
  \bibinfo {author} {\bibfnamefont {A.~H.~G.}\ \bibnamefont {Peter}},\ }\href
  {\doibase 10.1073/pnas.1308716112} {\bibfield  {journal} {\bibinfo  {journal}
  {Proc. Nat. Acad. Sci.}\ }\textbf {\bibinfo {volume} {112}},\ \bibinfo
  {pages} {12249} (\bibinfo {year} {2015})},\ \Eprint
  {http://arxiv.org/abs/1306.0913} {arXiv:1306.0913 [astro-ph.CO]} \BibitemShut
  {NoStop}%
\bibitem [{\citenamefont {Del~Popolo}\ and\ \citenamefont
  {Le~Delliou}(2017)}]{DelPopolo:2016emo}%
  \BibitemOpen
  \bibfield  {author} {\bibinfo {author} {\bibfnamefont {A.}~\bibnamefont
  {Del~Popolo}}\ and\ \bibinfo {author} {\bibfnamefont {M.}~\bibnamefont
  {Le~Delliou}},\ }\href {\doibase 10.3390/galaxies5010017} {\bibfield
  {journal} {\bibinfo  {journal} {Galaxies}\ }\textbf {\bibinfo {volume} {5}},\
  \bibinfo {pages} {17} (\bibinfo {year} {2017})},\ \Eprint
  {http://arxiv.org/abs/1606.07790} {arXiv:1606.07790 [astro-ph.CO]}
  \BibitemShut {NoStop}%
\bibitem [{\citenamefont {Bullock}\ and\ \citenamefont
  {Boylan-Kolchin}(2017)}]{Bullock:2017xww}%
  \BibitemOpen
  \bibfield  {author} {\bibinfo {author} {\bibfnamefont {J.~S.}\ \bibnamefont
  {Bullock}}\ and\ \bibinfo {author} {\bibfnamefont {M.}~\bibnamefont
  {Boylan-Kolchin}},\ }\href {\doibase 10.1146/annurev-astro-091916-055313}
  {\bibfield  {journal} {\bibinfo  {journal} {Ann. Rev. Astron. Astrophys.}\
  }\textbf {\bibinfo {volume} {55}},\ \bibinfo {pages} {343} (\bibinfo {year}
  {2017})},\ \Eprint {http://arxiv.org/abs/1707.04256} {arXiv:1707.04256
  [astro-ph.CO]} \BibitemShut {NoStop}%
\bibitem [{\citenamefont {de~Martino}\ \emph {et~al.}(2020)\citenamefont
  {de~Martino}, \citenamefont {Chakrabarty}, \citenamefont {Cesare},
  \citenamefont {Gallo}, \citenamefont {Ostorero},\ and\ \citenamefont
  {Diaferio}}]{deMartino:2020gfi}%
  \BibitemOpen
  \bibfield  {author} {\bibinfo {author} {\bibfnamefont {I.}~\bibnamefont
  {de~Martino}}, \bibinfo {author} {\bibfnamefont {S.~S.}\ \bibnamefont
  {Chakrabarty}}, \bibinfo {author} {\bibfnamefont {V.}~\bibnamefont {Cesare}},
  \bibinfo {author} {\bibfnamefont {A.}~\bibnamefont {Gallo}}, \bibinfo
  {author} {\bibfnamefont {L.}~\bibnamefont {Ostorero}}, \ and\ \bibinfo
  {author} {\bibfnamefont {A.}~\bibnamefont {Diaferio}},\ }\href {\doibase
  10.3390/universe6080107} {\bibfield  {journal} {\bibinfo  {journal}
  {Universe}\ }\textbf {\bibinfo {volume} {6}},\ \bibinfo {pages} {107}
  (\bibinfo {year} {2020})},\ \Eprint {http://arxiv.org/abs/2007.15539}
  {arXiv:2007.15539 [astro-ph.CO]} \BibitemShut {NoStop}%
\bibitem [{\citenamefont {Kay}\ \emph {et~al.}(2002)\citenamefont {Kay},
  \citenamefont {Pearce}, \citenamefont {Frenk},\ and\ \citenamefont
  {Jenkins}}]{Kay:2001hq}%
  \BibitemOpen
  \bibfield  {author} {\bibinfo {author} {\bibfnamefont {S.~T.}\ \bibnamefont
  {Kay}}, \bibinfo {author} {\bibfnamefont {F.~R.}\ \bibnamefont {Pearce}},
  \bibinfo {author} {\bibfnamefont {C.~S.}\ \bibnamefont {Frenk}}, \ and\
  \bibinfo {author} {\bibfnamefont {A.}~\bibnamefont {Jenkins}},\ }\href
  {\doibase 10.1046/j.1365-8711.2002.05070.x} {\bibfield  {journal} {\bibinfo
  {journal} {Mon. Not. Roy. Astron. Soc.}\ }\textbf {\bibinfo {volume} {330}},\
  \bibinfo {pages} {113} (\bibinfo {year} {2002})},\ \Eprint
  {http://arxiv.org/abs/astro-ph/0106462} {arXiv:astro-ph/0106462} \BibitemShut
  {NoStop}%
\bibitem [{\citenamefont {Efstathiou}(2000)}]{Efstathiou:2000xp}%
  \BibitemOpen
  \bibfield  {author} {\bibinfo {author} {\bibfnamefont {G.}~\bibnamefont
  {Efstathiou}},\ }\href {\doibase 10.1046/j.1365-8711.2000.03665.x} {\bibfield
   {journal} {\bibinfo  {journal} {Mon. Not. Roy. Astron. Soc.}\ }\textbf
  {\bibinfo {volume} {317}},\ \bibinfo {pages} {697} (\bibinfo {year}
  {2000})},\ \Eprint {http://arxiv.org/abs/astro-ph/0002245}
  {arXiv:astro-ph/0002245} \BibitemShut {NoStop}%
\bibitem [{\citenamefont {Naab}\ and\ \citenamefont
  {Ostriker}(2017)}]{Naab_2017}%
  \BibitemOpen
  \bibfield  {author} {\bibinfo {author} {\bibfnamefont {T.}~\bibnamefont
  {Naab}}\ and\ \bibinfo {author} {\bibfnamefont {J.~P.}\ \bibnamefont
  {Ostriker}},\ }\href {\doibase 10.1146/annurev-astro-081913-040019}
  {\bibfield  {journal} {\bibinfo  {journal} {Annual Review of Astronomy and
  Astrophysics}\ }\textbf {\bibinfo {volume} {55}},\ \bibinfo {pages}
  {59–109} (\bibinfo {year} {2017})}\BibitemShut {NoStop}%
\bibitem [{\citenamefont {Mashian}\ \emph {et~al.}(2015)\citenamefont
  {Mashian}, \citenamefont {Oesch},\ and\ \citenamefont
  {Loeb}}]{mashian2015empirical}%
  \BibitemOpen
  \bibfield  {author} {\bibinfo {author} {\bibfnamefont {N.}~\bibnamefont
  {Mashian}}, \bibinfo {author} {\bibfnamefont {P.}~\bibnamefont {Oesch}}, \
  and\ \bibinfo {author} {\bibfnamefont {A.}~\bibnamefont {Loeb}},\ }\href@noop
  {} {\  (\bibinfo {year} {2015})},\ \Eprint {http://arxiv.org/abs/1507.00999}
  {arXiv:1507.00999 [astro-ph.GA]} \BibitemShut {NoStop}%
\bibitem [{\citenamefont {Yung}\ \emph {et~al.}(2018)\citenamefont {Yung},
  \citenamefont {Somerville}, \citenamefont {Finkelstein}, \citenamefont
  {Popping},\ and\ \citenamefont {Dav\'{e}}}]{Yung_2018}%
  \BibitemOpen
  \bibfield  {author} {\bibinfo {author} {\bibfnamefont {L.~Y.~A.}\
  \bibnamefont {Yung}}, \bibinfo {author} {\bibfnamefont {R.~S.}\ \bibnamefont
  {Somerville}}, \bibinfo {author} {\bibfnamefont {S.~L.}\ \bibnamefont
  {Finkelstein}}, \bibinfo {author} {\bibfnamefont {G.}~\bibnamefont
  {Popping}}, \ and\ \bibinfo {author} {\bibfnamefont {R.}~\bibnamefont
  {Dav\'{e}}},\ }\href {\doibase 10.1093/mnras/sty3241} {\bibfield  {journal}
  {\bibinfo  {journal} {Monthly Notices of the Royal Astronomical Society}\
  }\textbf {\bibinfo {volume} {483}},\ \bibinfo {pages} {2983–3006} (\bibinfo
  {year} {2018})}\BibitemShut {NoStop}%
\bibitem [{\citenamefont {Park}\ \emph {et~al.}(2019)\citenamefont {Park},
  \citenamefont {Mesinger}, \citenamefont {Greig},\ and\ \citenamefont
  {Gillet}}]{Park:2018ljd}%
  \BibitemOpen
  \bibfield  {author} {\bibinfo {author} {\bibfnamefont {J.}~\bibnamefont
  {Park}}, \bibinfo {author} {\bibfnamefont {A.}~\bibnamefont {Mesinger}},
  \bibinfo {author} {\bibfnamefont {B.}~\bibnamefont {Greig}}, \ and\ \bibinfo
  {author} {\bibfnamefont {N.}~\bibnamefont {Gillet}},\ }\href {\doibase
  10.1093/mnras/stz032} {\bibfield  {journal} {\bibinfo  {journal} {Mon. Not.
  Roy. Astron. Soc.}\ }\textbf {\bibinfo {volume} {484}},\ \bibinfo {pages}
  {933} (\bibinfo {year} {2019})},\ \Eprint {http://arxiv.org/abs/1809.08995}
  {arXiv:1809.08995 [astro-ph.GA]} \BibitemShut {NoStop}%
\bibitem [{\citenamefont {Gillet}\ \emph {et~al.}(2020)\citenamefont {Gillet},
  \citenamefont {Mesinger},\ and\ \citenamefont {Park}}]{Gillet:2019fjd}%
  \BibitemOpen
  \bibfield  {author} {\bibinfo {author} {\bibfnamefont {N.~J.~F.}\
  \bibnamefont {Gillet}}, \bibinfo {author} {\bibfnamefont {A.}~\bibnamefont
  {Mesinger}}, \ and\ \bibinfo {author} {\bibfnamefont {J.}~\bibnamefont
  {Park}},\ }\href {\doibase 10.1093/mnras/stz2988} {\bibfield  {journal}
  {\bibinfo  {journal} {Mon. Not. Roy. Astron. Soc.}\ }\textbf {\bibinfo
  {volume} {491}},\ \bibinfo {pages} {1980} (\bibinfo {year} {2020})},\ \Eprint
  {http://arxiv.org/abs/1906.06296} {arXiv:1906.06296 [astro-ph.GA]}
  \BibitemShut {NoStop}%
%%CITATION = ARXIV:1906.06296;%%
\bibitem [{\citenamefont {Cai}\ \emph {et~al.}(2014)\citenamefont {Cai},
  \citenamefont {Lapi}, \citenamefont {Bressan}, \citenamefont {De~Zotti},
  \citenamefont {Negrello},\ and\ \citenamefont {Danese}}]{Cai:2014fja}%
  \BibitemOpen
  \bibfield  {author} {\bibinfo {author} {\bibfnamefont {Z.-Y.}\ \bibnamefont
  {Cai}}, \bibinfo {author} {\bibfnamefont {A.}~\bibnamefont {Lapi}}, \bibinfo
  {author} {\bibfnamefont {A.}~\bibnamefont {Bressan}}, \bibinfo {author}
  {\bibfnamefont {G.}~\bibnamefont {De~Zotti}}, \bibinfo {author}
  {\bibfnamefont {M.}~\bibnamefont {Negrello}}, \ and\ \bibinfo {author}
  {\bibfnamefont {L.}~\bibnamefont {Danese}},\ }\href {\doibase
  10.1088/0004-637X/785/1/65} {\bibfield  {journal} {\bibinfo  {journal}
  {Astrophys. J.}\ }\textbf {\bibinfo {volume} {785}},\ \bibinfo {pages} {65}
  (\bibinfo {year} {2014})},\ \Eprint {http://arxiv.org/abs/1403.0055}
  {arXiv:1403.0055 [astro-ph.CO]} \BibitemShut {NoStop}%
\bibitem [{\citenamefont {Sun}\ and\ \citenamefont
  {Furlanetto}(2016)}]{Sun_2016}%
  \BibitemOpen
  \bibfield  {author} {\bibinfo {author} {\bibfnamefont {G.}~\bibnamefont
  {Sun}}\ and\ \bibinfo {author} {\bibfnamefont {S.~R.}\ \bibnamefont
  {Furlanetto}},\ }\href {\doibase 10.1093/mnras/stw980} {\bibfield  {journal}
  {\bibinfo  {journal} {Monthly Notices of the Royal Astronomical Society}\
  }\textbf {\bibinfo {volume} {460}},\ \bibinfo {pages} {417–433} (\bibinfo
  {year} {2016})}\BibitemShut {NoStop}%
\bibitem [{\citenamefont {Tacchella}\ \emph {et~al.}(2018)\citenamefont
  {Tacchella}, \citenamefont {Bose}, \citenamefont {Conroy}, \citenamefont
  {Eisenstein},\ and\ \citenamefont {Johnson}}]{Tacchella:2018qny}%
  \BibitemOpen
  \bibfield  {author} {\bibinfo {author} {\bibfnamefont {S.}~\bibnamefont
  {Tacchella}}, \bibinfo {author} {\bibfnamefont {S.}~\bibnamefont {Bose}},
  \bibinfo {author} {\bibfnamefont {C.}~\bibnamefont {Conroy}}, \bibinfo
  {author} {\bibfnamefont {D.~J.}\ \bibnamefont {Eisenstein}}, \ and\ \bibinfo
  {author} {\bibfnamefont {B.~D.}\ \bibnamefont {Johnson}},\ }\href {\doibase
  10.3847/1538-4357/aae8e0} {\bibfield  {journal} {\bibinfo  {journal}
  {Astrophys. J.}\ }\textbf {\bibinfo {volume} {868}},\ \bibinfo {pages} {92}
  (\bibinfo {year} {2018})},\ \Eprint {http://arxiv.org/abs/1806.03299}
  {arXiv:1806.03299 [astro-ph.GA]} \BibitemShut {NoStop}%
\bibitem [{\citenamefont {Behroozi}\ \emph {et~al.}(2019)\citenamefont
  {Behroozi}, \citenamefont {Wechsler}, \citenamefont {Hearin},\ and\
  \citenamefont {Conroy}}]{Behroozi:2019kql}%
  \BibitemOpen
  \bibfield  {author} {\bibinfo {author} {\bibfnamefont {P.}~\bibnamefont
  {Behroozi}}, \bibinfo {author} {\bibfnamefont {R.~H.}\ \bibnamefont
  {Wechsler}}, \bibinfo {author} {\bibfnamefont {A.~P.}\ \bibnamefont
  {Hearin}}, \ and\ \bibinfo {author} {\bibfnamefont {C.}~\bibnamefont
  {Conroy}},\ }\href {\doibase 10.1093/mnras/stz1182} {\bibfield  {journal}
  {\bibinfo  {journal} {Mon. Not. Roy. Astron. Soc.}\ }\textbf {\bibinfo
  {volume} {488}},\ \bibinfo {pages} {3143} (\bibinfo {year} {2019})},\ \Eprint
  {http://arxiv.org/abs/1806.07893} {arXiv:1806.07893} \BibitemShut {NoStop}%
\bibitem [{\citenamefont {Salvaterra}\ \emph {et~al.}(2011)\citenamefont
  {Salvaterra}, \citenamefont {Ferrara},\ and\ \citenamefont
  {Dayal}}]{Salvaterra:2010nb}%
  \BibitemOpen
  \bibfield  {author} {\bibinfo {author} {\bibfnamefont {R.}~\bibnamefont
  {Salvaterra}}, \bibinfo {author} {\bibfnamefont {A.}~\bibnamefont {Ferrara}},
  \ and\ \bibinfo {author} {\bibfnamefont {P.}~\bibnamefont {Dayal}},\ }\href
  {\doibase 10.1111/j.1365-2966.2010.18155.x} {\bibfield  {journal} {\bibinfo
  {journal} {Mon. Not. Roy. Astron. Soc.}\ }\textbf {\bibinfo {volume} {414}},\
  \bibinfo {pages} {847} (\bibinfo {year} {2011})},\ \Eprint
  {http://arxiv.org/abs/1003.3873} {arXiv:1003.3873 [astro-ph.CO]} \BibitemShut
  {NoStop}%
\bibitem [{\citenamefont {Jaacks}\ \emph {et~al.}(2012)\citenamefont {Jaacks},
  \citenamefont {Choi}, \citenamefont {Nagamine}, \citenamefont {Thompson},\
  and\ \citenamefont {Varghese}}]{Jaacks}%
  \BibitemOpen
  \bibfield  {author} {\bibinfo {author} {\bibfnamefont {J.}~\bibnamefont
  {Jaacks}}, \bibinfo {author} {\bibfnamefont {J.-H.}\ \bibnamefont {Choi}},
  \bibinfo {author} {\bibfnamefont {K.}~\bibnamefont {Nagamine}}, \bibinfo
  {author} {\bibfnamefont {R.}~\bibnamefont {Thompson}}, \ and\ \bibinfo
  {author} {\bibfnamefont {S.}~\bibnamefont {Varghese}},\ }\href {\doibase
  10.1111/j.1365-2966.2011.20150.x} {\bibfield  {journal} {\bibinfo  {journal}
  {Monthly Notices of the Royal Astronomical Society}\ }\textbf {\bibinfo
  {volume} {420}},\ \bibinfo {pages} {1606} (\bibinfo {year} {2012})},\ \Eprint
  {http://arxiv.org/abs/https://academic.oup.com/mnras/article-pdf/420/2/1606/3077125/mnras0420-1606.pdf}
  {https://academic.oup.com/mnras/article-pdf/420/2/1606/3077125/mnras0420-1606.pdf}
  \BibitemShut {NoStop}%
\bibitem [{\citenamefont {Dayal}\ \emph {et~al.}(2013)\citenamefont {Dayal},
  \citenamefont {Dunlop}, \citenamefont {Maio},\ and\ \citenamefont
  {Ciardi}}]{Dayal:2012ah}%
  \BibitemOpen
  \bibfield  {author} {\bibinfo {author} {\bibfnamefont {P.}~\bibnamefont
  {Dayal}}, \bibinfo {author} {\bibfnamefont {J.~S.}\ \bibnamefont {Dunlop}},
  \bibinfo {author} {\bibfnamefont {U.}~\bibnamefont {Maio}}, \ and\ \bibinfo
  {author} {\bibfnamefont {B.}~\bibnamefont {Ciardi}},\ }\href {\doibase
  10.1093/mnras/stt1108} {\bibfield  {journal} {\bibinfo  {journal} {Mon. Not.
  Roy. Astron. Soc.}\ }\textbf {\bibinfo {volume} {434}},\ \bibinfo {pages}
  {1486} (\bibinfo {year} {2013})},\ \Eprint {http://arxiv.org/abs/1211.1034}
  {arXiv:1211.1034 [astro-ph.CO]} \BibitemShut {NoStop}%
\bibitem [{\citenamefont {O'Shea}\ \emph {et~al.}(2015)\citenamefont {O'Shea},
  \citenamefont {Wise}, \citenamefont {Xu},\ and\ \citenamefont
  {Norman}}]{oshea2015probing}%
  \BibitemOpen
  \bibfield  {author} {\bibinfo {author} {\bibfnamefont {B.~W.}\ \bibnamefont
  {O'Shea}}, \bibinfo {author} {\bibfnamefont {J.~H.}\ \bibnamefont {Wise}},
  \bibinfo {author} {\bibfnamefont {H.}~\bibnamefont {Xu}}, \ and\ \bibinfo
  {author} {\bibfnamefont {M.~L.}\ \bibnamefont {Norman}},\ }\href@noop {} {\
  (\bibinfo {year} {2015})},\ \Eprint {http://arxiv.org/abs/1503.01110}
  {arXiv:1503.01110 [astro-ph.GA]} \BibitemShut {NoStop}%
\bibitem [{\citenamefont {Vogelsberger}\ \emph {et~al.}(2020)\citenamefont
  {Vogelsberger} \emph {et~al.}}]{Vogelsberger_dust2020}%
  \BibitemOpen
  \bibfield  {author} {\bibinfo {author} {\bibfnamefont {M.}~\bibnamefont
  {Vogelsberger}} \emph {et~al.},\ }\href {\doibase 10.1093/mnras/staa137}
  {\bibfield  {journal} {\bibinfo  {journal} {Monthly Notices of the Royal
  Astronomical Society}\ }\textbf {\bibinfo {volume} {492}},\ \bibinfo {pages}
  {5167–5201} (\bibinfo {year} {2020})},\ \Eprint
  {http://arxiv.org/abs/1904.07238} {arXiv:1904.07238 [astro-ph.GA]}
  \BibitemShut {NoStop}%
\bibitem [{\citenamefont {Naidu}\ \emph {et~al.}(2019)\citenamefont {Naidu},
  \citenamefont {Tacchella}, \citenamefont {Mason}, \citenamefont {Bose},
  \citenamefont {Oesch},\ and\ \citenamefont {Conroy}}]{Naidu:2019gvi}%
  \BibitemOpen
  \bibfield  {author} {\bibinfo {author} {\bibfnamefont {R.~P.}\ \bibnamefont
  {Naidu}}, \bibinfo {author} {\bibfnamefont {S.}~\bibnamefont {Tacchella}},
  \bibinfo {author} {\bibfnamefont {C.~A.}\ \bibnamefont {Mason}}, \bibinfo
  {author} {\bibfnamefont {S.}~\bibnamefont {Bose}}, \bibinfo {author}
  {\bibfnamefont {P.~A.}\ \bibnamefont {Oesch}}, \ and\ \bibinfo {author}
  {\bibfnamefont {C.}~\bibnamefont {Conroy}},\ }\href {\doibase
  10.3847/1538-4357/ab7cc9} {\  (\bibinfo {year} {2019}),\
  10.3847/1538-4357/ab7cc9},\ \Eprint {http://arxiv.org/abs/1907.13130}
  {arXiv:1907.13130 [astro-ph.GA]} \BibitemShut {NoStop}%
\bibitem [{\citenamefont {Mason}\ \emph
  {et~al.}(2015{\natexlab{a}})\citenamefont {Mason}, \citenamefont {Trenti},\
  and\ \citenamefont {Treu}}]{Mason:2015cna}%
  \BibitemOpen
  \bibfield  {author} {\bibinfo {author} {\bibfnamefont {C.}~\bibnamefont
  {Mason}}, \bibinfo {author} {\bibfnamefont {M.}~\bibnamefont {Trenti}}, \
  and\ \bibinfo {author} {\bibfnamefont {T.}~\bibnamefont {Treu}},\ }\href
  {\doibase 10.1088/0004-637X/813/1/21} {\bibfield  {journal} {\bibinfo
  {journal} {Astrophys. J.}\ }\textbf {\bibinfo {volume} {813}},\ \bibinfo
  {pages} {21} (\bibinfo {year} {2015}{\natexlab{a}})},\ \Eprint
  {http://arxiv.org/abs/1508.01204} {arXiv:1508.01204 [astro-ph.GA]}
  \BibitemShut {NoStop}%
\bibitem [{\citenamefont {Sahl\'en}\ and\ \citenamefont
  {Zackrisson}(2021)}]{Sahlen:2021bqt}%
  \BibitemOpen
  \bibfield  {author} {\bibinfo {author} {\bibfnamefont {M.}~\bibnamefont
  {Sahl\'en}}\ and\ \bibinfo {author} {\bibfnamefont {E.}~\bibnamefont
  {Zackrisson}},\ }\href@noop {} {\  (\bibinfo {year} {2021})},\ \Eprint
  {http://arxiv.org/abs/2105.05098} {arXiv:2105.05098 [astro-ph.CO]}
  \BibitemShut {NoStop}%
\bibitem [{\citenamefont {Romanello}\ \emph {et~al.}(2021)\citenamefont
  {Romanello}, \citenamefont {Menci},\ and\ \citenamefont
  {Castellano}}]{Romanello:2021gnp}%
  \BibitemOpen
  \bibfield  {author} {\bibinfo {author} {\bibfnamefont {M.}~\bibnamefont
  {Romanello}}, \bibinfo {author} {\bibfnamefont {N.}~\bibnamefont {Menci}}, \
  and\ \bibinfo {author} {\bibfnamefont {M.}~\bibnamefont {Castellano}},\
  }\href {\doibase 10.3390/universe7100365} {\bibfield  {journal} {\bibinfo
  {journal} {Universe}\ }\textbf {\bibinfo {volume} {7}},\ \bibinfo {pages}
  {365} (\bibinfo {year} {2021})},\ \Eprint {http://arxiv.org/abs/2110.05262}
  {arXiv:2110.05262 [astro-ph.CO]} \BibitemShut {NoStop}%
\bibitem [{\citenamefont {Bozek}\ \emph {et~al.}(2015)\citenamefont {Bozek},
  \citenamefont {Marsh}, \citenamefont {Silk},\ and\ \citenamefont
  {Wyse}}]{Bozek:2014uqa}%
  \BibitemOpen
  \bibfield  {author} {\bibinfo {author} {\bibfnamefont {B.}~\bibnamefont
  {Bozek}}, \bibinfo {author} {\bibfnamefont {D.~J.~E.}\ \bibnamefont {Marsh}},
  \bibinfo {author} {\bibfnamefont {J.}~\bibnamefont {Silk}}, \ and\ \bibinfo
  {author} {\bibfnamefont {R.~F.~G.}\ \bibnamefont {Wyse}},\ }\href {\doibase
  10.1093/mnras/stv624} {\bibfield  {journal} {\bibinfo  {journal} {Mon. Not.
  Roy. Astron. Soc.}\ }\textbf {\bibinfo {volume} {450}},\ \bibinfo {pages}
  {209} (\bibinfo {year} {2015})},\ \Eprint {http://arxiv.org/abs/1409.3544}
  {arXiv:1409.3544 [astro-ph.CO]} \BibitemShut {NoStop}%
\bibitem [{\citenamefont {Schultz}\ \emph {et~al.}(2014)\citenamefont
  {Schultz}, \citenamefont {O\~norbe}, \citenamefont {Abazajian},\ and\
  \citenamefont {Bullock}}]{Schultz:2014eia}%
  \BibitemOpen
  \bibfield  {author} {\bibinfo {author} {\bibfnamefont {C.}~\bibnamefont
  {Schultz}}, \bibinfo {author} {\bibfnamefont {J.}~\bibnamefont {O\~norbe}},
  \bibinfo {author} {\bibfnamefont {K.~N.}\ \bibnamefont {Abazajian}}, \ and\
  \bibinfo {author} {\bibfnamefont {J.~S.}\ \bibnamefont {Bullock}},\ }\href
  {\doibase 10.1093/mnras/stu976} {\bibfield  {journal} {\bibinfo  {journal}
  {Mon. Not. Roy. Astron. Soc.}\ }\textbf {\bibinfo {volume} {442}},\ \bibinfo
  {pages} {1597} (\bibinfo {year} {2014})},\ \Eprint
  {http://arxiv.org/abs/1401.3769} {arXiv:1401.3769 [astro-ph.CO]} \BibitemShut
  {NoStop}%
\bibitem [{\citenamefont {Dayal}\ \emph {et~al.}(2015)\citenamefont {Dayal},
  \citenamefont {Mesinger},\ and\ \citenamefont {Pacucci}}]{Dayal:2014nva}%
  \BibitemOpen
  \bibfield  {author} {\bibinfo {author} {\bibfnamefont {P.}~\bibnamefont
  {Dayal}}, \bibinfo {author} {\bibfnamefont {A.}~\bibnamefont {Mesinger}}, \
  and\ \bibinfo {author} {\bibfnamefont {F.}~\bibnamefont {Pacucci}},\ }\href
  {\doibase 10.1088/0004-637X/806/1/67} {\bibfield  {journal} {\bibinfo
  {journal} {Astrophys. J.}\ }\textbf {\bibinfo {volume} {806}},\ \bibinfo
  {pages} {67} (\bibinfo {year} {2015})},\ \Eprint
  {http://arxiv.org/abs/1408.1102} {arXiv:1408.1102 [astro-ph.GA]} \BibitemShut
  {NoStop}%
\bibitem [{\citenamefont {Corasaniti}\ \emph {et~al.}(2017)\citenamefont
  {Corasaniti}, \citenamefont {Agarwal}, \citenamefont {Marsh},\ and\
  \citenamefont {Das}}]{Corasaniti:2016epp}%
  \BibitemOpen
  \bibfield  {author} {\bibinfo {author} {\bibfnamefont {P.}~\bibnamefont
  {Corasaniti}}, \bibinfo {author} {\bibfnamefont {S.}~\bibnamefont {Agarwal}},
  \bibinfo {author} {\bibfnamefont {D.}~\bibnamefont {Marsh}}, \ and\ \bibinfo
  {author} {\bibfnamefont {S.}~\bibnamefont {Das}},\ }\href {\doibase
  10.1103/PhysRevD.95.083512} {\bibfield  {journal} {\bibinfo  {journal} {Phys.
  Rev. D}\ }\textbf {\bibinfo {volume} {95}},\ \bibinfo {pages} {083512}
  (\bibinfo {year} {2017})},\ \Eprint {http://arxiv.org/abs/1611.05892}
  {arXiv:1611.05892 [astro-ph.CO]} \BibitemShut {NoStop}%
\bibitem [{\citenamefont {Menci}\ \emph {et~al.}(2017)\citenamefont {Menci},
  \citenamefont {Merle}, \citenamefont {Totzauer}, \citenamefont {Schneider},
  \citenamefont {Grazian}, \citenamefont {Castellano},\ and\ \citenamefont
  {Sanchez}}]{Menci:2017nsr}%
  \BibitemOpen
  \bibfield  {author} {\bibinfo {author} {\bibfnamefont {N.}~\bibnamefont
  {Menci}}, \bibinfo {author} {\bibfnamefont {A.}~\bibnamefont {Merle}},
  \bibinfo {author} {\bibfnamefont {M.}~\bibnamefont {Totzauer}}, \bibinfo
  {author} {\bibfnamefont {A.}~\bibnamefont {Schneider}}, \bibinfo {author}
  {\bibfnamefont {A.}~\bibnamefont {Grazian}}, \bibinfo {author} {\bibfnamefont
  {M.}~\bibnamefont {Castellano}}, \ and\ \bibinfo {author} {\bibfnamefont
  {N.~G.}\ \bibnamefont {Sanchez}},\ }\href {\doibase
  10.3847/1538-4357/836/1/61} {\bibfield  {journal} {\bibinfo  {journal}
  {Astrophys. J.}\ }\textbf {\bibinfo {volume} {836}},\ \bibinfo {pages} {61}
  (\bibinfo {year} {2017})},\ \Eprint {http://arxiv.org/abs/1701.01339}
  {arXiv:1701.01339 [astro-ph.CO]} \BibitemShut {NoStop}%
\bibitem [{\citenamefont {Menci}\ \emph {et~al.}(2018)\citenamefont {Menci},
  \citenamefont {Grazian}, \citenamefont {Lamastra}, \citenamefont {Calura},
  \citenamefont {Castellano},\ and\ \citenamefont {Santini}}]{Menci:2018lis}%
  \BibitemOpen
  \bibfield  {author} {\bibinfo {author} {\bibfnamefont {N.}~\bibnamefont
  {Menci}}, \bibinfo {author} {\bibfnamefont {A.}~\bibnamefont {Grazian}},
  \bibinfo {author} {\bibfnamefont {A.}~\bibnamefont {Lamastra}}, \bibinfo
  {author} {\bibfnamefont {F.}~\bibnamefont {Calura}}, \bibinfo {author}
  {\bibfnamefont {M.}~\bibnamefont {Castellano}}, \ and\ \bibinfo {author}
  {\bibfnamefont {P.}~\bibnamefont {Santini}},\ }\href {\doibase
  10.3847/1538-4357/aaa773} {\bibfield  {journal} {\bibinfo  {journal}
  {Astrophys. J.}\ }\textbf {\bibinfo {volume} {854}},\ \bibinfo {pages} {1}
  (\bibinfo {year} {2018})},\ \Eprint {http://arxiv.org/abs/1801.03697}
  {arXiv:1801.03697 [astro-ph.CO]} \BibitemShut {NoStop}%
\bibitem [{\citenamefont {Rudakovskyi}\ \emph {et~al.}(2021)\citenamefont
  {Rudakovskyi}, \citenamefont {Mesinger}, \citenamefont {Savchenko},\ and\
  \citenamefont {Gillet}}]{Rudakovskyi:2021jyf}%
  \BibitemOpen
  \bibfield  {author} {\bibinfo {author} {\bibfnamefont {A.}~\bibnamefont
  {Rudakovskyi}}, \bibinfo {author} {\bibfnamefont {A.}~\bibnamefont
  {Mesinger}}, \bibinfo {author} {\bibfnamefont {D.}~\bibnamefont {Savchenko}},
  \ and\ \bibinfo {author} {\bibfnamefont {N.}~\bibnamefont {Gillet}},\ }\href
  {\doibase 10.1093/mnras/stab2333} {\  (\bibinfo {year} {2021}),\
  10.1093/mnras/stab2333},\ \Eprint {http://arxiv.org/abs/2104.04481}
  {arXiv:2104.04481 [astro-ph.CO]} \BibitemShut {NoStop}%
\bibitem [{\citenamefont {Menci}\ \emph {et~al.}(2020)\citenamefont {Menci}
  \emph {et~al.}}]{Menci:2020ybl}%
  \BibitemOpen
  \bibfield  {author} {\bibinfo {author} {\bibfnamefont {N.}~\bibnamefont
  {Menci}} \emph {et~al.},\ }\href {\doibase 10.3847/1538-4357/aba9d2}
  {\bibfield  {journal} {\bibinfo  {journal} {Astrophys. J.}\ }\textbf
  {\bibinfo {volume} {900}},\ \bibinfo {pages} {108} (\bibinfo {year}
  {2020})},\ \Eprint {http://arxiv.org/abs/2007.12453} {arXiv:2007.12453
  [astro-ph.CO]} \BibitemShut {NoStop}%
\bibitem [{\citenamefont {Yoshiura}\ \emph {et~al.}(2020)\citenamefont
  {Yoshiura}, \citenamefont {Oguri}, \citenamefont {Takahashi},\ and\
  \citenamefont {Takahashi}}]{Yoshiura:2020soa}%
  \BibitemOpen
  \bibfield  {author} {\bibinfo {author} {\bibfnamefont {S.}~\bibnamefont
  {Yoshiura}}, \bibinfo {author} {\bibfnamefont {M.}~\bibnamefont {Oguri}},
  \bibinfo {author} {\bibfnamefont {K.}~\bibnamefont {Takahashi}}, \ and\
  \bibinfo {author} {\bibfnamefont {T.}~\bibnamefont {Takahashi}},\ }\href
  {\doibase 10.1103/PhysRevD.102.083515} {\bibfield  {journal} {\bibinfo
  {journal} {Phys. Rev. D}\ }\textbf {\bibinfo {volume} {102}},\ \bibinfo
  {pages} {083515} (\bibinfo {year} {2020})},\ \Eprint
  {http://arxiv.org/abs/2007.14695} {arXiv:2007.14695 [astro-ph.CO]}
  \BibitemShut {NoStop}%
\bibitem [{\citenamefont {Chevallard}\ \emph {et~al.}(2015)\citenamefont
  {Chevallard}, \citenamefont {Silk}, \citenamefont {Nishimichi}, \citenamefont
  {Habouzit}, \citenamefont {Mamon},\ and\ \citenamefont
  {Peirani}}]{Chevallard:2014sxa}%
  \BibitemOpen
  \bibfield  {author} {\bibinfo {author} {\bibfnamefont {J.}~\bibnamefont
  {Chevallard}}, \bibinfo {author} {\bibfnamefont {J.}~\bibnamefont {Silk}},
  \bibinfo {author} {\bibfnamefont {T.}~\bibnamefont {Nishimichi}}, \bibinfo
  {author} {\bibfnamefont {M.}~\bibnamefont {Habouzit}}, \bibinfo {author}
  {\bibfnamefont {G.~A.}\ \bibnamefont {Mamon}}, \ and\ \bibinfo {author}
  {\bibfnamefont {S.}~\bibnamefont {Peirani}},\ }\href {\doibase
  10.1093/mnras/stu2280} {\bibfield  {journal} {\bibinfo  {journal} {Mon. Not.
  Roy. Astron. Soc.}\ }\textbf {\bibinfo {volume} {446}},\ \bibinfo {pages}
  {3235} (\bibinfo {year} {2015})},\ \Eprint {http://arxiv.org/abs/1410.7768}
  {arXiv:1410.7768 [astro-ph.CO]} \BibitemShut {NoStop}%
\bibitem [{\citenamefont {Sabti}\ \emph
  {et~al.}(2021{\natexlab{a}})\citenamefont {Sabti}, \citenamefont {Mu\~noz},\
  and\ \citenamefont {Blas}}]{Sabti:2020ser}%
  \BibitemOpen
  \bibfield  {author} {\bibinfo {author} {\bibfnamefont {N.}~\bibnamefont
  {Sabti}}, \bibinfo {author} {\bibfnamefont {J.~B.}\ \bibnamefont {Mu\~noz}},
  \ and\ \bibinfo {author} {\bibfnamefont {D.}~\bibnamefont {Blas}},\ }\href
  {\doibase 10.1088/1475-7516/2021/01/010} {\bibfield  {journal} {\bibinfo
  {journal} {JCAP}\ }\textbf {\bibinfo {volume} {01}},\ \bibinfo {pages} {010}
  (\bibinfo {year} {2021}{\natexlab{a}})},\ \Eprint
  {http://arxiv.org/abs/2009.01245} {arXiv:2009.01245 [astro-ph.CO]}
  \BibitemShut {NoStop}%
\bibitem [{\citenamefont {Audren}\ \emph {et~al.}(2013)\citenamefont {Audren},
  \citenamefont {Lesgourgues}, \citenamefont {Benabed},\ and\ \citenamefont
  {Prunet}}]{Audren:2012wb}%
  \BibitemOpen
  \bibfield  {author} {\bibinfo {author} {\bibfnamefont {B.}~\bibnamefont
  {Audren}}, \bibinfo {author} {\bibfnamefont {J.}~\bibnamefont {Lesgourgues}},
  \bibinfo {author} {\bibfnamefont {K.}~\bibnamefont {Benabed}}, \ and\
  \bibinfo {author} {\bibfnamefont {S.}~\bibnamefont {Prunet}},\ }\href
  {\doibase 10.1088/1475-7516/2013/02/001} {\bibfield  {journal} {\bibinfo
  {journal} {JCAP}\ }\textbf {\bibinfo {volume} {1302}},\ \bibinfo {pages}
  {001} (\bibinfo {year} {2013})},\ \Eprint {http://arxiv.org/abs/1210.7183}
  {arXiv:1210.7183 [astro-ph.CO]} \BibitemShut {NoStop}%
%%CITATION = ARXIV:1210.7183;%%
\bibitem [{\citenamefont {Brinckmann}\ and\ \citenamefont
  {Lesgourgues}(2018)}]{Brinckmann:2018cvx}%
  \BibitemOpen
  \bibfield  {author} {\bibinfo {author} {\bibfnamefont {T.}~\bibnamefont
  {Brinckmann}}\ and\ \bibinfo {author} {\bibfnamefont {J.}~\bibnamefont
  {Lesgourgues}},\ }\href@noop {} {\  (\bibinfo {year} {2018})},\ \Eprint
  {http://arxiv.org/abs/1804.07261} {arXiv:1804.07261 [astro-ph.CO]}
  \BibitemShut {NoStop}%
%%CITATION = ARXIV:1804.07261;%%
\bibitem [{\citenamefont {Lesgourgues}(2011)}]{Lesgourgues:2011re}%
  \BibitemOpen
  \bibfield  {author} {\bibinfo {author} {\bibfnamefont {J.}~\bibnamefont
  {Lesgourgues}},\ }\href@noop {} {\  (\bibinfo {year} {2011})},\ \Eprint
  {http://arxiv.org/abs/1104.2932} {arXiv:1104.2932 [astro-ph.IM]} \BibitemShut
  {NoStop}%
\bibitem [{\citenamefont {Blas}\ \emph {et~al.}(2011)\citenamefont {Blas},
  \citenamefont {Lesgourgues},\ and\ \citenamefont {Tram}}]{Blas:2011rf}%
  \BibitemOpen
  \bibfield  {author} {\bibinfo {author} {\bibfnamefont {D.}~\bibnamefont
  {Blas}}, \bibinfo {author} {\bibfnamefont {J.}~\bibnamefont {Lesgourgues}}, \
  and\ \bibinfo {author} {\bibfnamefont {T.}~\bibnamefont {Tram}},\ }\href
  {\doibase 10.1088/1475-7516/2011/07/034} {\bibfield  {journal} {\bibinfo
  {journal} {JCAP}\ }\textbf {\bibinfo {volume} {07}},\ \bibinfo {pages} {034}
  (\bibinfo {year} {2011})},\ \Eprint {http://arxiv.org/abs/1104.2933}
  {arXiv:1104.2933 [astro-ph.CO]} \BibitemShut {NoStop}%
\bibitem [{\citenamefont {Sabti}\ \emph
  {et~al.}(2021{\natexlab{b}})\citenamefont {Sabti}, \citenamefont {Mu\~noz},\
  and\ \citenamefont {Blas}}]{Sabti:2021unj}%
  \BibitemOpen
  \bibfield  {author} {\bibinfo {author} {\bibfnamefont {N.}~\bibnamefont
  {Sabti}}, \bibinfo {author} {\bibfnamefont {J.~B.}\ \bibnamefont {Mu\~noz}},
  \ and\ \bibinfo {author} {\bibfnamefont {D.}~\bibnamefont {Blas}},\
  }\href@noop {} {\  (\bibinfo {year} {2021}{\natexlab{b}})},\ \Eprint
  {http://arxiv.org/abs/2110.13161} {arXiv:2110.13161 [astro-ph.CO]}
  \BibitemShut {NoStop}%
\bibitem [{\citenamefont {Williams}\ \emph {et~al.}(2018)\citenamefont
  {Williams} \emph {et~al.}}]{Williams_2018}%
  \BibitemOpen
  \bibfield  {author} {\bibinfo {author} {\bibfnamefont {C.~C.}\ \bibnamefont
  {Williams}} \emph {et~al.},\ }\href {\doibase 10.3847/1538-4365/aabcbb}
  {\bibfield  {journal} {\bibinfo  {journal} {The Astrophysical Journal
  Supplement Series}\ }\textbf {\bibinfo {volume} {236}},\ \bibinfo {pages}
  {33} (\bibinfo {year} {2018})}\BibitemShut {NoStop}%
\bibitem [{\citenamefont {Bhowmick}\ \emph {et~al.}(2019)\citenamefont
  {Bhowmick}, \citenamefont {Somerville}, \citenamefont {DiMatteo},
  \citenamefont {Wilkins}, \citenamefont {Feng},\ and\ \citenamefont
  {Tenneti}}]{Bhowmick:2019nnj}%
  \BibitemOpen
  \bibfield  {author} {\bibinfo {author} {\bibfnamefont {A.~K.}\ \bibnamefont
  {Bhowmick}}, \bibinfo {author} {\bibfnamefont {R.~S.}\ \bibnamefont
  {Somerville}}, \bibinfo {author} {\bibfnamefont {T.}~\bibnamefont
  {DiMatteo}}, \bibinfo {author} {\bibfnamefont {S.}~\bibnamefont {Wilkins}},
  \bibinfo {author} {\bibfnamefont {Y.}~\bibnamefont {Feng}}, \ and\ \bibinfo
  {author} {\bibfnamefont {A.}~\bibnamefont {Tenneti}},\ }\href {\doibase
  10.1093/mnras/staa1605} {\  (\bibinfo {year} {2019}),\
  10.1093/mnras/staa1605},\ \Eprint {http://arxiv.org/abs/1908.02787}
  {arXiv:1908.02787 [astro-ph.CO]} \BibitemShut {NoStop}%
\bibitem [{\citenamefont {Jenkins}\ \emph {et~al.}(2001)\citenamefont
  {Jenkins}, \citenamefont {Frenk}, \citenamefont {White}, \citenamefont
  {Colberg}, \citenamefont {Cole}, \citenamefont {Evrard}, \citenamefont
  {Couchman},\ and\ \citenamefont {Yoshida}}]{Jenkins:2000bv}%
  \BibitemOpen
  \bibfield  {author} {\bibinfo {author} {\bibfnamefont {A.}~\bibnamefont
  {Jenkins}}, \bibinfo {author} {\bibfnamefont {C.}~\bibnamefont {Frenk}},
  \bibinfo {author} {\bibfnamefont {S.~D.}\ \bibnamefont {White}}, \bibinfo
  {author} {\bibfnamefont {J.}~\bibnamefont {Colberg}}, \bibinfo {author}
  {\bibfnamefont {S.}~\bibnamefont {Cole}}, \bibinfo {author} {\bibfnamefont
  {A.~E.}\ \bibnamefont {Evrard}}, \bibinfo {author} {\bibfnamefont
  {H.}~\bibnamefont {Couchman}}, \ and\ \bibinfo {author} {\bibfnamefont
  {N.}~\bibnamefont {Yoshida}},\ }\href {\doibase
  10.1046/j.1365-8711.2001.04029.x} {\bibfield  {journal} {\bibinfo  {journal}
  {Mon. Not. Roy. Astron. Soc.}\ }\textbf {\bibinfo {volume} {321}},\ \bibinfo
  {pages} {372} (\bibinfo {year} {2001})},\ \Eprint
  {http://arxiv.org/abs/astro-ph/0005260} {arXiv:astro-ph/0005260} \BibitemShut
  {NoStop}%
\bibitem [{\citenamefont {Lewis}\ \emph {et~al.}(2000)\citenamefont {Lewis},
  \citenamefont {Challinor},\ and\ \citenamefont {Lasenby}}]{Lewis:1999bs}%
  \BibitemOpen
  \bibfield  {author} {\bibinfo {author} {\bibfnamefont {A.}~\bibnamefont
  {Lewis}}, \bibinfo {author} {\bibfnamefont {A.}~\bibnamefont {Challinor}}, \
  and\ \bibinfo {author} {\bibfnamefont {A.}~\bibnamefont {Lasenby}},\ }\href
  {\doibase 10.1086/309179} {\bibfield  {journal} {\bibinfo  {journal} {\apj}\
  }\textbf {\bibinfo {volume} {538}},\ \bibinfo {pages} {473} (\bibinfo {year}
  {2000})},\ \Eprint {http://arxiv.org/abs/astro-ph/9911177}
  {arXiv:astro-ph/9911177 [astro-ph]} \BibitemShut {NoStop}%
%%CITATION = ASTRO-PH/9911177;%%
\bibitem [{\citenamefont {Press}\ and\ \citenamefont
  {Schechter}(1974)}]{Press:1973iz}%
  \BibitemOpen
  \bibfield  {author} {\bibinfo {author} {\bibfnamefont {W.~H.}\ \bibnamefont
  {Press}}\ and\ \bibinfo {author} {\bibfnamefont {P.}~\bibnamefont
  {Schechter}},\ }\href {\doibase 10.1086/152650} {\bibfield  {journal}
  {\bibinfo  {journal} {Astrophys. J.}\ }\textbf {\bibinfo {volume} {187}},\
  \bibinfo {pages} {425} (\bibinfo {year} {1974})}\BibitemShut {NoStop}%
\bibitem [{\citenamefont {Bond}\ \emph {et~al.}(1991)\citenamefont {Bond},
  \citenamefont {Cole}, \citenamefont {Efstathiou},\ and\ \citenamefont
  {Kaiser}}]{Bond:1990iw}%
  \BibitemOpen
  \bibfield  {author} {\bibinfo {author} {\bibfnamefont {J.~R.}\ \bibnamefont
  {Bond}}, \bibinfo {author} {\bibfnamefont {S.}~\bibnamefont {Cole}}, \bibinfo
  {author} {\bibfnamefont {G.}~\bibnamefont {Efstathiou}}, \ and\ \bibinfo
  {author} {\bibfnamefont {N.}~\bibnamefont {Kaiser}},\ }\href {\doibase
  10.1086/170520} {\bibfield  {journal} {\bibinfo  {journal} {Astrophys. J.}\
  }\textbf {\bibinfo {volume} {379}},\ \bibinfo {pages} {440} (\bibinfo {year}
  {1991})}\BibitemShut {NoStop}%
\bibitem [{\citenamefont {Sheth}\ and\ \citenamefont
  {Tormen}(1999)}]{Sheth:1999mn}%
  \BibitemOpen
  \bibfield  {author} {\bibinfo {author} {\bibfnamefont {R.~K.}\ \bibnamefont
  {Sheth}}\ and\ \bibinfo {author} {\bibfnamefont {G.}~\bibnamefont {Tormen}},\
  }\href {\doibase 10.1046/j.1365-8711.1999.02692.x} {\bibfield  {journal}
  {\bibinfo  {journal} {Mon. Not. Roy. Astron. Soc.}\ }\textbf {\bibinfo
  {volume} {308}},\ \bibinfo {pages} {119} (\bibinfo {year} {1999})},\ \Eprint
  {http://arxiv.org/abs/astro-ph/9901122} {arXiv:astro-ph/9901122} \BibitemShut
  {NoStop}%
\bibitem [{\citenamefont {Sheth}\ \emph {et~al.}(2001)\citenamefont {Sheth},
  \citenamefont {Mo},\ and\ \citenamefont {Tormen}}]{Sheth:1999su}%
  \BibitemOpen
  \bibfield  {author} {\bibinfo {author} {\bibfnamefont {R.~K.}\ \bibnamefont
  {Sheth}}, \bibinfo {author} {\bibfnamefont {H.~J.}\ \bibnamefont {Mo}}, \
  and\ \bibinfo {author} {\bibfnamefont {G.}~\bibnamefont {Tormen}},\ }\href
  {\doibase 10.1046/j.1365-8711.2001.04006.x} {\bibfield  {journal} {\bibinfo
  {journal} {Mon. Not. Roy. Astron. Soc.}\ }\textbf {\bibinfo {volume} {323}},\
  \bibinfo {pages} {1} (\bibinfo {year} {2001})},\ \Eprint
  {http://arxiv.org/abs/astro-ph/9907024} {arXiv:astro-ph/9907024} \BibitemShut
  {NoStop}%
\bibitem [{\citenamefont {Sheth}\ and\ \citenamefont
  {Tormen}(2002)}]{Sheth:2001dp}%
  \BibitemOpen
  \bibfield  {author} {\bibinfo {author} {\bibfnamefont {R.~K.}\ \bibnamefont
  {Sheth}}\ and\ \bibinfo {author} {\bibfnamefont {G.}~\bibnamefont {Tormen}},\
  }\href {\doibase 10.1046/j.1365-8711.2002.04950.x} {\bibfield  {journal}
  {\bibinfo  {journal} {Mon. Not. Roy. Astron. Soc.}\ }\textbf {\bibinfo
  {volume} {329}},\ \bibinfo {pages} {61} (\bibinfo {year} {2002})},\ \Eprint
  {http://arxiv.org/abs/astro-ph/0105113} {arXiv:astro-ph/0105113} \BibitemShut
  {NoStop}%
\bibitem [{\citenamefont {Lukic}\ \emph {et~al.}(2007)\citenamefont {Lukic},
  \citenamefont {Heitmann}, \citenamefont {Habib}, \citenamefont {Bashinsky},\
  and\ \citenamefont {Ricker}}]{Lukic:2007fc}%
  \BibitemOpen
  \bibfield  {author} {\bibinfo {author} {\bibfnamefont {Z.}~\bibnamefont
  {Lukic}}, \bibinfo {author} {\bibfnamefont {K.}~\bibnamefont {Heitmann}},
  \bibinfo {author} {\bibfnamefont {S.}~\bibnamefont {Habib}}, \bibinfo
  {author} {\bibfnamefont {S.}~\bibnamefont {Bashinsky}}, \ and\ \bibinfo
  {author} {\bibfnamefont {P.~M.}\ \bibnamefont {Ricker}},\ }\href {\doibase
  10.1086/523083} {\bibfield  {journal} {\bibinfo  {journal} {Astrophys. J.}\
  }\textbf {\bibinfo {volume} {671}},\ \bibinfo {pages} {1160} (\bibinfo {year}
  {2007})},\ \Eprint {http://arxiv.org/abs/astro-ph/0702360}
  {arXiv:astro-ph/0702360} \BibitemShut {NoStop}%
\bibitem [{\citenamefont {Schneider}(2015)}]{Schneider:2014rda}%
  \BibitemOpen
  \bibfield  {author} {\bibinfo {author} {\bibfnamefont {A.}~\bibnamefont
  {Schneider}},\ }\href {\doibase 10.1093/mnras/stv1169} {\bibfield  {journal}
  {\bibinfo  {journal} {Mon. Not. Roy. Astron. Soc.}\ }\textbf {\bibinfo
  {volume} {451}},\ \bibinfo {pages} {3117} (\bibinfo {year} {2015})},\ \Eprint
  {http://arxiv.org/abs/1412.2133} {arXiv:1412.2133 [astro-ph.CO]} \BibitemShut
  {NoStop}%
\bibitem [{\citenamefont {Mirocha}\ \emph
  {et~al.}(2020{\natexlab{a}})\citenamefont {Mirocha}, \citenamefont
  {Lamarre},\ and\ \citenamefont {Liu}}]{Mirocha:2020qto}%
  \BibitemOpen
  \bibfield  {author} {\bibinfo {author} {\bibfnamefont {J.}~\bibnamefont
  {Mirocha}}, \bibinfo {author} {\bibfnamefont {H.}~\bibnamefont {Lamarre}}, \
  and\ \bibinfo {author} {\bibfnamefont {A.}~\bibnamefont {Liu}},\ }\href
  {\doibase 10.1093/mnras/stab949} {\  (\bibinfo {year} {2020}{\natexlab{a}}),\
  10.1093/mnras/stab949},\ \Eprint {http://arxiv.org/abs/2012.06588}
  {arXiv:2012.06588 [astro-ph.CO]} \BibitemShut {NoStop}%
\bibitem [{\citenamefont {Behroozi}\ \emph {et~al.}(2021)\citenamefont
  {Behroozi}, \citenamefont {Hearin},\ and\ \citenamefont
  {Moster}}]{behroozi2021observational}%
  \BibitemOpen
  \bibfield  {author} {\bibinfo {author} {\bibfnamefont {P.}~\bibnamefont
  {Behroozi}}, \bibinfo {author} {\bibfnamefont {A.}~\bibnamefont {Hearin}}, \
  and\ \bibinfo {author} {\bibfnamefont {B.~P.}\ \bibnamefont {Moster}},\
  }\href@noop {} {\  (\bibinfo {year} {2021})},\ \Eprint
  {http://arxiv.org/abs/2101.05280} {arXiv:2101.05280 [astro-ph.GA]}
  \BibitemShut {NoStop}%
\bibitem [{\citenamefont {Zhang}\ \emph {et~al.}(2021)\citenamefont {Zhang},
  \citenamefont {Yang},\ and\ \citenamefont {Guo}}]{Zhang:2021aau}%
  \BibitemOpen
  \bibfield  {author} {\bibinfo {author} {\bibfnamefont {Y.}~\bibnamefont
  {Zhang}}, \bibinfo {author} {\bibfnamefont {X.}~\bibnamefont {Yang}}, \ and\
  \bibinfo {author} {\bibfnamefont {H.}~\bibnamefont {Guo}},\ }\href@noop {} {\
   (\bibinfo {year} {2021})},\ \Eprint {http://arxiv.org/abs/2108.11565}
  {arXiv:2108.11565 [astro-ph.GA]} \BibitemShut {NoStop}%
\bibitem [{\citenamefont {{Kennicutt}}(1983)}]{Kennicutt1983}%
  \BibitemOpen
  \bibfield  {author} {\bibinfo {author} {\bibfnamefont {J.}~\bibnamefont
  {{Kennicutt}}, \bibfnamefont {R.~C.}},\ }\href {\doibase 10.1086/161261}
  {\bibfield  {journal} {\bibinfo  {journal} {\apj}\ }\textbf {\bibinfo
  {volume} {272}},\ \bibinfo {pages} {54} (\bibinfo {year} {1983})}\BibitemShut
  {NoStop}%
\bibitem [{\citenamefont {Salim}\ \emph {et~al.}(2007)\citenamefont {Salim}
  \emph {et~al.}}]{Salim:2007is}%
  \BibitemOpen
  \bibfield  {author} {\bibinfo {author} {\bibfnamefont {S.}~\bibnamefont
  {Salim}} \emph {et~al.},\ }\href {\doibase 10.1086/519218} {\bibfield
  {journal} {\bibinfo  {journal} {Astrophys. J. Suppl.}\ }\textbf {\bibinfo
  {volume} {173}},\ \bibinfo {pages} {267} (\bibinfo {year} {2007})},\ \Eprint
  {http://arxiv.org/abs/0704.3611} {arXiv:0704.3611 [astro-ph]} \BibitemShut
  {NoStop}%
\bibitem [{\citenamefont {Moster}\ \emph {et~al.}(2018)\citenamefont {Moster},
  \citenamefont {Naab},\ and\ \citenamefont {White}}]{Moster_2018}%
  \BibitemOpen
  \bibfield  {author} {\bibinfo {author} {\bibfnamefont {B.~P.}\ \bibnamefont
  {Moster}}, \bibinfo {author} {\bibfnamefont {T.}~\bibnamefont {Naab}}, \ and\
  \bibinfo {author} {\bibfnamefont {S.~D.~M.}\ \bibnamefont {White}},\ }\href
  {\doibase 10.1093/mnras/sty655} {\bibfield  {journal} {\bibinfo  {journal}
  {Monthly Notices of the Royal Astronomical Society}\ }\textbf {\bibinfo
  {volume} {477}},\ \bibinfo {pages} {1822–1852} (\bibinfo {year}
  {2018})}\BibitemShut {NoStop}%
\bibitem [{\citenamefont {Fabian}(2012)}]{Fabian:2012xr}%
  \BibitemOpen
  \bibfield  {author} {\bibinfo {author} {\bibfnamefont {A.~C.}\ \bibnamefont
  {Fabian}},\ }\href {\doibase 10.1146/annurev-astro-081811-125521} {\bibfield
  {journal} {\bibinfo  {journal} {Ann. Rev. Astron. Astrophys.}\ }\textbf
  {\bibinfo {volume} {50}},\ \bibinfo {pages} {455} (\bibinfo {year} {2012})},\
  \Eprint {http://arxiv.org/abs/1204.4114} {arXiv:1204.4114 [astro-ph.CO]}
  \BibitemShut {NoStop}%
\bibitem [{\citenamefont {Madau}\ \emph {et~al.}(1998)\citenamefont {Madau},
  \citenamefont {Pozzetti},\ and\ \citenamefont {Dickinson}}]{Madau:1997pg}%
  \BibitemOpen
  \bibfield  {author} {\bibinfo {author} {\bibfnamefont {P.}~\bibnamefont
  {Madau}}, \bibinfo {author} {\bibfnamefont {L.}~\bibnamefont {Pozzetti}}, \
  and\ \bibinfo {author} {\bibfnamefont {M.}~\bibnamefont {Dickinson}},\ }\href
  {\doibase 10.1086/305523} {\bibfield  {journal} {\bibinfo  {journal}
  {Astrophys. J.}\ }\textbf {\bibinfo {volume} {498}},\ \bibinfo {pages} {106}
  (\bibinfo {year} {1998})},\ \Eprint {http://arxiv.org/abs/astro-ph/9708220}
  {arXiv:astro-ph/9708220} \BibitemShut {NoStop}%
\bibitem [{\citenamefont {Kennicutt}(1998)}]{Kennicutt:1998zb}%
  \BibitemOpen
  \bibfield  {author} {\bibinfo {author} {\bibfnamefont {J.}~\bibnamefont
  {Kennicutt}, \bibfnamefont {Robert~C.}},\ }\href {\doibase
  10.1146/annurev.astro.36.1.189} {\bibfield  {journal} {\bibinfo  {journal}
  {Ann. Rev. Astron. Astrophys.}\ }\textbf {\bibinfo {volume} {36}},\ \bibinfo
  {pages} {189} (\bibinfo {year} {1998})},\ \Eprint
  {http://arxiv.org/abs/astro-ph/9807187} {arXiv:astro-ph/9807187} \BibitemShut
  {NoStop}%
\bibitem [{\citenamefont {Madau}\ and\ \citenamefont
  {Dickinson}(2014)}]{Madau:2014bja}%
  \BibitemOpen
  \bibfield  {author} {\bibinfo {author} {\bibfnamefont {P.}~\bibnamefont
  {Madau}}\ and\ \bibinfo {author} {\bibfnamefont {M.}~\bibnamefont
  {Dickinson}},\ }\href {\doibase 10.1146/annurev-astro-081811-125615}
  {\bibfield  {journal} {\bibinfo  {journal} {Ann. Rev. Astron. Astrophys.}\
  }\textbf {\bibinfo {volume} {52}},\ \bibinfo {pages} {415} (\bibinfo {year}
  {2014})},\ \Eprint {http://arxiv.org/abs/1403.0007} {arXiv:1403.0007
  [astro-ph.CO]} \BibitemShut {NoStop}%
\bibitem [{\citenamefont {Furlanetto}\ \emph {et~al.}(2017)\citenamefont
  {Furlanetto}, \citenamefont {Mirocha}, \citenamefont {Mebane},\ and\
  \citenamefont {Sun}}]{Furlanetto_2017}%
  \BibitemOpen
  \bibfield  {author} {\bibinfo {author} {\bibfnamefont {S.~R.}\ \bibnamefont
  {Furlanetto}}, \bibinfo {author} {\bibfnamefont {J.}~\bibnamefont {Mirocha}},
  \bibinfo {author} {\bibfnamefont {R.~H.}\ \bibnamefont {Mebane}}, \ and\
  \bibinfo {author} {\bibfnamefont {G.}~\bibnamefont {Sun}},\ }\href {\doibase
  10.1093/mnras/stx2132} {\bibfield  {journal} {\bibinfo  {journal} {Monthly
  Notices of the Royal Astronomical Society}\ }\textbf {\bibinfo {volume}
  {472}},\ \bibinfo {pages} {1576–1592} (\bibinfo {year} {2017})}\BibitemShut
  {NoStop}%
\bibitem [{\citenamefont {Oke}\ and\ \citenamefont {Gunn}(1983)}]{Oke:1983nt}%
  \BibitemOpen
  \bibfield  {author} {\bibinfo {author} {\bibfnamefont {J.~B.}\ \bibnamefont
  {Oke}}\ and\ \bibinfo {author} {\bibfnamefont {J.~E.}\ \bibnamefont {Gunn}},\
  }\href {\doibase 10.1086/160817} {\bibfield  {journal} {\bibinfo  {journal}
  {Astrophys. J.}\ }\textbf {\bibinfo {volume} {266}},\ \bibinfo {pages} {713}
  (\bibinfo {year} {1983})}\BibitemShut {NoStop}%
\bibitem [{\citenamefont {Neistein}\ and\ \citenamefont {van~den
  Bosch}(2006)}]{Neistein:2006ak}%
  \BibitemOpen
  \bibfield  {author} {\bibinfo {author} {\bibfnamefont {E.}~\bibnamefont
  {Neistein}}\ and\ \bibinfo {author} {\bibfnamefont {F.~C.}\ \bibnamefont
  {van~den Bosch}},\ }\href {\doibase 10.1111/j.1365-2966.2006.10918.x}
  {\bibfield  {journal} {\bibinfo  {journal} {Mon. Not. Roy. Astron. Soc.}\
  }\textbf {\bibinfo {volume} {372}},\ \bibinfo {pages} {933} (\bibinfo {year}
  {2006})},\ \Eprint {http://arxiv.org/abs/astro-ph/0605045}
  {arXiv:astro-ph/0605045} \BibitemShut {NoStop}%
\bibitem [{\citenamefont {Correa}\ \emph {et~al.}(2015)\citenamefont {Correa},
  \citenamefont {Wyithe}, \citenamefont {Schaye},\ and\ \citenamefont
  {Duffy}}]{Correa:2014xma}%
  \BibitemOpen
  \bibfield  {author} {\bibinfo {author} {\bibfnamefont {C.~A.}\ \bibnamefont
  {Correa}}, \bibinfo {author} {\bibfnamefont {J.~S.~B.}\ \bibnamefont
  {Wyithe}}, \bibinfo {author} {\bibfnamefont {J.}~\bibnamefont {Schaye}}, \
  and\ \bibinfo {author} {\bibfnamefont {A.~R.}\ \bibnamefont {Duffy}},\ }\href
  {\doibase 10.1093/mnras/stv689} {\bibfield  {journal} {\bibinfo  {journal}
  {Mon. Not. Roy. Astron. Soc.}\ }\textbf {\bibinfo {volume} {450}},\ \bibinfo
  {pages} {1514} (\bibinfo {year} {2015})},\ \Eprint
  {http://arxiv.org/abs/1409.5228} {arXiv:1409.5228 [astro-ph.GA]} \BibitemShut
  {NoStop}%
\bibitem [{\citenamefont {White}\ and\ \citenamefont
  {Frenk}(1991)}]{White:1991mr}%
  \BibitemOpen
  \bibfield  {author} {\bibinfo {author} {\bibfnamefont {S.~D.~M.}\
  \bibnamefont {White}}\ and\ \bibinfo {author} {\bibfnamefont {C.~S.}\
  \bibnamefont {Frenk}},\ }\href {\doibase 10.1086/170483} {\bibfield
  {journal} {\bibinfo  {journal} {Astrophys. J.}\ }\textbf {\bibinfo {volume}
  {379}},\ \bibinfo {pages} {52} (\bibinfo {year} {1991})}\BibitemShut
  {NoStop}%
\bibitem [{\citenamefont {Ren}\ \emph {et~al.}(2019)\citenamefont {Ren},
  \citenamefont {Trenti},\ and\ \citenamefont {Mason}}]{ren2019}%
  \BibitemOpen
  \bibfield  {author} {\bibinfo {author} {\bibfnamefont {K.}~\bibnamefont
  {Ren}}, \bibinfo {author} {\bibfnamefont {M.}~\bibnamefont {Trenti}}, \ and\
  \bibinfo {author} {\bibfnamefont {C.~A.}\ \bibnamefont {Mason}},\ }\href
  {\doibase 10.3847/1538-4357/ab2117} {\bibfield  {journal} {\bibinfo
  {journal} {The Astrophysical Journal}\ }\textbf {\bibinfo {volume} {878}},\
  \bibinfo {pages} {114} (\bibinfo {year} {2019})}\BibitemShut {NoStop}%
\bibitem [{\citenamefont {Mirocha}(2020)}]{Mirocha:2020sid}%
  \BibitemOpen
  \bibfield  {author} {\bibinfo {author} {\bibfnamefont {J.}~\bibnamefont
  {Mirocha}},\ }\href {\doibase 10.1093/mnras/staa3150} {\bibfield  {journal}
  {\bibinfo  {journal} {Mon. Not. Roy. Astron. Soc.}\ }\textbf {\bibinfo
  {volume} {499}},\ \bibinfo {pages} {4534} (\bibinfo {year} {2020})},\ \Eprint
  {http://arxiv.org/abs/2008.04322} {arXiv:2008.04322 [astro-ph.GA]}
  \BibitemShut {NoStop}%
\bibitem [{\citenamefont {Khusanova}\ \emph {et~al.}(2020)\citenamefont
  {Khusanova} \emph {et~al.}}]{Khusanova:2019cxr}%
  \BibitemOpen
  \bibfield  {author} {\bibinfo {author} {\bibfnamefont {Y.}~\bibnamefont
  {Khusanova}} \emph {et~al.},\ }\href {\doibase 10.1051/0004-6361/201935400}
  {\bibfield  {journal} {\bibinfo  {journal} {Astron. Astrophys.}\ }\textbf
  {\bibinfo {volume} {634}},\ \bibinfo {pages} {A97} (\bibinfo {year}
  {2020})},\ \Eprint {http://arxiv.org/abs/1903.01884} {arXiv:1903.01884
  [astro-ph.GA]} \BibitemShut {NoStop}%
\bibitem [{\citenamefont {van~der Burg}\ \emph {et~al.}(2010)\citenamefont
  {van~der Burg}, \citenamefont {Hildebrandt},\ and\ \citenamefont
  {Erben}}]{van_der_Burg_2010}%
  \BibitemOpen
  \bibfield  {author} {\bibinfo {author} {\bibfnamefont {R.~F.~J.}\
  \bibnamefont {van~der Burg}}, \bibinfo {author} {\bibfnamefont
  {H.}~\bibnamefont {Hildebrandt}}, \ and\ \bibinfo {author} {\bibfnamefont
  {T.}~\bibnamefont {Erben}},\ }\href {\doibase 10.1051/0004-6361/200913812}
  {\bibfield  {journal} {\bibinfo  {journal} {Astronomy \& Astrophysics}\
  }\textbf {\bibinfo {volume} {523}},\ \bibinfo {pages} {A74} (\bibinfo {year}
  {2010})}\BibitemShut {NoStop}%
\bibitem [{\citenamefont {Harikane}\ \emph {et~al.}(2021)\citenamefont
  {Harikane}, \citenamefont {Ono}, \citenamefont {Ouchi}, \citenamefont {Liu},
  \citenamefont {Sawicki}, \citenamefont {Shibuya}, \citenamefont {Behroozi},
  \citenamefont {He}, \citenamefont {Shimasaku}, \citenamefont {Arnouts},
  \citenamefont {Coupon}, \citenamefont {Fujimoto}, \citenamefont {Gwyn},
  \citenamefont {Huang}, \citenamefont {Inoue}, \citenamefont {Kashikawa},
  \citenamefont {Komiyama}, \citenamefont {Matsuoka},\ and\ \citenamefont
  {Willott}}]{Harikane2021}%
  \BibitemOpen
  \bibfield  {author} {\bibinfo {author} {\bibfnamefont {Y.}~\bibnamefont
  {Harikane}}, \bibinfo {author} {\bibfnamefont {Y.}~\bibnamefont {Ono}},
  \bibinfo {author} {\bibfnamefont {M.}~\bibnamefont {Ouchi}}, \bibinfo
  {author} {\bibfnamefont {C.}~\bibnamefont {Liu}}, \bibinfo {author}
  {\bibfnamefont {M.}~\bibnamefont {Sawicki}}, \bibinfo {author} {\bibfnamefont
  {T.}~\bibnamefont {Shibuya}}, \bibinfo {author} {\bibfnamefont {P.~S.}\
  \bibnamefont {Behroozi}}, \bibinfo {author} {\bibfnamefont {W.}~\bibnamefont
  {He}}, \bibinfo {author} {\bibfnamefont {K.}~\bibnamefont {Shimasaku}},
  \bibinfo {author} {\bibfnamefont {S.}~\bibnamefont {Arnouts}}, \bibinfo
  {author} {\bibfnamefont {J.}~\bibnamefont {Coupon}}, \bibinfo {author}
  {\bibfnamefont {S.}~\bibnamefont {Fujimoto}}, \bibinfo {author}
  {\bibfnamefont {S.}~\bibnamefont {Gwyn}}, \bibinfo {author} {\bibfnamefont
  {J.}~\bibnamefont {Huang}}, \bibinfo {author} {\bibfnamefont {A.~K.}\
  \bibnamefont {Inoue}}, \bibinfo {author} {\bibfnamefont {N.}~\bibnamefont
  {Kashikawa}}, \bibinfo {author} {\bibfnamefont {Y.}~\bibnamefont {Komiyama}},
  \bibinfo {author} {\bibfnamefont {Y.}~\bibnamefont {Matsuoka}}, \ and\
  \bibinfo {author} {\bibfnamefont {C.~J.}\ \bibnamefont {Willott}},\
  }\href@noop {} {\  (\bibinfo {year} {2021})},\ \Eprint
  {http://arxiv.org/abs/2108.01090} {arXiv:2108.01090 [astro-ph.GA]}
  \BibitemShut {NoStop}%
\bibitem [{\citenamefont {Santos}\ \emph {et~al.}(2021)\citenamefont {Santos},
  \citenamefont {Sobral}, \citenamefont {Butterworth}, \citenamefont
  {Paulino-Afonso}, \citenamefont {Ribeiro}, \citenamefont {da~Cunha},
  \citenamefont {Calhau}, \citenamefont {Khostovan}, \citenamefont {Matthee},\
  and\ \citenamefont {Arrabal Haro}}]{Santos_2021}%
  \BibitemOpen
  \bibfield  {author} {\bibinfo {author} {\bibfnamefont {S.}~\bibnamefont
  {Santos}}, \bibinfo {author} {\bibfnamefont {D.}~\bibnamefont {Sobral}},
  \bibinfo {author} {\bibfnamefont {J.}~\bibnamefont {Butterworth}}, \bibinfo
  {author} {\bibfnamefont {A.}~\bibnamefont {Paulino-Afonso}}, \bibinfo
  {author} {\bibfnamefont {B.}~\bibnamefont {Ribeiro}}, \bibinfo {author}
  {\bibfnamefont {E.}~\bibnamefont {da~Cunha}}, \bibinfo {author}
  {\bibfnamefont {J.}~\bibnamefont {Calhau}}, \bibinfo {author} {\bibfnamefont
  {A.~A.}\ \bibnamefont {Khostovan}}, \bibinfo {author} {\bibfnamefont
  {J.}~\bibnamefont {Matthee}}, \ and\ \bibinfo {author} {\bibfnamefont
  {P.}~\bibnamefont {Arrabal Haro}},\ }\href {\doibase 10.1093/mnras/stab1218}
  {\bibfield  {journal} {\bibinfo  {journal} {Monthly Notices of the Royal
  Astronomical Society}\ }\textbf {\bibinfo {volume} {505}},\ \bibinfo {pages}
  {1117–1134} (\bibinfo {year} {2021})}\BibitemShut {NoStop}%
\bibitem [{\citenamefont {Maizy}\ \emph {et~al.}(2010)\citenamefont {Maizy},
  \citenamefont {Richard}, \citenamefont {De~Leo}, \citenamefont {Pello},\ and\
  \citenamefont {Kneib}}]{Maizy:2009df}%
  \BibitemOpen
  \bibfield  {author} {\bibinfo {author} {\bibfnamefont {A.}~\bibnamefont
  {Maizy}}, \bibinfo {author} {\bibfnamefont {J.}~\bibnamefont {Richard}},
  \bibinfo {author} {\bibfnamefont {M.~A.}\ \bibnamefont {De~Leo}}, \bibinfo
  {author} {\bibfnamefont {R.}~\bibnamefont {Pello}}, \ and\ \bibinfo {author}
  {\bibfnamefont {J.~P.}\ \bibnamefont {Kneib}},\ }\href {\doibase
  10.1051/0004-6361/200911829} {\bibfield  {journal} {\bibinfo  {journal}
  {Astron. Astrophys.}\ }\textbf {\bibinfo {volume} {509}},\ \bibinfo {pages}
  {A105} (\bibinfo {year} {2010})},\ \Eprint {http://arxiv.org/abs/0910.4910}
  {arXiv:0910.4910 [astro-ph.CO]} \BibitemShut {NoStop}%
%%CITATION = ARXIV:0910.4910;%%
\bibitem [{\citenamefont {Wyithe}\ \emph {et~al.}(2011)\citenamefont {Wyithe},
  \citenamefont {Yan}, \citenamefont {Windhorst},\ and\ \citenamefont
  {Mao}}]{Wyithe:2011gh}%
  \BibitemOpen
  \bibfield  {author} {\bibinfo {author} {\bibfnamefont {J.~S.~B.}\
  \bibnamefont {Wyithe}}, \bibinfo {author} {\bibfnamefont {H.}~\bibnamefont
  {Yan}}, \bibinfo {author} {\bibfnamefont {R.~A.}\ \bibnamefont {Windhorst}},
  \ and\ \bibinfo {author} {\bibfnamefont {S.}~\bibnamefont {Mao}},\ }\href
  {\doibase 10.1038/nature09619} {\bibfield  {journal} {\bibinfo  {journal}
  {Nature}\ }\textbf {\bibinfo {volume} {469}},\ \bibinfo {pages} {181}
  (\bibinfo {year} {2011})},\ \Eprint {http://arxiv.org/abs/1101.2291}
  {arXiv:1101.2291 [astro-ph.CO]} \BibitemShut {NoStop}%
\bibitem [{\citenamefont {Mason}\ \emph
  {et~al.}(2015{\natexlab{b}})\citenamefont {Mason}, \citenamefont {Treu},
  \citenamefont {Schmidt}, \citenamefont {Collett}, \citenamefont {Trenti},
  \citenamefont {Marshall}, \citenamefont {Barone-Nugent}, \citenamefont
  {Bradley}, \citenamefont {Stiavelli},\ and\ \citenamefont
  {Wyithe}}]{Mason:2015wla}%
  \BibitemOpen
  \bibfield  {author} {\bibinfo {author} {\bibfnamefont {C.~A.}\ \bibnamefont
  {Mason}}, \bibinfo {author} {\bibfnamefont {T.}~\bibnamefont {Treu}},
  \bibinfo {author} {\bibfnamefont {K.~B.}\ \bibnamefont {Schmidt}}, \bibinfo
  {author} {\bibfnamefont {T.~E.}\ \bibnamefont {Collett}}, \bibinfo {author}
  {\bibfnamefont {M.}~\bibnamefont {Trenti}}, \bibinfo {author} {\bibfnamefont
  {P.~J.}\ \bibnamefont {Marshall}}, \bibinfo {author} {\bibfnamefont
  {R.}~\bibnamefont {Barone-Nugent}}, \bibinfo {author} {\bibfnamefont {L.~D.}\
  \bibnamefont {Bradley}}, \bibinfo {author} {\bibfnamefont {M.}~\bibnamefont
  {Stiavelli}}, \ and\ \bibinfo {author} {\bibfnamefont {S.}~\bibnamefont
  {Wyithe}},\ }\href {\doibase 10.1088/0004-637X/805/1/79} {\bibfield
  {journal} {\bibinfo  {journal} {Astrophys. J.}\ }\textbf {\bibinfo {volume}
  {805}},\ \bibinfo {pages} {79} (\bibinfo {year} {2015}{\natexlab{b}})},\
  \Eprint {http://arxiv.org/abs/1502.03795} {arXiv:1502.03795 [astro-ph.CO]}
  \BibitemShut {NoStop}%
\bibitem [{\citenamefont {Trapp}\ and\ \citenamefont
  {Furlanetto}(2020)}]{Trapp_2020}%
  \BibitemOpen
  \bibfield  {author} {\bibinfo {author} {\bibfnamefont {A.~C.}\ \bibnamefont
  {Trapp}}\ and\ \bibinfo {author} {\bibfnamefont {S.~R.}\ \bibnamefont
  {Furlanetto}},\ }\href {\doibase 10.1093/mnras/staa2828} {\bibfield
  {journal} {\bibinfo  {journal} {Monthly Notices of the Royal Astronomical
  Society}\ }\textbf {\bibinfo {volume} {499}},\ \bibinfo {pages} {2401–2415}
  (\bibinfo {year} {2020})}\BibitemShut {NoStop}%
\bibitem [{\citenamefont {Newman}\ and\ \citenamefont
  {Davis}(2002)}]{Newman:2001ca}%
  \BibitemOpen
  \bibfield  {author} {\bibinfo {author} {\bibfnamefont {J.~A.}\ \bibnamefont
  {Newman}}\ and\ \bibinfo {author} {\bibfnamefont {M.}~\bibnamefont {Davis}},\
  }\href {\doibase 10.1086/324148} {\bibfield  {journal} {\bibinfo  {journal}
  {Astrophys. J.}\ }\textbf {\bibinfo {volume} {564}},\ \bibinfo {pages} {567}
  (\bibinfo {year} {2002})},\ \Eprint {http://arxiv.org/abs/astro-ph/0109130}
  {arXiv:astro-ph/0109130} \BibitemShut {NoStop}%
\bibitem [{\citenamefont {Somerville}\ \emph {et~al.}(2004)\citenamefont
  {Somerville}, \citenamefont {Lee}, \citenamefont {Ferguson}, \citenamefont
  {Gardner}, \citenamefont {Moustakas},\ and\ \citenamefont
  {Giavalisco}}]{Somerville:2003bq}%
  \BibitemOpen
  \bibfield  {author} {\bibinfo {author} {\bibfnamefont {R.~S.}\ \bibnamefont
  {Somerville}}, \bibinfo {author} {\bibfnamefont {K.}~\bibnamefont {Lee}},
  \bibinfo {author} {\bibfnamefont {H.~C.}\ \bibnamefont {Ferguson}}, \bibinfo
  {author} {\bibfnamefont {J.~P.}\ \bibnamefont {Gardner}}, \bibinfo {author}
  {\bibfnamefont {L.~A.}\ \bibnamefont {Moustakas}}, \ and\ \bibinfo {author}
  {\bibfnamefont {M.}~\bibnamefont {Giavalisco}},\ }\href {\doibase
  10.1086/378628} {\bibfield  {journal} {\bibinfo  {journal} {Astrophys. J.
  Lett.}\ }\textbf {\bibinfo {volume} {600}},\ \bibinfo {pages} {L171}
  (\bibinfo {year} {2004})},\ \Eprint {http://arxiv.org/abs/astro-ph/0309071}
  {arXiv:astro-ph/0309071} \BibitemShut {NoStop}%
\bibitem [{\citenamefont {Stark}\ \emph {et~al.}(2007)\citenamefont {Stark},
  \citenamefont {Loeb},\ and\ \citenamefont {Ellis}}]{Stark:2007pv}%
  \BibitemOpen
  \bibfield  {author} {\bibinfo {author} {\bibfnamefont {D.~P.}\ \bibnamefont
  {Stark}}, \bibinfo {author} {\bibfnamefont {A.}~\bibnamefont {Loeb}}, \ and\
  \bibinfo {author} {\bibfnamefont {R.~S.}\ \bibnamefont {Ellis}},\ }\href
  {\doibase 10.1086/520947} {\bibfield  {journal} {\bibinfo  {journal}
  {Astrophys. J.}\ }\textbf {\bibinfo {volume} {668}},\ \bibinfo {pages} {627}
  (\bibinfo {year} {2007})},\ \Eprint {http://arxiv.org/abs/astro-ph/0701882}
  {arXiv:astro-ph/0701882} \BibitemShut {NoStop}%
\bibitem [{\citenamefont {Moster}\ \emph {et~al.}(2011)\citenamefont {Moster},
  \citenamefont {Somerville}, \citenamefont {Newman},\ and\ \citenamefont
  {Rix}}]{Moster:2010hf}%
  \BibitemOpen
  \bibfield  {author} {\bibinfo {author} {\bibfnamefont {B.~P.}\ \bibnamefont
  {Moster}}, \bibinfo {author} {\bibfnamefont {R.~S.}\ \bibnamefont
  {Somerville}}, \bibinfo {author} {\bibfnamefont {J.~A.}\ \bibnamefont
  {Newman}}, \ and\ \bibinfo {author} {\bibfnamefont {H.-W.}\ \bibnamefont
  {Rix}},\ }\href {\doibase 10.1088/0004-637X/731/2/113} {\bibfield  {journal}
  {\bibinfo  {journal} {Astrophys. J.}\ }\textbf {\bibinfo {volume} {731}},\
  \bibinfo {pages} {113} (\bibinfo {year} {2011})},\ \Eprint
  {http://arxiv.org/abs/1001.1737} {arXiv:1001.1737 [astro-ph.CO]} \BibitemShut
  {NoStop}%
\bibitem [{\citenamefont {Kitzbichler}\ and\ \citenamefont
  {White}(2007)}]{Kitzbichler:2006ec}%
  \BibitemOpen
  \bibfield  {author} {\bibinfo {author} {\bibfnamefont {M.~G.}\ \bibnamefont
  {Kitzbichler}}\ and\ \bibinfo {author} {\bibfnamefont {S.~D.~M.}\
  \bibnamefont {White}},\ }\href {\doibase 10.1111/j.1365-2966.2007.11458.x}
  {\bibfield  {journal} {\bibinfo  {journal} {Mon. Not. Roy. Astron. Soc.}\
  }\textbf {\bibinfo {volume} {376}},\ \bibinfo {pages} {2} (\bibinfo {year}
  {2007})},\ \Eprint {http://arxiv.org/abs/astro-ph/0609636}
  {arXiv:astro-ph/0609636} \BibitemShut {NoStop}%
\bibitem [{\citenamefont {Trenti}\ and\ \citenamefont
  {Stiavelli}(2008)}]{Trenti:2007dh}%
  \BibitemOpen
  \bibfield  {author} {\bibinfo {author} {\bibfnamefont {M.}~\bibnamefont
  {Trenti}}\ and\ \bibinfo {author} {\bibfnamefont {M.}~\bibnamefont
  {Stiavelli}},\ }\href {\doibase 10.1086/528674} {\bibfield  {journal}
  {\bibinfo  {journal} {Astrophys. J.}\ }\textbf {\bibinfo {volume} {676}},\
  \bibinfo {pages} {767} (\bibinfo {year} {2008})},\ \Eprint
  {http://arxiv.org/abs/0712.0398} {arXiv:0712.0398 [astro-ph]} \BibitemShut
  {NoStop}%
\bibitem [{\citenamefont {Ucci}\ \emph {et~al.}(2021)\citenamefont {Ucci},
  \citenamefont {Dayal}, \citenamefont {Hutter}, \citenamefont {Yepes},
  \citenamefont {Gottlöber}, \citenamefont {Legrand}, \citenamefont
  {Pentericci}, \citenamefont {Castellano},\ and\ \citenamefont
  {Choudhury}}]{Ucci_2021}%
  \BibitemOpen
  \bibfield  {author} {\bibinfo {author} {\bibfnamefont {G.}~\bibnamefont
  {Ucci}}, \bibinfo {author} {\bibfnamefont {P.}~\bibnamefont {Dayal}},
  \bibinfo {author} {\bibfnamefont {A.}~\bibnamefont {Hutter}}, \bibinfo
  {author} {\bibfnamefont {G.}~\bibnamefont {Yepes}}, \bibinfo {author}
  {\bibfnamefont {S.}~\bibnamefont {Gottlöber}}, \bibinfo {author}
  {\bibfnamefont {L.}~\bibnamefont {Legrand}}, \bibinfo {author} {\bibfnamefont
  {L.}~\bibnamefont {Pentericci}}, \bibinfo {author} {\bibfnamefont
  {M.}~\bibnamefont {Castellano}}, \ and\ \bibinfo {author} {\bibfnamefont
  {T.~R.}\ \bibnamefont {Choudhury}},\ }\href {\doibase 10.1093/mnras/stab1229}
  {\bibfield  {journal} {\bibinfo  {journal} {Monthly Notices of the Royal
  Astronomical Society}\ }\textbf {\bibinfo {volume} {506}},\ \bibinfo {pages}
  {202–214} (\bibinfo {year} {2021})}\BibitemShut {NoStop}%
\bibitem [{\citenamefont {Meurer}\ \emph {et~al.}(1999)\citenamefont {Meurer},
  \citenamefont {Heckman},\ and\ \citenamefont {Calzetti}}]{Meurer:1999jj}%
  \BibitemOpen
  \bibfield  {author} {\bibinfo {author} {\bibfnamefont {G.~R.}\ \bibnamefont
  {Meurer}}, \bibinfo {author} {\bibfnamefont {T.~M.}\ \bibnamefont {Heckman}},
  \ and\ \bibinfo {author} {\bibfnamefont {D.}~\bibnamefont {Calzetti}},\
  }\href {\doibase 10.1086/307523} {\bibfield  {journal} {\bibinfo  {journal}
  {Astrophys. J.}\ }\textbf {\bibinfo {volume} {521}},\ \bibinfo {pages} {64}
  (\bibinfo {year} {1999})},\ \Eprint {http://arxiv.org/abs/astro-ph/9903054}
  {arXiv:astro-ph/9903054} \BibitemShut {NoStop}%
\bibitem [{\citenamefont {Ma}\ \emph {et~al.}(2019)\citenamefont {Ma},
  \citenamefont {Hayward}, \citenamefont {Casey}, \citenamefont {Hopkins},
  \citenamefont {Quataert}, \citenamefont {Liang}, \citenamefont
  {Faucher-Gigu\`ere}, \citenamefont {Feldmann},\ and\ \citenamefont
  {Kere\v{s}}}]{Ma:2019hwa}%
  \BibitemOpen
  \bibfield  {author} {\bibinfo {author} {\bibfnamefont {X.}~\bibnamefont
  {Ma}}, \bibinfo {author} {\bibfnamefont {C.~C.}\ \bibnamefont {Hayward}},
  \bibinfo {author} {\bibfnamefont {C.~M.}\ \bibnamefont {Casey}}, \bibinfo
  {author} {\bibfnamefont {P.~F.}\ \bibnamefont {Hopkins}}, \bibinfo {author}
  {\bibfnamefont {E.}~\bibnamefont {Quataert}}, \bibinfo {author}
  {\bibfnamefont {L.}~\bibnamefont {Liang}}, \bibinfo {author} {\bibfnamefont
  {C.-A.}\ \bibnamefont {Faucher-Gigu\`ere}}, \bibinfo {author} {\bibfnamefont
  {R.}~\bibnamefont {Feldmann}}, \ and\ \bibinfo {author} {\bibfnamefont
  {D.}~\bibnamefont {Kere\v{s}}},\ }\href {\doibase 10.1093/mnras/stz1324}
  {\bibfield  {journal} {\bibinfo  {journal} {Mon. Not. Roy. Astron. Soc.}\
  }\textbf {\bibinfo {volume} {487}},\ \bibinfo {pages} {1844} (\bibinfo {year}
  {2019})},\ \Eprint {http://arxiv.org/abs/1902.10152} {arXiv:1902.10152
  [astro-ph.GA]} \BibitemShut {NoStop}%
\bibitem [{\citenamefont {Mirocha}\ \emph
  {et~al.}(2020{\natexlab{b}})\citenamefont {Mirocha}, \citenamefont {Mason},\
  and\ \citenamefont {Stark}}]{Mirocha_2020}%
  \BibitemOpen
  \bibfield  {author} {\bibinfo {author} {\bibfnamefont {J.}~\bibnamefont
  {Mirocha}}, \bibinfo {author} {\bibfnamefont {C.}~\bibnamefont {Mason}}, \
  and\ \bibinfo {author} {\bibfnamefont {D.~P.}\ \bibnamefont {Stark}},\ }\href
  {\doibase 10.1093/mnras/staa2586} {\bibfield  {journal} {\bibinfo  {journal}
  {Monthly Notices of the Royal Astronomical Society}\ }\textbf {\bibinfo
  {volume} {498}},\ \bibinfo {pages} {2645–2661} (\bibinfo {year}
  {2020}{\natexlab{b}})}\BibitemShut {NoStop}%
\bibitem [{\citenamefont {Liang}\ \emph {et~al.}(2021)\citenamefont {Liang},
  \citenamefont {Feldmann}, \citenamefont {Hayward}, \citenamefont {Narayanan},
  \citenamefont {\c{C}atmabacak}, \citenamefont {Kere\v{s}}, \citenamefont
  {Faucher-Gigu\`ere},\ and\ \citenamefont {Hopkins}}]{Liang:2020aay}%
  \BibitemOpen
  \bibfield  {author} {\bibinfo {author} {\bibfnamefont {L.}~\bibnamefont
  {Liang}}, \bibinfo {author} {\bibfnamefont {R.}~\bibnamefont {Feldmann}},
  \bibinfo {author} {\bibfnamefont {C.~C.}\ \bibnamefont {Hayward}}, \bibinfo
  {author} {\bibfnamefont {D.}~\bibnamefont {Narayanan}}, \bibinfo {author}
  {\bibfnamefont {O.}~\bibnamefont {\c{C}atmabacak}}, \bibinfo {author}
  {\bibfnamefont {D.}~\bibnamefont {Kere\v{s}}}, \bibinfo {author}
  {\bibfnamefont {C.-A.}\ \bibnamefont {Faucher-Gigu\`ere}}, \ and\ \bibinfo
  {author} {\bibfnamefont {P.~F.}\ \bibnamefont {Hopkins}},\ }\href {\doibase
  10.1093/mnras/stab096} {\bibfield  {journal} {\bibinfo  {journal} {Mon. Not.
  Roy. Astron. Soc.}\ }\textbf {\bibinfo {volume} {502}},\ \bibinfo {pages}
  {3210} (\bibinfo {year} {2021})},\ \Eprint {http://arxiv.org/abs/2009.13522}
  {arXiv:2009.13522 [astro-ph.GA]} \BibitemShut {NoStop}%
\bibitem [{\citenamefont {Smit}\ \emph {et~al.}(2012)\citenamefont {Smit} \emph
  {et~al.}}]{Smit:2012nf}%
  \BibitemOpen
  \bibfield  {author} {\bibinfo {author} {\bibfnamefont {R.}~\bibnamefont
  {Smit}} \emph {et~al.},\ }\href {\doibase 10.1088/0004-637X/756/1/14}
  {\bibfield  {journal} {\bibinfo  {journal} {Astrophys. J.}\ }\textbf
  {\bibinfo {volume} {756}},\ \bibinfo {pages} {14} (\bibinfo {year} {2012})},\
  \Eprint {http://arxiv.org/abs/1204.3626} {arXiv:1204.3626 [astro-ph.CO]}
  \BibitemShut {NoStop}%
\bibitem [{\citenamefont {Bouwens}\ \emph {et~al.}(2012)\citenamefont {Bouwens}
  \emph {et~al.}}]{Bouwens:2011yy}%
  \BibitemOpen
  \bibfield  {author} {\bibinfo {author} {\bibfnamefont {R.}~\bibnamefont
  {Bouwens}} \emph {et~al.},\ }\href {\doibase 10.1088/0004-637X/754/2/83}
  {\bibfield  {journal} {\bibinfo  {journal} {Astrophys. J.}\ }\textbf
  {\bibinfo {volume} {754}},\ \bibinfo {pages} {83} (\bibinfo {year} {2012})},\
  \Eprint {http://arxiv.org/abs/1109.0994} {arXiv:1109.0994 [astro-ph.CO]}
  \BibitemShut {NoStop}%
\bibitem [{\citenamefont {Trenti}\ \emph {et~al.}(2015)\citenamefont {Trenti},
  \citenamefont {Perna},\ and\ \citenamefont {Jimenez}}]{Trenti:2014hka}%
  \BibitemOpen
  \bibfield  {author} {\bibinfo {author} {\bibfnamefont {M.}~\bibnamefont
  {Trenti}}, \bibinfo {author} {\bibfnamefont {R.}~\bibnamefont {Perna}}, \
  and\ \bibinfo {author} {\bibfnamefont {R.}~\bibnamefont {Jimenez}},\ }\href
  {\doibase 10.1088/0004-637X/802/2/103} {\bibfield  {journal} {\bibinfo
  {journal} {Astrophys. J.}\ }\textbf {\bibinfo {volume} {802}},\ \bibinfo
  {pages} {103} (\bibinfo {year} {2015})},\ \Eprint
  {http://arxiv.org/abs/1406.1503} {arXiv:1406.1503 [astro-ph.GA]} \BibitemShut
  {NoStop}%
\bibitem [{\citenamefont {Bouwens}\ \emph {et~al.}(2014)\citenamefont {Bouwens}
  \emph {et~al.}}]{Bouwens:2013hxa}%
  \BibitemOpen
  \bibfield  {author} {\bibinfo {author} {\bibfnamefont {R.}~\bibnamefont
  {Bouwens}} \emph {et~al.},\ }\href {\doibase 10.1088/0004-637X/793/2/115}
  {\bibfield  {journal} {\bibinfo  {journal} {Astrophys. J.}\ }\textbf
  {\bibinfo {volume} {793}},\ \bibinfo {pages} {115} (\bibinfo {year}
  {2014})},\ \Eprint {http://arxiv.org/abs/1306.2950} {arXiv:1306.2950
  [astro-ph.CO]} \BibitemShut {NoStop}%
\bibitem [{\citenamefont {Siana}\ \emph {et~al.}(2009)\citenamefont {Siana}
  \emph {et~al.}}]{Siana:2009um}%
  \BibitemOpen
  \bibfield  {author} {\bibinfo {author} {\bibfnamefont {B.}~\bibnamefont
  {Siana}} \emph {et~al.},\ }\href {\doibase 10.1088/0004-637X/698/2/1273}
  {\bibfield  {journal} {\bibinfo  {journal} {Astrophys. J.}\ }\textbf
  {\bibinfo {volume} {698}},\ \bibinfo {pages} {1273} (\bibinfo {year}
  {2009})},\ \Eprint {http://arxiv.org/abs/0904.1742} {arXiv:0904.1742
  [astro-ph.CO]} \BibitemShut {NoStop}%
\bibitem [{\citenamefont {Overzier}\ \emph {et~al.}(2011)\citenamefont
  {Overzier} \emph {et~al.}}]{Overzier:2010aa}%
  \BibitemOpen
  \bibfield  {author} {\bibinfo {author} {\bibfnamefont {R.}~\bibnamefont
  {Overzier}} \emph {et~al.},\ }\href {\doibase 10.1088/2041-8205/726/1/L7}
  {\bibfield  {journal} {\bibinfo  {journal} {Astrophys. J. Lett.}\ }\textbf
  {\bibinfo {volume} {726}},\ \bibinfo {pages} {L7} (\bibinfo {year} {2011})},\
  \Eprint {http://arxiv.org/abs/1011.6098} {arXiv:1011.6098 [astro-ph.CO]}
  \BibitemShut {NoStop}%
\bibitem [{\citenamefont {Casey}\ \emph {et~al.}(2014)\citenamefont {Casey}
  \emph {et~al.}}]{Casey:2014cqa}%
  \BibitemOpen
  \bibfield  {author} {\bibinfo {author} {\bibfnamefont {C.}~\bibnamefont
  {Casey}} \emph {et~al.},\ }\href {\doibase 10.1088/0004-637X/796/2/95}
  {\bibfield  {journal} {\bibinfo  {journal} {Astrophys. J.}\ }\textbf
  {\bibinfo {volume} {796}},\ \bibinfo {pages} {95} (\bibinfo {year} {2014})},\
  \Eprint {http://arxiv.org/abs/1410.0702} {arXiv:1410.0702 [astro-ph.GA]}
  \BibitemShut {NoStop}%
\bibitem [{\citenamefont {Castellano}\ \emph {et~al.}(2014)\citenamefont
  {Castellano} \emph {et~al.}}]{Castellano:2014lua}%
  \BibitemOpen
  \bibfield  {author} {\bibinfo {author} {\bibfnamefont {M.}~\bibnamefont
  {Castellano}} \emph {et~al.},\ }\href {\doibase 10.1051/0004-6361/201322704}
  {\bibfield  {journal} {\bibinfo  {journal} {Astron. Astrophys.}\ }\textbf
  {\bibinfo {volume} {566}},\ \bibinfo {pages} {A19} (\bibinfo {year}
  {2014})},\ \Eprint {http://arxiv.org/abs/1403.0743} {arXiv:1403.0743
  [astro-ph.GA]} \BibitemShut {NoStop}%
\bibitem [{\citenamefont {Reddy}\ \emph {et~al.}(2015)\citenamefont {Reddy}
  \emph {et~al.}}]{reddy2015mosdef}%
  \BibitemOpen
  \bibfield  {author} {\bibinfo {author} {\bibfnamefont {N.~A.}\ \bibnamefont
  {Reddy}} \emph {et~al.},\ }\href@noop {} {\  (\bibinfo {year} {2015})},\
  \Eprint {http://arxiv.org/abs/1504.02782} {arXiv:1504.02782 [astro-ph.GA]}
  \BibitemShut {NoStop}%
\bibitem [{\citenamefont {J.~Bouwens}\ \emph {et~al.}(2016)\citenamefont
  {J.~Bouwens} \emph {et~al.}}]{J_Bouwens_2016}%
  \BibitemOpen
  \bibfield  {author} {\bibinfo {author} {\bibfnamefont {R.}~\bibnamefont
  {J.~Bouwens}} \emph {et~al.},\ }\href {\doibase 10.3847/1538-4357/833/1/72}
  {\bibfield  {journal} {\bibinfo  {journal} {The Astrophysical Journal}\
  }\textbf {\bibinfo {volume} {833}},\ \bibinfo {pages} {72} (\bibinfo {year}
  {2016})},\ \Eprint {http://arxiv.org/abs/1606.05280} {arXiv:1606.05280
  [astro-ph.GA]} \BibitemShut {NoStop}%
\bibitem [{\citenamefont {Alcock}\ and\ \citenamefont
  {Paczynski}(1979)}]{Alcock:1979mp}%
  \BibitemOpen
  \bibfield  {author} {\bibinfo {author} {\bibfnamefont {C.}~\bibnamefont
  {Alcock}}\ and\ \bibinfo {author} {\bibfnamefont {B.}~\bibnamefont
  {Paczynski}},\ }\href {\doibase 10.1038/281358a0} {\bibfield  {journal}
  {\bibinfo  {journal} {Nature}\ }\textbf {\bibinfo {volume} {281}},\ \bibinfo
  {pages} {358} (\bibinfo {year} {1979})}\BibitemShut {NoStop}%
\bibitem [{\citenamefont {Marinacci}\ \emph {et~al.}(2018)\citenamefont
  {Marinacci} \emph {et~al.}}]{Marinacci:2017wew}%
  \BibitemOpen
  \bibfield  {author} {\bibinfo {author} {\bibfnamefont {F.}~\bibnamefont
  {Marinacci}} \emph {et~al.},\ }\href {\doibase 10.1093/mnras/sty2206}
  {\bibfield  {journal} {\bibinfo  {journal} {Mon. Not. Roy. Astron. Soc.}\
  }\textbf {\bibinfo {volume} {480}},\ \bibinfo {pages} {5113} (\bibinfo {year}
  {2018})},\ \Eprint {http://arxiv.org/abs/1707.03396} {arXiv:1707.03396
  [astro-ph.CO]} \BibitemShut {NoStop}%
\bibitem [{\citenamefont {{Naiman}}\ \emph {et~al.}(2018)\citenamefont
  {{Naiman}}, \citenamefont {{Pillepich}}, \citenamefont {{Springel}},
  \citenamefont {{Ramirez-Ruiz}}, \citenamefont {{Torrey}}, \citenamefont
  {{Vogelsberger}}, \citenamefont {{Pakmor}}, \citenamefont {{Nelson}},
  \citenamefont {{Marinacci}}, \citenamefont {{Hernquist}}, \citenamefont
  {{Weinberger}},\ and\ \citenamefont {{Genel}}}]{Naiman2018}%
  \BibitemOpen
  \bibfield  {author} {\bibinfo {author} {\bibfnamefont {J.~P.}\ \bibnamefont
  {{Naiman}}}, \bibinfo {author} {\bibfnamefont {A.}~\bibnamefont
  {{Pillepich}}}, \bibinfo {author} {\bibfnamefont {V.}~\bibnamefont
  {{Springel}}}, \bibinfo {author} {\bibfnamefont {E.}~\bibnamefont
  {{Ramirez-Ruiz}}}, \bibinfo {author} {\bibfnamefont {P.}~\bibnamefont
  {{Torrey}}}, \bibinfo {author} {\bibfnamefont {M.}~\bibnamefont
  {{Vogelsberger}}}, \bibinfo {author} {\bibfnamefont {R.}~\bibnamefont
  {{Pakmor}}}, \bibinfo {author} {\bibfnamefont {D.}~\bibnamefont {{Nelson}}},
  \bibinfo {author} {\bibfnamefont {F.}~\bibnamefont {{Marinacci}}}, \bibinfo
  {author} {\bibfnamefont {L.}~\bibnamefont {{Hernquist}}}, \bibinfo {author}
  {\bibfnamefont {R.}~\bibnamefont {{Weinberger}}}, \ and\ \bibinfo {author}
  {\bibfnamefont {S.}~\bibnamefont {{Genel}}},\ }\href {\doibase
  10.1093/mnras/sty618} {\bibfield  {journal} {\bibinfo  {journal} {Mon. Not.
  Roy. Astron. Soc.}\ }\textbf {\bibinfo {volume} {477}},\ \bibinfo {pages}
  {1206} (\bibinfo {year} {2018})},\ \Eprint {http://arxiv.org/abs/1707.03401}
  {arXiv:1707.03401 [astro-ph.GA]} \BibitemShut {NoStop}%
\bibitem [{\citenamefont {Nelson}\ \emph {et~al.}(2018)\citenamefont {Nelson}
  \emph {et~al.}}]{Nelson:2017cxy}%
  \BibitemOpen
  \bibfield  {author} {\bibinfo {author} {\bibfnamefont {D.}~\bibnamefont
  {Nelson}} \emph {et~al.},\ }\href {\doibase 10.1093/mnras/stx3040} {\bibfield
   {journal} {\bibinfo  {journal} {Mon. Not. Roy. Astron. Soc.}\ }\textbf
  {\bibinfo {volume} {475}},\ \bibinfo {pages} {624} (\bibinfo {year}
  {2018})},\ \Eprint {http://arxiv.org/abs/1707.03395} {arXiv:1707.03395
  [astro-ph.GA]} \BibitemShut {NoStop}%
\bibitem [{\citenamefont {Pillepich}\ \emph {et~al.}(2018)\citenamefont
  {Pillepich} \emph {et~al.}}]{Pillepich:2017fcc}%
  \BibitemOpen
  \bibfield  {author} {\bibinfo {author} {\bibfnamefont {A.}~\bibnamefont
  {Pillepich}} \emph {et~al.},\ }\href {\doibase 10.1093/mnras/stx3112}
  {\bibfield  {journal} {\bibinfo  {journal} {Mon. Not. Roy. Astron. Soc.}\
  }\textbf {\bibinfo {volume} {475}},\ \bibinfo {pages} {648} (\bibinfo {year}
  {2018})},\ \Eprint {http://arxiv.org/abs/1707.03406} {arXiv:1707.03406
  [astro-ph.GA]} \BibitemShut {NoStop}%
\bibitem [{\citenamefont {Springel}\ \emph {et~al.}(2018)\citenamefont
  {Springel} \emph {et~al.}}]{Springel:2017tpz}%
  \BibitemOpen
  \bibfield  {author} {\bibinfo {author} {\bibfnamefont {V.}~\bibnamefont
  {Springel}} \emph {et~al.},\ }\href {\doibase 10.1093/mnras/stx3304}
  {\bibfield  {journal} {\bibinfo  {journal} {Mon. Not. Roy. Astron. Soc.}\
  }\textbf {\bibinfo {volume} {475}},\ \bibinfo {pages} {676} (\bibinfo {year}
  {2018})},\ \Eprint {http://arxiv.org/abs/1707.03397} {arXiv:1707.03397
  [astro-ph.GA]} \BibitemShut {NoStop}%
\bibitem [{\citenamefont {Nelson}\ \emph {et~al.}(2019)\citenamefont {Nelson},
  \citenamefont {Pillepich}, \citenamefont {Springel}, \citenamefont {Pakmor},
  \citenamefont {Weinberger}, \citenamefont {Genel}, \citenamefont {Torrey},
  \citenamefont {Vogelsberger}, \citenamefont {Marinacci},\ and\ \citenamefont
  {Hernquist}}]{Nelson:2019jkf}%
  \BibitemOpen
  \bibfield  {author} {\bibinfo {author} {\bibfnamefont {D.}~\bibnamefont
  {Nelson}}, \bibinfo {author} {\bibfnamefont {A.}~\bibnamefont {Pillepich}},
  \bibinfo {author} {\bibfnamefont {V.}~\bibnamefont {Springel}}, \bibinfo
  {author} {\bibfnamefont {R.}~\bibnamefont {Pakmor}}, \bibinfo {author}
  {\bibfnamefont {R.}~\bibnamefont {Weinberger}}, \bibinfo {author}
  {\bibfnamefont {S.}~\bibnamefont {Genel}}, \bibinfo {author} {\bibfnamefont
  {P.}~\bibnamefont {Torrey}}, \bibinfo {author} {\bibfnamefont
  {M.}~\bibnamefont {Vogelsberger}}, \bibinfo {author} {\bibfnamefont
  {F.}~\bibnamefont {Marinacci}}, \ and\ \bibinfo {author} {\bibfnamefont
  {L.}~\bibnamefont {Hernquist}},\ }\href {\doibase 10.1093/mnras/stz2306}
  {\bibfield  {journal} {\bibinfo  {journal} {Mon. Not. Roy. Astron. Soc.}\
  }\textbf {\bibinfo {volume} {490}},\ \bibinfo {pages} {3234} (\bibinfo {year}
  {2019})},\ \Eprint {http://arxiv.org/abs/1902.05554} {arXiv:1902.05554
  [astro-ph.GA]} \BibitemShut {NoStop}%
\bibitem [{\citenamefont {Pillepich}\ \emph {et~al.}(2019)\citenamefont
  {Pillepich} \emph {et~al.}}]{Pillepich:2019bmb}%
  \BibitemOpen
  \bibfield  {author} {\bibinfo {author} {\bibfnamefont {A.}~\bibnamefont
  {Pillepich}} \emph {et~al.},\ }\href {\doibase 10.1093/mnras/stz2338}
  {\bibfield  {journal} {\bibinfo  {journal} {Mon. Not. Roy. Astron. Soc.}\
  }\textbf {\bibinfo {volume} {490}},\ \bibinfo {pages} {3196} (\bibinfo {year}
  {2019})},\ \Eprint {http://arxiv.org/abs/1902.05553} {arXiv:1902.05553
  [astro-ph.GA]} \BibitemShut {NoStop}%
\bibitem [{\citenamefont {Aghanim}\ \emph {et~al.}(2020)\citenamefont {Aghanim}
  \emph {et~al.}}]{Planck:2019nip}%
  \BibitemOpen
  \bibfield  {author} {\bibinfo {author} {\bibfnamefont {N.}~\bibnamefont
  {Aghanim}} \emph {et~al.} (\bibinfo {collaboration} {Planck}),\ }\href
  {\doibase 10.1051/0004-6361/201936386} {\bibfield  {journal} {\bibinfo
  {journal} {Astron. Astrophys.}\ }\textbf {\bibinfo {volume} {641}},\ \bibinfo
  {pages} {A5} (\bibinfo {year} {2020})},\ \Eprint
  {http://arxiv.org/abs/1907.12875} {arXiv:1907.12875 [astro-ph.CO]}
  \BibitemShut {NoStop}%
\bibitem [{\citenamefont {Scolnic}\ \emph {et~al.}(2018)\citenamefont {Scolnic}
  \emph {et~al.}}]{Scolnic:2017caz}%
  \BibitemOpen
  \bibfield  {author} {\bibinfo {author} {\bibfnamefont {D.~M.}\ \bibnamefont
  {Scolnic}} \emph {et~al.},\ }\href {\doibase 10.3847/1538-4357/aab9bb}
  {\bibfield  {journal} {\bibinfo  {journal} {Astrophys. J.}\ }\textbf
  {\bibinfo {volume} {859}},\ \bibinfo {pages} {101} (\bibinfo {year}
  {2018})},\ \Eprint {http://arxiv.org/abs/1710.00845} {arXiv:1710.00845
  [astro-ph.CO]} \BibitemShut {NoStop}%
\bibitem [{\citenamefont {Jones}\ \emph {et~al.}(2018)\citenamefont {Jones}
  \emph {et~al.}}]{Jones:2017udy}%
  \BibitemOpen
  \bibfield  {author} {\bibinfo {author} {\bibfnamefont {D.~O.}\ \bibnamefont
  {Jones}} \emph {et~al.},\ }\href {\doibase 10.3847/1538-4357/aab6b1}
  {\bibfield  {journal} {\bibinfo  {journal} {Astrophys. J.}\ }\textbf
  {\bibinfo {volume} {857}},\ \bibinfo {pages} {51} (\bibinfo {year} {2018})},\
  \Eprint {http://arxiv.org/abs/1710.00846} {arXiv:1710.00846 [astro-ph.CO]}
  \BibitemShut {NoStop}%
\bibitem [{\citenamefont {Betoule}\ \emph {et~al.}(2014)\citenamefont {Betoule}
  \emph {et~al.}}]{SDSS:2014iwm}%
  \BibitemOpen
  \bibfield  {author} {\bibinfo {author} {\bibfnamefont {M.}~\bibnamefont
  {Betoule}} \emph {et~al.} (\bibinfo {collaboration} {SDSS}),\ }\href
  {\doibase 10.1051/0004-6361/201423413} {\bibfield  {journal} {\bibinfo
  {journal} {Astron. Astrophys.}\ }\textbf {\bibinfo {volume} {568}},\ \bibinfo
  {pages} {A22} (\bibinfo {year} {2014})},\ \Eprint
  {http://arxiv.org/abs/1401.4064} {arXiv:1401.4064 [astro-ph.CO]} \BibitemShut
  {NoStop}%
\bibitem [{\citenamefont {Aghanim}\ \emph {et~al.}(2018)\citenamefont {Aghanim}
  \emph {et~al.}}]{Aghanim:2018eyx}%
  \BibitemOpen
  \bibfield  {author} {\bibinfo {author} {\bibfnamefont {N.}~\bibnamefont
  {Aghanim}} \emph {et~al.} (\bibinfo {collaboration} {Planck}),\ }\href@noop
  {} {\  (\bibinfo {year} {2018})},\ \Eprint {http://arxiv.org/abs/1807.06209}
  {arXiv:1807.06209 [astro-ph.CO]} \BibitemShut {NoStop}%
\bibitem [{\citenamefont {Hikage}\ \emph {et~al.}(2019)\citenamefont {Hikage}
  \emph {et~al.}}]{HSC:2018mrq}%
  \BibitemOpen
  \bibfield  {author} {\bibinfo {author} {\bibfnamefont {C.}~\bibnamefont
  {Hikage}} \emph {et~al.} (\bibinfo {collaboration} {HSC}),\ }\href {\doibase
  10.1093/pasj/psz010} {\bibfield  {journal} {\bibinfo  {journal} {Publ.
  Astron. Soc. Jap.}\ }\textbf {\bibinfo {volume} {71}},\ \bibinfo {pages} {43}
  (\bibinfo {year} {2019})},\ \Eprint {http://arxiv.org/abs/1809.09148}
  {arXiv:1809.09148 [astro-ph.CO]} \BibitemShut {NoStop}%
\bibitem [{\citenamefont {Asgari}\ \emph {et~al.}(2021)\citenamefont {Asgari}
  \emph {et~al.}}]{KiDS:2020suj}%
  \BibitemOpen
  \bibfield  {author} {\bibinfo {author} {\bibfnamefont {M.}~\bibnamefont
  {Asgari}} \emph {et~al.} (\bibinfo {collaboration} {KiDS}),\ }\href {\doibase
  10.1051/0004-6361/202039070} {\bibfield  {journal} {\bibinfo  {journal}
  {Astron. Astrophys.}\ }\textbf {\bibinfo {volume} {645}},\ \bibinfo {pages}
  {A104} (\bibinfo {year} {2021})},\ \Eprint {http://arxiv.org/abs/2007.15633}
  {arXiv:2007.15633 [astro-ph.CO]} \BibitemShut {NoStop}%
\bibitem [{\citenamefont {Abbott}\ \emph {et~al.}(2021)\citenamefont {Abbott}
  \emph {et~al.}}]{DES:2021wwk}%
  \BibitemOpen
  \bibfield  {author} {\bibinfo {author} {\bibfnamefont {T.~M.~C.}\
  \bibnamefont {Abbott}} \emph {et~al.} (\bibinfo {collaboration} {DES}),\
  }\href@noop {} {\  (\bibinfo {year} {2021})},\ \Eprint
  {http://arxiv.org/abs/2105.13549} {arXiv:2105.13549 [astro-ph.CO]}
  \BibitemShut {NoStop}%
\bibitem [{\citenamefont {Pisanti}\ \emph {et~al.}(2021)\citenamefont
  {Pisanti}, \citenamefont {Mangano}, \citenamefont {Miele},\ and\
  \citenamefont {Mazzella}}]{Pisanti:2020efz}%
  \BibitemOpen
  \bibfield  {author} {\bibinfo {author} {\bibfnamefont {O.}~\bibnamefont
  {Pisanti}}, \bibinfo {author} {\bibfnamefont {G.}~\bibnamefont {Mangano}},
  \bibinfo {author} {\bibfnamefont {G.}~\bibnamefont {Miele}}, \ and\ \bibinfo
  {author} {\bibfnamefont {P.}~\bibnamefont {Mazzella}},\ }\href {\doibase
  10.1088/1475-7516/2021/04/020} {\bibfield  {journal} {\bibinfo  {journal}
  {JCAP}\ }\textbf {\bibinfo {volume} {04}},\ \bibinfo {pages} {020} (\bibinfo
  {year} {2021})},\ \Eprint {http://arxiv.org/abs/2011.11537} {arXiv:2011.11537
  [astro-ph.CO]} \BibitemShut {NoStop}%
\bibitem [{\citenamefont {Riess}\ \emph {et~al.}(2019)\citenamefont {Riess},
  \citenamefont {Casertano}, \citenamefont {Yuan}, \citenamefont {Macri},\ and\
  \citenamefont {Scolnic}}]{Riess:2019cxk}%
  \BibitemOpen
  \bibfield  {author} {\bibinfo {author} {\bibfnamefont {A.~G.}\ \bibnamefont
  {Riess}}, \bibinfo {author} {\bibfnamefont {S.}~\bibnamefont {Casertano}},
  \bibinfo {author} {\bibfnamefont {W.}~\bibnamefont {Yuan}}, \bibinfo {author}
  {\bibfnamefont {L.~M.}\ \bibnamefont {Macri}}, \ and\ \bibinfo {author}
  {\bibfnamefont {D.}~\bibnamefont {Scolnic}},\ }\href {\doibase
  10.3847/1538-4357/ab1422} {\bibfield  {journal} {\bibinfo  {journal}
  {Astrophys. J.}\ }\textbf {\bibinfo {volume} {876}},\ \bibinfo {pages} {85}
  (\bibinfo {year} {2019})},\ \Eprint {http://arxiv.org/abs/1903.07603}
  {arXiv:1903.07603 [astro-ph.CO]} \BibitemShut {NoStop}%
\bibitem [{\citenamefont {Percival}\ \emph {et~al.}(2002)\citenamefont
  {Percival} \emph {et~al.}}]{2dFGRSTeam:2002tzq}%
  \BibitemOpen
  \bibfield  {author} {\bibinfo {author} {\bibfnamefont {W.~J.}\ \bibnamefont
  {Percival}} \emph {et~al.} (\bibinfo {collaboration} {2dFGRS Team}),\ }\href
  {\doibase 10.1046/j.1365-8711.2002.06001.x} {\bibfield  {journal} {\bibinfo
  {journal} {Mon. Not. Roy. Astron. Soc.}\ }\textbf {\bibinfo {volume} {337}},\
  \bibinfo {pages} {1068} (\bibinfo {year} {2002})},\ \Eprint
  {http://arxiv.org/abs/astro-ph/0206256} {arXiv:astro-ph/0206256} \BibitemShut
  {NoStop}%
\bibitem [{\citenamefont {Kable}\ \emph {et~al.}(2019)\citenamefont {Kable},
  \citenamefont {Addison},\ and\ \citenamefont {Bennett}}]{Kable:2018bgg}%
  \BibitemOpen
  \bibfield  {author} {\bibinfo {author} {\bibfnamefont {J.~A.}\ \bibnamefont
  {Kable}}, \bibinfo {author} {\bibfnamefont {G.~E.}\ \bibnamefont {Addison}},
  \ and\ \bibinfo {author} {\bibfnamefont {C.~L.}\ \bibnamefont {Bennett}},\
  }\href {\doibase 10.3847/1538-4357/aaf56d} {\bibfield  {journal} {\bibinfo
  {journal} {Astrophys. J.}\ }\textbf {\bibinfo {volume} {871}},\ \bibinfo
  {pages} {77} (\bibinfo {year} {2019})},\ \Eprint
  {http://arxiv.org/abs/1809.03983} {arXiv:1809.03983 [astro-ph.CO]}
  \BibitemShut {NoStop}%
\bibitem [{\citenamefont {Schneider}\ \emph {et~al.}(2021)\citenamefont
  {Schneider}, \citenamefont {Giri},\ and\ \citenamefont
  {Mirocha}}]{Schneider:2020xmf}%
  \BibitemOpen
  \bibfield  {author} {\bibinfo {author} {\bibfnamefont {A.}~\bibnamefont
  {Schneider}}, \bibinfo {author} {\bibfnamefont {S.~K.}\ \bibnamefont {Giri}},
  \ and\ \bibinfo {author} {\bibfnamefont {J.}~\bibnamefont {Mirocha}},\ }\href
  {\doibase 10.1103/PhysRevD.103.083025} {\bibfield  {journal} {\bibinfo
  {journal} {Phys. Rev. D}\ }\textbf {\bibinfo {volume} {103}},\ \bibinfo
  {pages} {083025} (\bibinfo {year} {2021})},\ \Eprint
  {http://arxiv.org/abs/2011.12308} {arXiv:2011.12308 [astro-ph.CO]}
  \BibitemShut {NoStop}%
\bibitem [{\citenamefont {Harikane}\ \emph {et~al.}(2017)\citenamefont
  {Harikane} \emph {et~al.}}]{Harikane:2017lcw}%
  \BibitemOpen
  \bibfield  {author} {\bibinfo {author} {\bibfnamefont {Y.}~\bibnamefont
  {Harikane}} \emph {et~al.},\ }\href {\doibase 10.1093/pasj/psx097} {\
  (\bibinfo {year} {2017}),\ 10.1093/pasj/psx097},\ \Eprint
  {http://arxiv.org/abs/1704.06535} {arXiv:1704.06535 [astro-ph.GA]}
  \BibitemShut {NoStop}%
\bibitem [{\citenamefont {Spergel}\ \emph {et~al.}(2015)\citenamefont {Spergel}
  \emph {et~al.}}]{Spergel:2015sza}%
  \BibitemOpen
  \bibfield  {author} {\bibinfo {author} {\bibfnamefont {D.}~\bibnamefont
  {Spergel}} \emph {et~al.},\ }\href@noop {} {\  (\bibinfo {year} {2015})},\
  \Eprint {http://arxiv.org/abs/1503.03757} {arXiv:1503.03757 [astro-ph.IM]}
  \BibitemShut {NoStop}%
\bibitem [{\citenamefont {Bunker}(2019)}]{bunker_2019}%
  \BibitemOpen
  \bibfield  {author} {\bibinfo {author} {\bibfnamefont {A.~J.}\ \bibnamefont
  {Bunker}},\ }\href {\doibase 10.1017/S1743921319009463} {\bibfield  {journal}
  {\bibinfo  {journal} {Proceedings of the International Astronomical Union}\
  }\textbf {\bibinfo {volume} {15}},\ \bibinfo {pages} {342–346} (\bibinfo
  {year} {2019})}\BibitemShut {NoStop}%
\bibitem [{\citenamefont {{Williams}}\ \emph {et~al.}(2021)\citenamefont
  {{Williams}}, \citenamefont {{Oesch}}, \citenamefont {{Barrufet}},
  \citenamefont {{Bezanson}}, \citenamefont {{Bowler}}, \citenamefont
  {{Brammer}}, \citenamefont {{Dayal}}, \citenamefont {{Franx}}, \citenamefont
  {{Hutter}}, \citenamefont {{Labbe}}, \citenamefont {{Maseda}}, \citenamefont
  {{Ucci}},\ and\ \citenamefont {{Whitaker}}}]{PANORAMIC_JWST}%
  \BibitemOpen
  \bibfield  {author} {\bibinfo {author} {\bibfnamefont {C.~C.}\ \bibnamefont
  {{Williams}}}, \bibinfo {author} {\bibfnamefont {P.}~\bibnamefont {{Oesch}}},
  \bibinfo {author} {\bibfnamefont {L.}~\bibnamefont {{Barrufet}}}, \bibinfo
  {author} {\bibfnamefont {R.}~\bibnamefont {{Bezanson}}}, \bibinfo {author}
  {\bibfnamefont {R.~A.~A.}\ \bibnamefont {{Bowler}}}, \bibinfo {author}
  {\bibfnamefont {G.}~\bibnamefont {{Brammer}}}, \bibinfo {author}
  {\bibfnamefont {P.}~\bibnamefont {{Dayal}}}, \bibinfo {author} {\bibfnamefont
  {M.}~\bibnamefont {{Franx}}}, \bibinfo {author} {\bibfnamefont
  {A.}~\bibnamefont {{Hutter}}}, \bibinfo {author} {\bibfnamefont
  {I.}~\bibnamefont {{Labbe}}}, \bibinfo {author} {\bibfnamefont
  {M.}~\bibnamefont {{Maseda}}}, \bibinfo {author} {\bibfnamefont
  {G.}~\bibnamefont {{Ucci}}}, \ and\ \bibinfo {author} {\bibfnamefont {K.~E.}\
  \bibnamefont {{Whitaker}}},\ }\href@noop {} {\enquote {\bibinfo {title}
  {{PANORAMIC - A Pure Parallel Wide Area Legacy Imaging Survey at 1-5
  Micron}},}\ }\bibinfo {howpublished} {JWST Proposal. Cycle 1} (\bibinfo
  {year} {2021})\BibitemShut {NoStop}%
\bibitem [{\citenamefont {{Rix}}\ and\ \citenamefont
  {{Rieke}}(1993)}]{Rix_Rieke}%
  \BibitemOpen
  \bibfield  {author} {\bibinfo {author} {\bibfnamefont {H.-W.}\ \bibnamefont
  {{Rix}}}\ and\ \bibinfo {author} {\bibfnamefont {M.~J.}\ \bibnamefont
  {{Rieke}}},\ }\href {\doibase 10.1086/173376} {\bibfield  {journal} {\bibinfo
   {journal} {\apj}\ }\textbf {\bibinfo {volume} {418}},\ \bibinfo {pages}
  {123} (\bibinfo {year} {1993})}\BibitemShut {NoStop}%
\bibitem [{\citenamefont {Kauffmann}\ and\ \citenamefont
  {Charlot}(1998)}]{Kauffmann:1998gm}%
  \BibitemOpen
  \bibfield  {author} {\bibinfo {author} {\bibfnamefont {G.}~\bibnamefont
  {Kauffmann}}\ and\ \bibinfo {author} {\bibfnamefont {S.}~\bibnamefont
  {Charlot}},\ }\href {\doibase 10.1046/j.1365-8711.1998.01708.x} {\bibfield
  {journal} {\bibinfo  {journal} {Mon. Not. Roy. Astron. Soc.}\ }\textbf
  {\bibinfo {volume} {297}},\ \bibinfo {pages} {L23} (\bibinfo {year}
  {1998})},\ \Eprint {http://arxiv.org/abs/astro-ph/9802233}
  {arXiv:astro-ph/9802233} \BibitemShut {NoStop}%
\bibitem [{\citenamefont {Bell}\ and\ \citenamefont
  {de~Jong}(2001)}]{Bell:2000jt}%
  \BibitemOpen
  \bibfield  {author} {\bibinfo {author} {\bibfnamefont {E.~F.}\ \bibnamefont
  {Bell}}\ and\ \bibinfo {author} {\bibfnamefont {R.~S.}\ \bibnamefont
  {de~Jong}},\ }\href {\doibase 10.1086/319728} {\bibfield  {journal} {\bibinfo
   {journal} {Astrophys. J.}\ }\textbf {\bibinfo {volume} {550}},\ \bibinfo
  {pages} {212} (\bibinfo {year} {2001})},\ \Eprint
  {http://arxiv.org/abs/astro-ph/0011493} {arXiv:astro-ph/0011493} \BibitemShut
  {NoStop}%
\bibitem [{\citenamefont {Bell}\ \emph {et~al.}(2003)\citenamefont {Bell},
  \citenamefont {McIntosh}, \citenamefont {Katz},\ and\ \citenamefont
  {Weinberg}}]{Bell:2003cj}%
  \BibitemOpen
  \bibfield  {author} {\bibinfo {author} {\bibfnamefont {E.~F.}\ \bibnamefont
  {Bell}}, \bibinfo {author} {\bibfnamefont {D.~H.}\ \bibnamefont {McIntosh}},
  \bibinfo {author} {\bibfnamefont {N.}~\bibnamefont {Katz}}, \ and\ \bibinfo
  {author} {\bibfnamefont {M.~D.}\ \bibnamefont {Weinberg}},\ }\href {\doibase
  10.1086/378847} {\bibfield  {journal} {\bibinfo  {journal} {Astrophys. J.
  Suppl.}\ }\textbf {\bibinfo {volume} {149}},\ \bibinfo {pages} {289}
  (\bibinfo {year} {2003})},\ \Eprint {http://arxiv.org/abs/astro-ph/0302543}
  {arXiv:astro-ph/0302543} \BibitemShut {NoStop}%
\bibitem [{\citenamefont {Drory}\ \emph {et~al.}(2004)\citenamefont {Drory},
  \citenamefont {Bender}, \citenamefont {Feulner}, \citenamefont {Hopp},
  \citenamefont {Maraston}, \citenamefont {Snigula},\ and\ \citenamefont
  {Hill}}]{Drory:2004eh}%
  \BibitemOpen
  \bibfield  {author} {\bibinfo {author} {\bibfnamefont {N.}~\bibnamefont
  {Drory}}, \bibinfo {author} {\bibfnamefont {R.}~\bibnamefont {Bender}},
  \bibinfo {author} {\bibfnamefont {G.}~\bibnamefont {Feulner}}, \bibinfo
  {author} {\bibfnamefont {U.}~\bibnamefont {Hopp}}, \bibinfo {author}
  {\bibfnamefont {C.}~\bibnamefont {Maraston}}, \bibinfo {author}
  {\bibfnamefont {J.}~\bibnamefont {Snigula}}, \ and\ \bibinfo {author}
  {\bibfnamefont {G.~J.}\ \bibnamefont {Hill}},\ }\href {\doibase
  10.1086/420781} {\bibfield  {journal} {\bibinfo  {journal} {Astrophys. J.}\
  }\textbf {\bibinfo {volume} {608}},\ \bibinfo {pages} {742} (\bibinfo {year}
  {2004})},\ \Eprint {http://arxiv.org/abs/astro-ph/0403041}
  {arXiv:astro-ph/0403041} \BibitemShut {NoStop}%
\bibitem [{\citenamefont {{Stark}}\ \emph {et~al.}(2009)\citenamefont
  {{Stark}}, \citenamefont {{Ellis}}, \citenamefont {{Bunker}}, \citenamefont
  {{Bundy}}, \citenamefont {{Targett}}, \citenamefont {{Benson}},\ and\
  \citenamefont {{Lacy}}}]{stark2009}%
  \BibitemOpen
  \bibfield  {author} {\bibinfo {author} {\bibfnamefont {D.~P.}\ \bibnamefont
  {{Stark}}}, \bibinfo {author} {\bibfnamefont {R.~S.}\ \bibnamefont
  {{Ellis}}}, \bibinfo {author} {\bibfnamefont {A.}~\bibnamefont {{Bunker}}},
  \bibinfo {author} {\bibfnamefont {K.}~\bibnamefont {{Bundy}}}, \bibinfo
  {author} {\bibfnamefont {T.}~\bibnamefont {{Targett}}}, \bibinfo {author}
  {\bibfnamefont {A.}~\bibnamefont {{Benson}}}, \ and\ \bibinfo {author}
  {\bibfnamefont {M.}~\bibnamefont {{Lacy}}},\ }\href {\doibase
  10.1088/0004-637X/697/2/1493} {\bibfield  {journal} {\bibinfo  {journal}
  {\apj}\ }\textbf {\bibinfo {volume} {697}},\ \bibinfo {pages} {1493}
  (\bibinfo {year} {2009})},\ \Eprint {http://arxiv.org/abs/0902.2907}
  {arXiv:0902.2907 [astro-ph.CO]} \BibitemShut {NoStop}%
\bibitem [{\citenamefont {{Gonz{\'a}lez}}\ \emph {et~al.}(2011)\citenamefont
  {{Gonz{\'a}lez}}, \citenamefont {{Labb{\'e}}}, \citenamefont {{Bouwens}},
  \citenamefont {{Illingworth}}, \citenamefont {{Franx}},\ and\ \citenamefont
  {{Kriek}}}]{gonzalez2011}%
  \BibitemOpen
  \bibfield  {author} {\bibinfo {author} {\bibfnamefont {V.}~\bibnamefont
  {{Gonz{\'a}lez}}}, \bibinfo {author} {\bibfnamefont {I.}~\bibnamefont
  {{Labb{\'e}}}}, \bibinfo {author} {\bibfnamefont {R.~J.}\ \bibnamefont
  {{Bouwens}}}, \bibinfo {author} {\bibfnamefont {G.}~\bibnamefont
  {{Illingworth}}}, \bibinfo {author} {\bibfnamefont {M.}~\bibnamefont
  {{Franx}}}, \ and\ \bibinfo {author} {\bibfnamefont {M.}~\bibnamefont
  {{Kriek}}},\ }\href {\doibase 10.1088/2041-8205/735/2/L34} {\bibfield
  {journal} {\bibinfo  {journal} {The Astrophysical Journal}\ }\textbf
  {\bibinfo {volume} {735}},\ \bibinfo {eid} {L34} (\bibinfo {year} {2011})},\
  \Eprint {http://arxiv.org/abs/1008.3901} {arXiv:1008.3901 [astro-ph.CO]}
  \BibitemShut {NoStop}%
\bibitem [{\citenamefont {{Duncan}}\ \emph {et~al.}(2014)\citenamefont
  {{Duncan}}, \citenamefont {{Conselice}}, \citenamefont {{Mortlock}},
  \citenamefont {{Hartley}}, \citenamefont {{Guo}}, \citenamefont {{Ferguson}},
  \citenamefont {{Dav{\'e}}}, \citenamefont {{Lu}}, \citenamefont
  {{Ownsworth}}, \citenamefont {{Ashby}}, \citenamefont {{Dekel}},
  \citenamefont {{Dickinson}}, \citenamefont {{Faber}}, \citenamefont
  {{Giavalisco}}, \citenamefont {{Grogin}}, \citenamefont {{Kocevski}},
  \citenamefont {{Koekemoer}}, \citenamefont {{Somerville}},\ and\
  \citenamefont {{White}}}]{duncan2014}%
  \BibitemOpen
  \bibfield  {author} {\bibinfo {author} {\bibfnamefont {K.}~\bibnamefont
  {{Duncan}}}, \bibinfo {author} {\bibfnamefont {C.~J.}\ \bibnamefont
  {{Conselice}}}, \bibinfo {author} {\bibfnamefont {A.}~\bibnamefont
  {{Mortlock}}}, \bibinfo {author} {\bibfnamefont {W.~G.}\ \bibnamefont
  {{Hartley}}}, \bibinfo {author} {\bibfnamefont {Y.}~\bibnamefont {{Guo}}},
  \bibinfo {author} {\bibfnamefont {H.~C.}\ \bibnamefont {{Ferguson}}},
  \bibinfo {author} {\bibfnamefont {R.}~\bibnamefont {{Dav{\'e}}}}, \bibinfo
  {author} {\bibfnamefont {Y.}~\bibnamefont {{Lu}}}, \bibinfo {author}
  {\bibfnamefont {J.}~\bibnamefont {{Ownsworth}}}, \bibinfo {author}
  {\bibfnamefont {M.~L.~N.}\ \bibnamefont {{Ashby}}}, \bibinfo {author}
  {\bibfnamefont {A.}~\bibnamefont {{Dekel}}}, \bibinfo {author} {\bibfnamefont
  {M.}~\bibnamefont {{Dickinson}}}, \bibinfo {author} {\bibfnamefont
  {S.}~\bibnamefont {{Faber}}}, \bibinfo {author} {\bibfnamefont
  {M.}~\bibnamefont {{Giavalisco}}}, \bibinfo {author} {\bibfnamefont
  {N.}~\bibnamefont {{Grogin}}}, \bibinfo {author} {\bibfnamefont
  {D.}~\bibnamefont {{Kocevski}}}, \bibinfo {author} {\bibfnamefont
  {A.}~\bibnamefont {{Koekemoer}}}, \bibinfo {author} {\bibfnamefont {R.~S.}\
  \bibnamefont {{Somerville}}}, \ and\ \bibinfo {author} {\bibfnamefont
  {C.~E.}\ \bibnamefont {{White}}},\ }\href {\doibase 10.1093/mnras/stu1622}
  {\bibfield  {journal} {\bibinfo  {journal} {Monthly Notices of the Royal
  Astronomical Society}\ }\textbf {\bibinfo {volume} {444}},\ \bibinfo {pages}
  {2960} (\bibinfo {year} {2014})},\ \Eprint {http://arxiv.org/abs/1408.2527}
  {arXiv:1408.2527 [astro-ph.GA]} \BibitemShut {NoStop}%
\bibitem [{\citenamefont {Grazian}\ \emph {et~al.}(2015)\citenamefont
  {Grazian}, \citenamefont {Fontana}, \citenamefont {Santini}, \citenamefont
  {Dunlop}, \citenamefont {Ferguson}, \citenamefont {Castellano}, \citenamefont
  {Amorin}, \citenamefont {Ashby}, \citenamefont {Barro}, \citenamefont
  {Behroozi},\ and\ \citenamefont {et~al.}}]{grazian2015}%
  \BibitemOpen
  \bibfield  {author} {\bibinfo {author} {\bibfnamefont {A.}~\bibnamefont
  {Grazian}}, \bibinfo {author} {\bibfnamefont {A.}~\bibnamefont {Fontana}},
  \bibinfo {author} {\bibfnamefont {P.}~\bibnamefont {Santini}}, \bibinfo
  {author} {\bibfnamefont {J.~S.}\ \bibnamefont {Dunlop}}, \bibinfo {author}
  {\bibfnamefont {H.~C.}\ \bibnamefont {Ferguson}}, \bibinfo {author}
  {\bibfnamefont {M.}~\bibnamefont {Castellano}}, \bibinfo {author}
  {\bibfnamefont {R.}~\bibnamefont {Amorin}}, \bibinfo {author} {\bibfnamefont
  {M.~L.~N.}\ \bibnamefont {Ashby}}, \bibinfo {author} {\bibfnamefont
  {G.}~\bibnamefont {Barro}}, \bibinfo {author} {\bibfnamefont
  {P.}~\bibnamefont {Behroozi}}, \ and\ \bibinfo {author} {\bibnamefont
  {et~al.}},\ }\href {\doibase 10.1051/0004-6361/201424750} {\bibfield
  {journal} {\bibinfo  {journal} {Astronomy \& Astrophysics}\ }\textbf
  {\bibinfo {volume} {575}},\ \bibinfo {pages} {A96} (\bibinfo {year}
  {2015})}\BibitemShut {NoStop}%
\bibitem [{\citenamefont {{Song}}\ \emph {et~al.}(2016)\citenamefont {{Song}},
  \citenamefont {{Finkelstein}}, \citenamefont {{Ashby}}, \citenamefont
  {{Grazian}}, \citenamefont {{Lu}}, \citenamefont {{Papovich}}, \citenamefont
  {{Salmon}}, \citenamefont {{Somerville}}, \citenamefont {{Dickinson}},
  \citenamefont {{Duncan}}, \citenamefont {{Faber}}, \citenamefont {{Fazio}},
  \citenamefont {{Ferguson}}, \citenamefont {{Fontana}}, \citenamefont {{Guo}},
  \citenamefont {{Hathi}}, \citenamefont {{Lee}}, \citenamefont {{Merlin}},\
  and\ \citenamefont {{Willner}}}]{song2016}%
  \BibitemOpen
  \bibfield  {author} {\bibinfo {author} {\bibfnamefont {M.}~\bibnamefont
  {{Song}}}, \bibinfo {author} {\bibfnamefont {S.~L.}\ \bibnamefont
  {{Finkelstein}}}, \bibinfo {author} {\bibfnamefont {M.~L.~N.}\ \bibnamefont
  {{Ashby}}}, \bibinfo {author} {\bibfnamefont {A.}~\bibnamefont {{Grazian}}},
  \bibinfo {author} {\bibfnamefont {Y.}~\bibnamefont {{Lu}}}, \bibinfo {author}
  {\bibfnamefont {C.}~\bibnamefont {{Papovich}}}, \bibinfo {author}
  {\bibfnamefont {B.}~\bibnamefont {{Salmon}}}, \bibinfo {author}
  {\bibfnamefont {R.~S.}\ \bibnamefont {{Somerville}}}, \bibinfo {author}
  {\bibfnamefont {M.}~\bibnamefont {{Dickinson}}}, \bibinfo {author}
  {\bibfnamefont {K.}~\bibnamefont {{Duncan}}}, \bibinfo {author}
  {\bibfnamefont {S.~M.}\ \bibnamefont {{Faber}}}, \bibinfo {author}
  {\bibfnamefont {G.~G.}\ \bibnamefont {{Fazio}}}, \bibinfo {author}
  {\bibfnamefont {H.~C.}\ \bibnamefont {{Ferguson}}}, \bibinfo {author}
  {\bibfnamefont {A.}~\bibnamefont {{Fontana}}}, \bibinfo {author}
  {\bibfnamefont {Y.}~\bibnamefont {{Guo}}}, \bibinfo {author} {\bibfnamefont
  {N.}~\bibnamefont {{Hathi}}}, \bibinfo {author} {\bibfnamefont {S.-K.}\
  \bibnamefont {{Lee}}}, \bibinfo {author} {\bibfnamefont {E.}~\bibnamefont
  {{Merlin}}}, \ and\ \bibinfo {author} {\bibfnamefont {S.~P.}\ \bibnamefont
  {{Willner}}},\ }\href {\doibase 10.3847/0004-637X/825/1/5} {\bibfield
  {journal} {\bibinfo  {journal} {\apj}\ }\textbf {\bibinfo {volume} {825}},\
  \bibinfo {eid} {5} (\bibinfo {year} {2016})},\ \Eprint
  {http://arxiv.org/abs/1507.05636} {arXiv:1507.05636 [astro-ph.GA]}
  \BibitemShut {NoStop}%
\bibitem [{\citenamefont {{Bhatawdekar}}\ \emph {et~al.}(2019)\citenamefont
  {{Bhatawdekar}}, \citenamefont {{Conselice}}, \citenamefont
  {{Margalef-Bentabol}},\ and\ \citenamefont {{Duncan}}}]{Bhatawdekar2019}%
  \BibitemOpen
  \bibfield  {author} {\bibinfo {author} {\bibfnamefont {R.}~\bibnamefont
  {{Bhatawdekar}}}, \bibinfo {author} {\bibfnamefont {C.~J.}\ \bibnamefont
  {{Conselice}}}, \bibinfo {author} {\bibfnamefont {B.}~\bibnamefont
  {{Margalef-Bentabol}}}, \ and\ \bibinfo {author} {\bibfnamefont
  {K.}~\bibnamefont {{Duncan}}},\ }\href {\doibase 10.1093/mnras/stz866}
  {\bibfield  {journal} {\bibinfo  {journal} {Monthly Notices of the Royal
  Astronomical Society}\ }\textbf {\bibinfo {volume} {486}},\ \bibinfo {pages}
  {3805} (\bibinfo {year} {2019})},\ \Eprint {http://arxiv.org/abs/1807.07580}
  {arXiv:1807.07580 [astro-ph.GA]} \BibitemShut {NoStop}%
\bibitem [{\citenamefont {{Kikuchihara}}\ \emph {et~al.}(2020)\citenamefont
  {{Kikuchihara}}, \citenamefont {{Ouchi}}, \citenamefont {{Ono}},
  \citenamefont {{Mawatari}}, \citenamefont {{Chevallard}}, \citenamefont
  {{Harikane}}, \citenamefont {{Kojima}}, \citenamefont {{Oguri}},
  \citenamefont {{Bruzual}},\ and\ \citenamefont
  {{Charlot}}}]{Kikuchihara2020}%
  \BibitemOpen
  \bibfield  {author} {\bibinfo {author} {\bibfnamefont {S.}~\bibnamefont
  {{Kikuchihara}}}, \bibinfo {author} {\bibfnamefont {M.}~\bibnamefont
  {{Ouchi}}}, \bibinfo {author} {\bibfnamefont {Y.}~\bibnamefont {{Ono}}},
  \bibinfo {author} {\bibfnamefont {K.}~\bibnamefont {{Mawatari}}}, \bibinfo
  {author} {\bibfnamefont {J.}~\bibnamefont {{Chevallard}}}, \bibinfo {author}
  {\bibfnamefont {Y.}~\bibnamefont {{Harikane}}}, \bibinfo {author}
  {\bibfnamefont {T.}~\bibnamefont {{Kojima}}}, \bibinfo {author}
  {\bibfnamefont {M.}~\bibnamefont {{Oguri}}}, \bibinfo {author} {\bibfnamefont
  {G.}~\bibnamefont {{Bruzual}}}, \ and\ \bibinfo {author} {\bibfnamefont
  {S.}~\bibnamefont {{Charlot}}},\ }\href {\doibase 10.3847/1538-4357/ab7dbe}
  {\bibfield  {journal} {\bibinfo  {journal} {\apj}\ }\textbf {\bibinfo
  {volume} {893}},\ \bibinfo {eid} {60} (\bibinfo {year} {2020})},\ \Eprint
  {http://arxiv.org/abs/1905.06927} {arXiv:1905.06927 [astro-ph.GA]}
  \BibitemShut {NoStop}%
\bibitem [{\citenamefont {Stefanon}\ \emph {et~al.}(2021)\citenamefont
  {Stefanon}, \citenamefont {Bouwens}, \citenamefont {Labbé}, \citenamefont
  {Illingworth}, \citenamefont {Gonzalez},\ and\ \citenamefont
  {Oesch}}]{stefanon2021galaxy}%
  \BibitemOpen
  \bibfield  {author} {\bibinfo {author} {\bibfnamefont {M.}~\bibnamefont
  {Stefanon}}, \bibinfo {author} {\bibfnamefont {R.~J.}\ \bibnamefont
  {Bouwens}}, \bibinfo {author} {\bibfnamefont {I.}~\bibnamefont {Labbé}},
  \bibinfo {author} {\bibfnamefont {G.~D.}\ \bibnamefont {Illingworth}},
  \bibinfo {author} {\bibfnamefont {V.}~\bibnamefont {Gonzalez}}, \ and\
  \bibinfo {author} {\bibfnamefont {P.~A.}\ \bibnamefont {Oesch}},\ }\href@noop
  {} {\  (\bibinfo {year} {2021})},\ \Eprint {http://arxiv.org/abs/2103.16571}
  {arXiv:2103.16571 [astro-ph.GA]} \BibitemShut {NoStop}%
\bibitem [{\citenamefont {Reed}\ \emph {et~al.}(2007)\citenamefont {Reed},
  \citenamefont {Bower}, \citenamefont {Frenk}, \citenamefont {Jenkins},\ and\
  \citenamefont {Theuns}}]{Reed:2006rw}%
  \BibitemOpen
  \bibfield  {author} {\bibinfo {author} {\bibfnamefont {D.}~\bibnamefont
  {Reed}}, \bibinfo {author} {\bibfnamefont {R.}~\bibnamefont {Bower}},
  \bibinfo {author} {\bibfnamefont {C.}~\bibnamefont {Frenk}}, \bibinfo
  {author} {\bibfnamefont {A.}~\bibnamefont {Jenkins}}, \ and\ \bibinfo
  {author} {\bibfnamefont {T.}~\bibnamefont {Theuns}},\ }\href {\doibase
  10.1111/j.1365-2966.2006.11204.x} {\bibfield  {journal} {\bibinfo  {journal}
  {Mon. Not. Roy. Astron. Soc.}\ }\textbf {\bibinfo {volume} {374}},\ \bibinfo
  {pages} {2} (\bibinfo {year} {2007})},\ \Eprint
  {http://arxiv.org/abs/astro-ph/0607150} {arXiv:astro-ph/0607150} \BibitemShut
  {NoStop}%
\end{thebibliography}%

\phantomsection
%%%%%%%%%%%%%%%%%%%%%%%%%%%%%%%%%%%%%%%%%%%%%%%%%%%%%%
\appendix

\vspace{-0.2cm}
\section{Robustness Checks}
\label{app:robustness_checks}
Here we will check how our posteriors for $\sigma_8$, $\Omega_\mathrm{m}$ and $n_\mathrm{s}$ depend on the choice of the halo mass function and the calibration of the dust correction.

\vspace{-0.3cm}
\subsection*{Alternative Halo Mass Function}
\label{app:other_HMF}
\begin{figure}[t!]
    \centering
    \includegraphics[width=\textwidth]{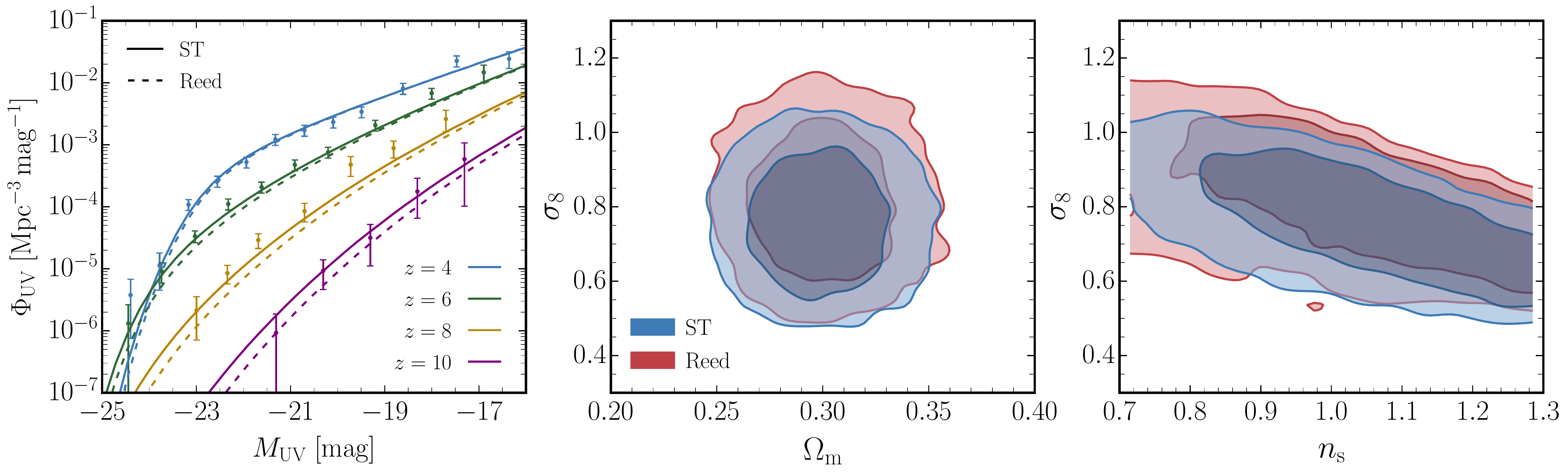}
    \caption{\emph{Left panel}: UV LFs obtained using the Sheth-Tormen HMF (solid lines) and the Reed HMF (dashed lines). Included in this panel are the HST data from~\cite{Oesch_2018, Bouwens_2021}. At high redshifts, the two HMFs give different predictions for the UV LF. We use this difference to assess the impact of the HMF choice on our results. \emph{Middle and Right panels}: Posteriors on cosmological parameters, where the inner (outer) contours indicate the 68\% (95\%) CL.}
    \label{fig:HMF_comparison}
\end{figure}

In our main analysis, we made use of the Sheth-Tormen mass function, see Eq.~\eqref{eq:ST_HMF}. Here we will rerun our analysis with the Reed mass function, given by~\cite{Reed:2006rw}:
\begin{align}
    \label{eq:Reed_HMF}
    f_\mathrm{Rd}(\sigma_{M_\mathrm{h}}) = & A_\mathrm{Rd}\sqrt{\frac{2a_\mathrm{Rd}}{\pi}}\left[1+\left(\frac{\sigma_{M_\mathrm{h}}^2}{a_\mathrm{Rd}\delta_\mathrm{Rd}^2}\right)^{p_\mathrm{Rd}} + 0.2\exp\left(-\frac{\left(\ln\sigma_{M_\mathrm{h}}^{-1}-0.4\right)^2}{2(0.6)^2}\right)\right]\frac{\delta_\mathrm{Rd}}{\sigma_{M_\mathrm{h}}}\exp\left(-\frac{c_\mathrm{Rd}a_\mathrm{Rd}\delta_\mathrm{Rd}^2}{2\sigma_{M_\mathrm{h}}^2}\right)\ ,
\end{align}

where $A_\mathrm{Rd} = 0.3235$, $a_\mathrm{Rd} = 0.707$, $p_\mathrm{Rd} = 0.3$, $c_\mathrm{Rd} = 1.081$ and $\delta_\mathrm{Rd} = 1.686$. The UV LFs are plotted in the left panel of Fig.~\ref{fig:HMF_comparison}, where we see that using the Reed HMF results in a smaller LF, especially at high redshifts. With the current HST data, however, this does not seem to have a significant impact on $\sigma_8$ (middle panel), where the main effect is a small shift in its central value. With regards to the spectral tilt (right panel), the HMF choice has pretty much no influence.

\vspace{-0.4cm}
\subsection*{Alternative Dust Calibration}
\label{app:alternative_dust}

\begin{figure}[t!]
    \centering
    \includegraphics[width=\textwidth]{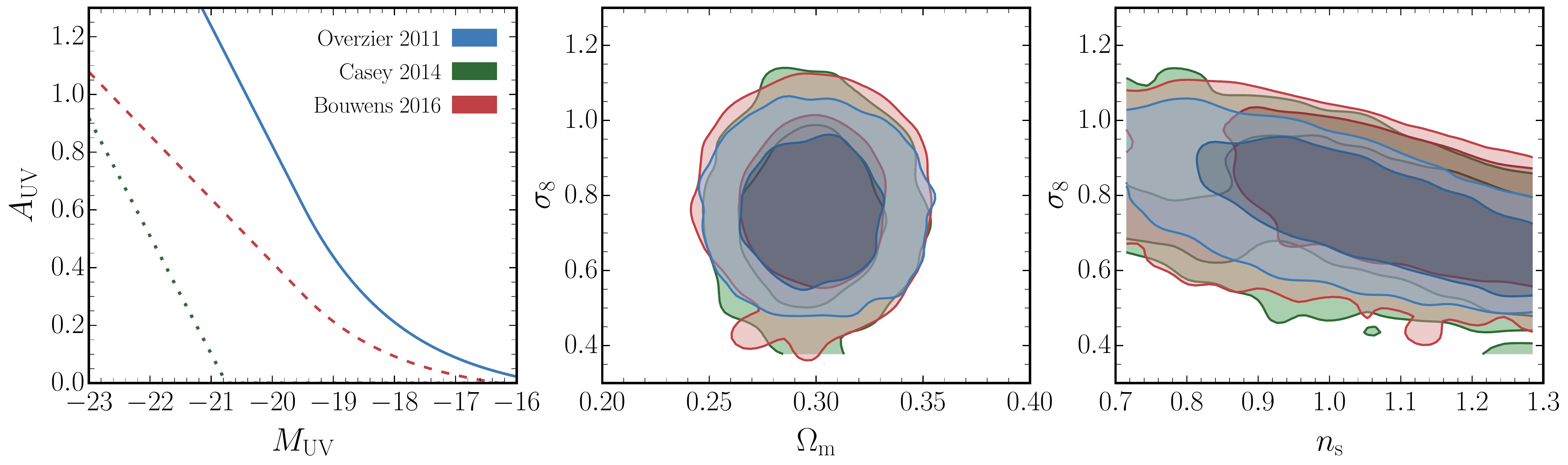}
    \caption{\emph{Left panel}: Dust attenuation parameter at redshift $z = 6$ as a function of absolute UV magnitude using the calibrations of~\cite{Overzier:2010aa} (blue, our fiducial one), \cite{Casey:2014cqa} (green) and~\cite{J_Bouwens_2016} (red). \emph{Middle and Right panels}: Posteriors on cosmological parameters, where the inner (outer) contours indicate the 68\% (95\%) CL.}
    \label{fig:alt_dust}
\end{figure}

The dust attenuation relation in Eq.~\eqref{eq:dust_general} has been calibrated over the years by multiple groups using different instruments and systems. In our main analysis we used the calibration parameters $C_0 = 4.54$ and $C_1 = 2.07$ by~\cite{Overzier:2010aa} (see Eq.~\eqref{eq:dust}), which are obtained from observations of starburst galaxies. Besides this one, we rerun our analysis with the calibrations of~\cite{Casey:2014cqa}  and~\cite{J_Bouwens_2016}. The former is based on observations of dusty star-forming galaxies at $0 < z < 3.5$ and reports $C_0 = 3.36$ and $C_1 = 2.04$, while the latter is based on measurements of the Small Magellanic Cloud, where $C_0 = 2.45$ and $C_1 = 1.1$. The dust attenuation is shown in the left panel of Fig.~\ref{fig:alt_dust} and the posteriors on $\sigma_8$ and $n_\mathrm{s}$ in the middle and right panels, respectively. We find no appreciable difference between the three choices, which is as expected given that the dust attenuation mostly affects the bright side of the UV LF, where Poisson errors are large. 

\section{Results for General Redshift Evolution of Astrophysical Parameters}
\label{app:results_z_general}

\begin{figure}[h!]
    \centering
    \includegraphics[width=0.8\textwidth]{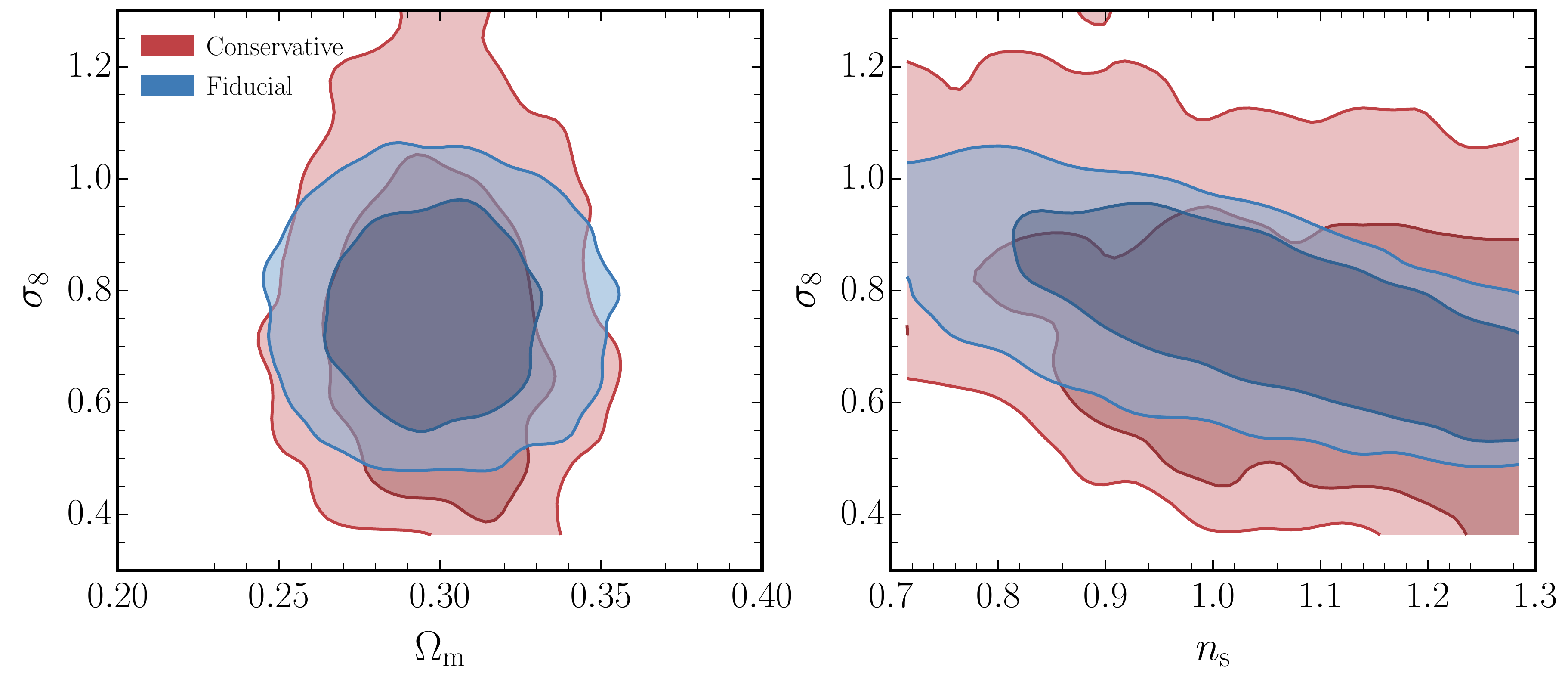}
    \caption{Impact of the redshift evolution parametrisation of the astrophysical parameters on measurements of cosmological parameters. The blue colour indicates the fiducial case (Eq.~\eqref{eq:fiducial_astro}), while the red indicates the conservative parametrisation (Eq.~\eqref{eq:conservative_astro}). The inner (outer) contours correspond to the 68\% (95\%) confidence levels. See Sec.~\ref{subsec:z_evol_astro} and the text surrounding this figure for more details.}
    \label{fig:sigmas_conservative}
\end{figure}

\enlargethispage{0.5cm}
In this appendix, we will show our results for $\sigma_8$ assuming the conservative approach in parameterising the redshift evolution of the astrophysical parameters, see Eq.~\eqref{eq:conservative_astro}. Here, we assign an independent set of astrophysical parameters at each redshift slice in the range $z = 4-10$. From Figs.~\ref{fig:UVLF_param_depend} and~\ref{fig:v2_posteriors}, it is evident that many of the astrophysical parameters correlate with $\sigma_8$. The conservative approach therefore simply increases the number of free parameters that can (strongly) correlate with $\sigma_8$, hence relaxing the strength of our measurements. In Fig.~\ref{fig:sigmas_conservative}, we show the posteriors in this scenario. Marginalising over all parameters, we find a measurements of $\sigma_8 = 0.74^{+0.14}_{-0.27}$ at 68\% CL. This is roughly a factor of 2 increase in the lower error as compared to the result in the fiducial case, see Eq.~\eqref{eq:sigma8_fiducial}. 
Nevertheless, we are still able to measure $\sigma_8$ in this case.
That is because the shape of the HMF (and its redshift evolution) are different than that of the halo-galaxy connection.
For all cosmologies considered, the HMF follows an exponentially decaying function in the halo mass range of interest, whereas 
data from both observations and simulations suggest the halo-galaxy connection follows a smoother (power law-like) function.
Thus, the cosmological and astrophysical parameters will impact the UV LF in a different way, as shown in the lower-right panel of Fig.~\ref{fig:UVLF_param_depend}. If a Gaussian prior on $n_\mathrm{s}$ from Planck CMB observations~\cite{Aghanim:2018eyx} is included in the analysis, then the conservative case gives $\sigma_8 = 0.84^{+0.16}_{-0.21}$ at 68\% CL, a modest improvement.

\section{UV LF Data, Best-Fits and Scatter}
\label{app:Hubble_TNG_data}

\begin{figure}[h!]
    \centering
    \includegraphics[width=0.73\textwidth]{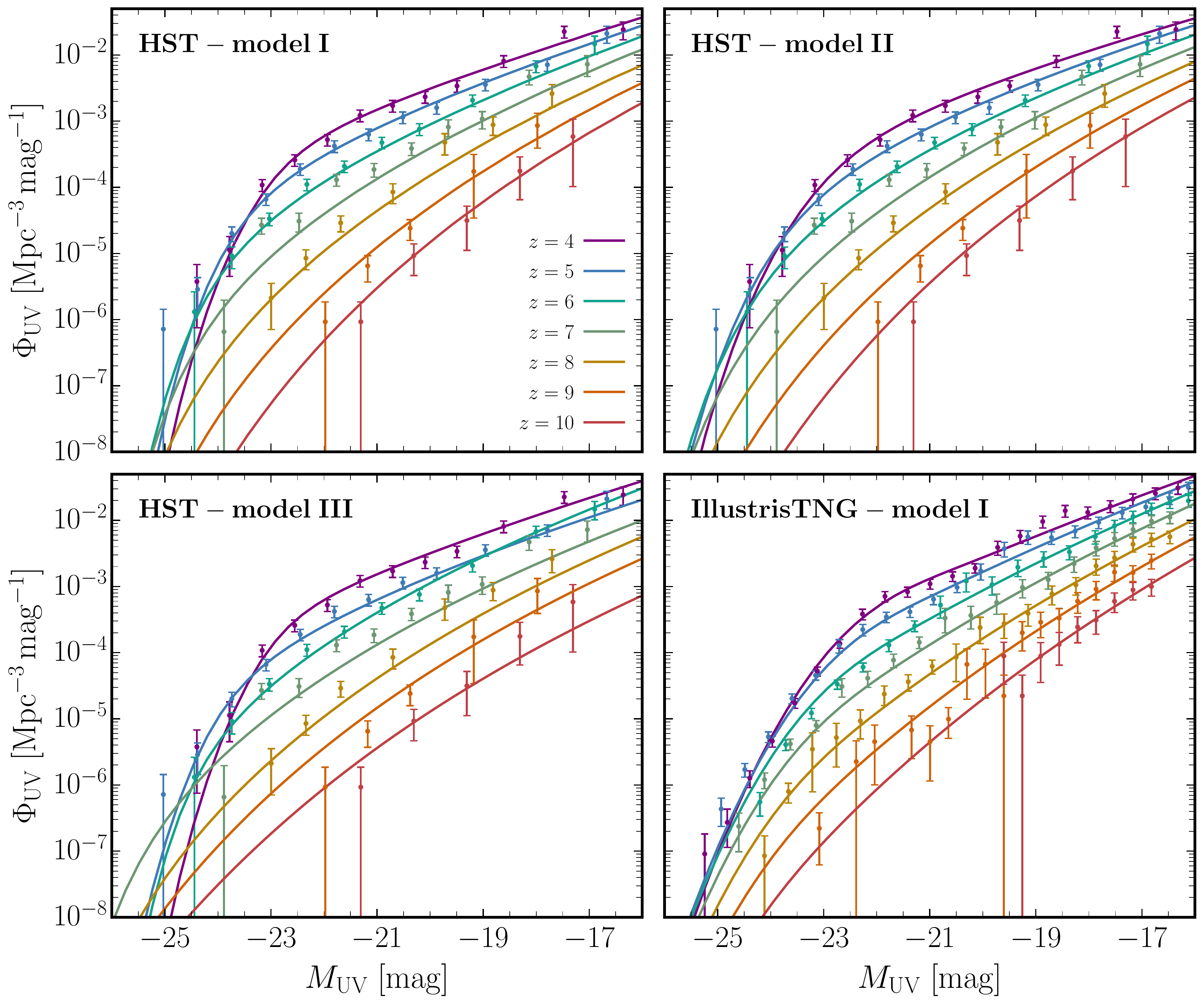}
    \caption{UV luminosity function data from the HST and IllustrisTNG suite of hydrodynamical simulations. The solid lines represent the best-fits obtained using our models I$-$III for the halo-galaxy connection, see Secs.~\ref{subsubsec:Mhalo_MUV_connection} and~\ref{sec:astro_modelling}. Corrections due to cosmic variance, dust attenuation and the Alcock-Paczy\'{n}ski effect are accounted for here.}
    \label{fig:TNG_data_bestfit}
\end{figure}

We show the HST data and IllustrisTNG mock data used in our analysis in Fig.~\ref{fig:TNG_data_bestfit} (see also Sec.~\ref{sec:data} for a description). Included in this figure are the best-fit curves within our models I$-$III (Secs.~\ref{subsubsec:Mhalo_MUV_connection} and~\ref{sec:astro_modelling}) at the corresponding redshifts. Corrections due to dust attenuation, cosmic variance and the Alcock-Paczy\'{n}ski effect are accounted for in the data here.
\enlargethispage{0.8cm}

\begin{figure}[h!]
    \centering
    \includegraphics[width=0.51\textwidth]{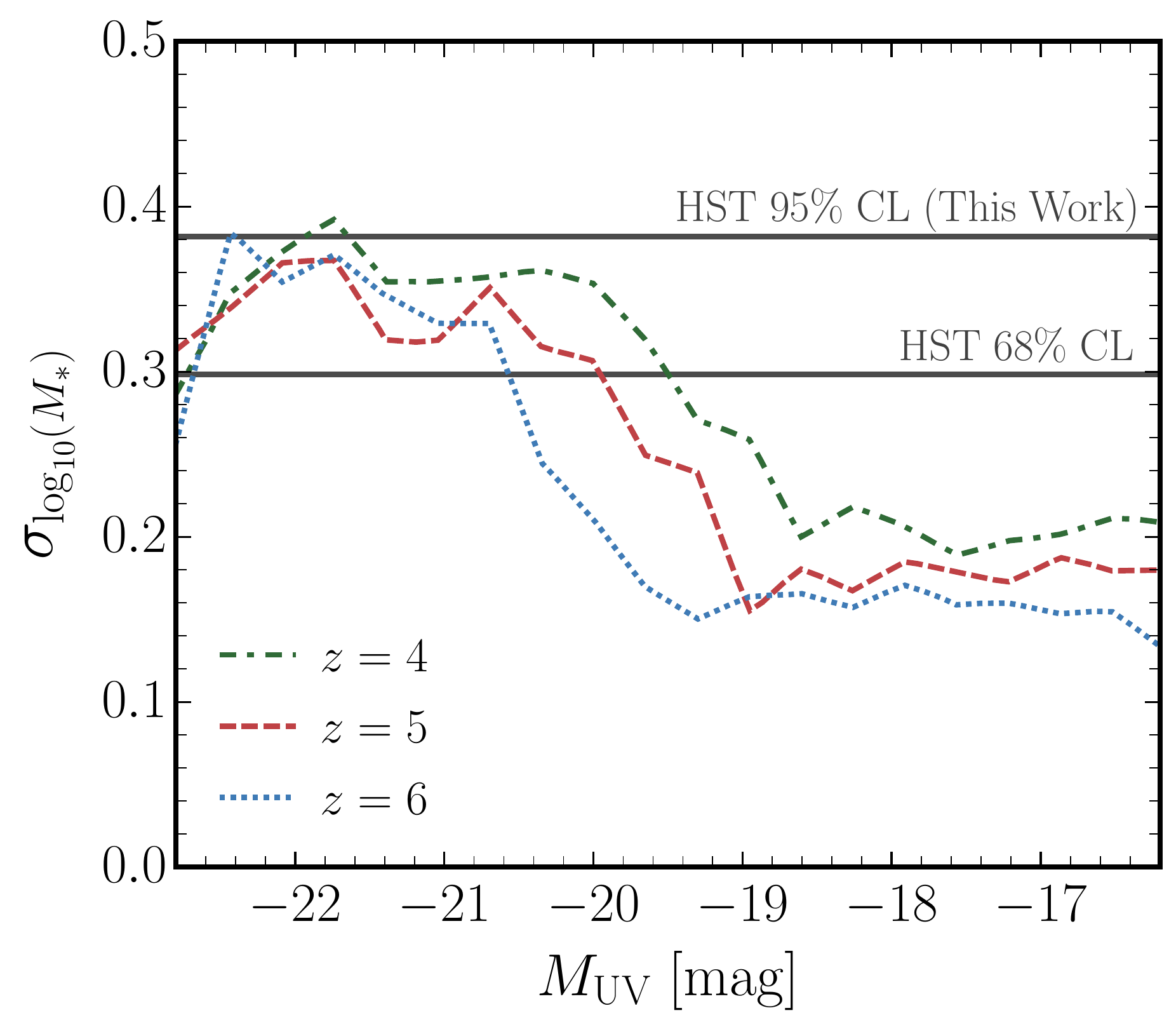}
    \caption{Scatter in the $M_*-M_\mathrm{UV}$ relation found in the IllustrisTNG simulations at $z=4-6$ (coloured lines) and in our analysis using HST data together with our model II (horizontal, black lines).}
    \label{fig:TNG_scatter}
\end{figure}

We also compare with the scatter in the $M_*-M_\mathrm{UV}$ relation found in the IllustrisTNG simulations. To this end, we use our model II (see Sec.~\ref{sec:astro_modelling}) to obtain the parameter $\sigma_{M_\mathrm{UV}}$ that accounts for stochasticity in our simulations, see Eq.~\eqref{eq:gaussian_MUV}. We show the comparison in Fig.~\ref{fig:TNG_scatter}. A caveat that should be mentioned here is that the parameter $\sigma_{M_\mathrm{UV}}$ includes scatter in both the $M_\mathrm{h}-M_*$ and the $M_*-M_\mathrm{UV}$ relation.

\end{document}